\documentclass{article}

\usepackage[eandd,final]{neurips_2026}

\usepackage{iftex}
\ifPDFTeX
  \usepackage[utf8]{inputenc}
\fi
\usepackage[T1]{fontenc}
\usepackage[pdfversion=1.7]{hyperref}
\usepackage{url}
\usepackage{booktabs}
\usepackage{amsfonts}
\usepackage{amsmath}
\usepackage{amssymb}
\usepackage{amsthm}
\usepackage{mathtools}
\usepackage{nicefrac}
\usepackage{microtype}
\usepackage{xcolor}
\usepackage{graphicx}
\usepackage{caption}
\usepackage{subcaption}
\usepackage{tcolorbox}
\usepackage{wrapfig}
\usepackage{multirow}
\usepackage{enumitem}
\usepackage{latexsym}
\usepackage{algorithm}
\usepackage{algorithmic}
\usepackage{pgfplots}
\pgfplotsset{compat=1.18}
\usepackage{tabularx}
\newcolumntype{Y}{>{\raggedright\arraybackslash}X}
\usepackage{makecell}
\usepackage{placeins}
\usepackage[capitalize,noabbrev]{cleveref}
\hypersetup{
  pdftitle={MHRC-Bench: A Multilingual Hardware Repository-Level Code Completion Benchmark},
  pdfauthor={Qingyun Zou, Jiahao Cui, Nuo Chen, Bingsheng He, Weng-Fai Wong},
}

\theoremstyle{plain}

\theoremstyle{definition}

\theoremstyle{remark}

\title{MHRC-Bench: A Multilingual Hardware Repository-Level Code Completion Benchmark}

\author{%
  Qingyun Zou \\
  National University of Singapore\\
  \texttt{qingyunzou@u.nus.edu} \\
  \And
  Jiahao Cui\thanks{Corresponding author.} \\
  National University of Singapore\\
  \texttt{cuijiahao@u.nus.edu} \\
  \And
  Nuo Chen \\
  National University of Singapore\\
  \texttt{nuochen@u.nus.edu} \\
  \And
  Bingsheng He \\
  National University of Singapore\\
  \texttt{dcsheb@nus.edu.sg} \\
  \And
  Weng-Fai Wong \\
  National University of Singapore\\
  \texttt{wongwf@nus.edu.sg} \\
}

\begin{document}
\raggedbottom

\maketitle

\begin{abstract}
Large language models (LLMs) have achieved strong performance on code completion in general-purpose programming languages, but existing repository-level benchmarks focus almost exclusively on software code and largely overlook hardware description languages (HDLs). We present \textbf{MHRC-Bench}, consisting of \textbf{MHRC-Bench-Train} and \textbf{MHRC-Bench-Eval}, the first benchmark for multilingual hardware code completion at the repository level with functional evaluation. MHRC-Bench covers three major hardware design coding styles and annotates each completion target with code-structure-level and hardware-oriented semantic labels derived from concrete syntax tree (CST) analysis. Post-training billion-scale code LLMs on MHRC-Bench-Train enables smaller open-weight models to outperform much larger general-purpose models under matched context windows. Beyond aggregate accuracy, our analysis surfaces four findings that distinguish hardware from software code completion: \emph{(i)}~performance varies sharply across HDLs, with high-level synthesis (HLS) easiest and VHDL hardest in most settings across functional pass rate, exact match (EM), and edit similarity (ES); \emph{(ii)}~hardware accuracy is non-monotonic in code-structure depth, in contrast with the monotonic depth--accuracy curve known for software; \emph{(iii)}~accuracy plateaus once the input context exceeds 2{,}048 tokens, while software completion continues to benefit from longer context; and \emph{(iv)}~certain repository-level retrievers tuned on software underperform a no-retrieval baseline on HDL repositories because their structural and dependency assumptions are violated by hardware code.
\end{abstract}

\section{Introduction}

Recent advancements in LLMs have led to significant progress in code-related tasks such as code understanding, generation, and completion \citep{liu2023repobench,nam2024using,liu2024exploring,pan2024lost, raihan2025mhumaneval,yang2024execrepobench}.

\begin{figure*}[!t]
  \centering
  \includegraphics[width=0.9\textwidth]{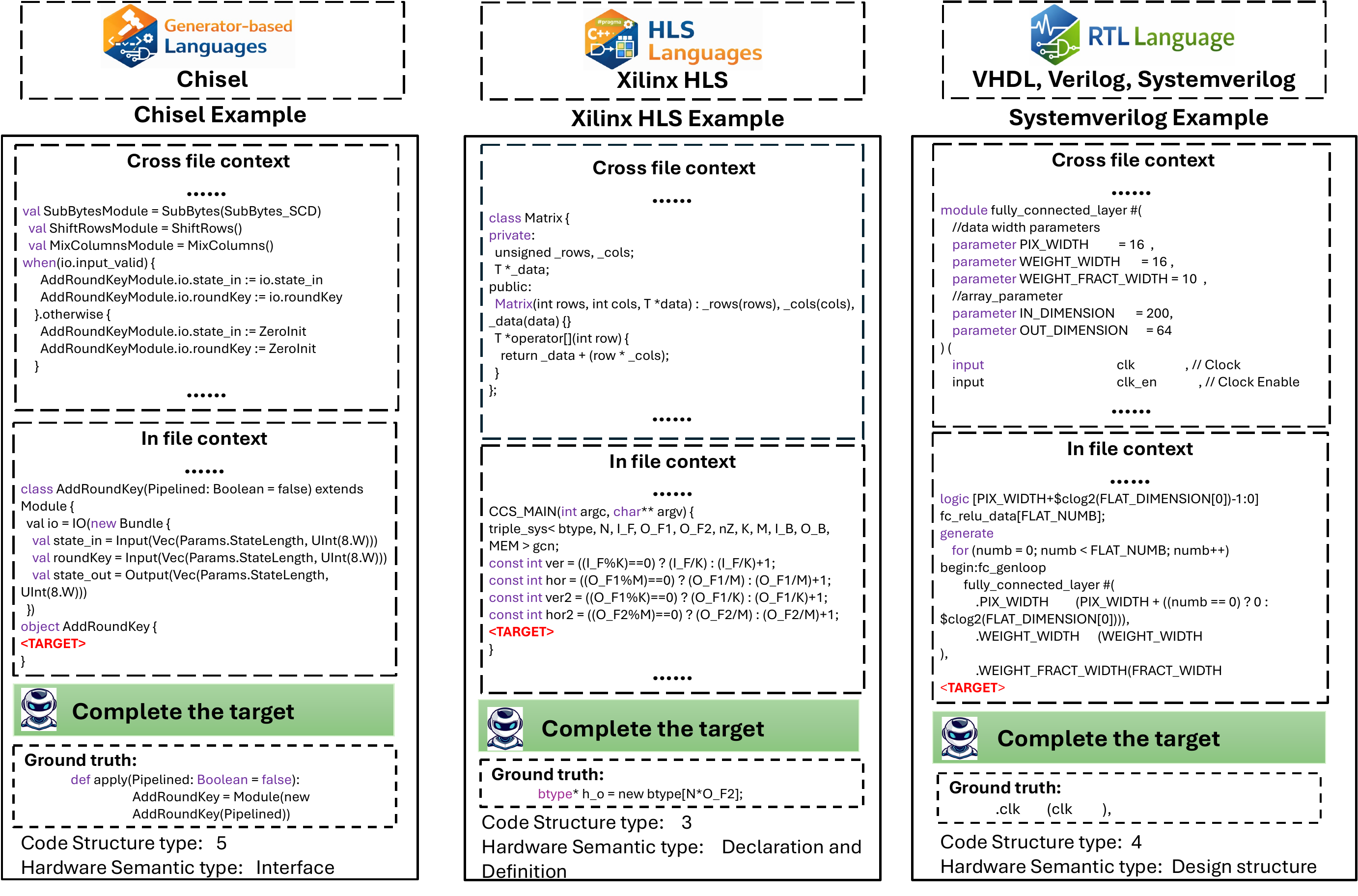}
  \caption{\textbf{Overview of our hardware code completion benchmark across multiple hardware description paradigms}. We present three representative examples from domain-specific language (DSL) (e.g., Chisel), HLS languages (e.g., Xilinx HLS), and Register-Transfer Level (RTL) design languages (SystemVerilog). For each example, both in-file and cross-file contexts are provided, with the completion target explicitly marked. Models are required to predict the correct completion given the provided contexts, where ``\texttt{\textless TARGET\textgreater}'' indicates the completion position. More examples are introduced in Appendix \ref{app:gemini_failure_example}.}
  \label{fig:three_types}
\end{figure*}

\begin{samepage}
These models, trained on vast datasets encompassing both natural language and source code, have demonstrated remarkable proficiency in modeling programming languages as structured sequences and predicting logical code continuations \citep{liang2024repofuse, li2025coderag, deepseek-coder, hui2024qwen2,liu2024graphcoder, abedu2025repochat, hua2025researchcodebench}. Consequently, LLM-based code completion has emerged as a valuable tool in real-world software development environments.
\par
\end{samepage}

More recently, research efforts have shifted toward repository-level code completion, where models are required to leverage cross-file information such as imports, symbol definitions, and project structure. \citet{wang2024rlcoder} train a dedicated retriever to identify and retrieve relevant code fragments that can be incorporated into LLM-based code completion. In a complementary direction, \citet{deng2025enhancing} equip LLMs with external tool APIs to more effectively utilize repository-level context during the code completion process. Despite their success in software code, LLM performance on hardware code remains unclear, as hardware differs fundamentally in execution semantics, structural parallelism, and cross-file dependency patterns. In particular, hardware code completion requires understanding static structure, signal connectivity, and design intent rather than purely sequential execution. We introduce \textit{MHRC-Bench}, a multilingual repository-level benchmark with both \textit{MHRC-Bench-Train} and \textit{MHRC-Bench-Eval}, designed to expose structural and semantic failure modes of LLMs in hardware code completion.

Using MHRC-Bench-Train to fine-tune code LLMs, we observe substantial gains across complementary evaluation signals, including exact/edit similarity and functional pass rate. Billion-scale post-trained models such as Qwen2.5-Coder-7B and Qwen2.5-Coder-14B substantially improve over their pretrained counterparts, whose exact-match scores are near zero in preceding-context-only settings, and can outperform much larger general-purpose models under matched context windows. This demonstrates that MHRC-Bench-Train effectively enhances the hardware code completion capabilities of smaller-scale code LLMs.

Beyond the gains from post-training, our analysis surfaces four observations that distinguish hardware from software code completion. \emph{(i)~Strong language dependence}: In most settings, HLS is easiest and VHDL hardest across both functional pass rate and EM/ES; VHDL is the only language where full-context fill-in-the-middle (FIM) does not outperform preceding-context-only completion. \emph{(ii)~Non-monotonic structural sensitivity}: unlike software code, where deeper CST nodes monotonically correlate with higher LLM accuracy~\cite{liu2025m2rc}, hardware accuracy varies non-monotonically with structural depth. \emph{(iii)~Context-length plateau}: extending the input window from 2{,}048 to 4{,}096 tokens yields diminishing or negative returns on hardware code, whereas software code completion continues to benefit. \emph{(iv)~Software-tuned retrieval breaks on hardware}: repository-level retrievers designed for software (e.g., GraphCoder) underperform a no-retrieval baseline on HDL, because their structural and dependency assumptions are violated by hardware code (Section~\ref{subsec:graph_coder}).

As shown in Figure~\ref{fig:three_types}, MHRC-Bench covers three main categories of hardware programming languages. To enable fine-grained analysis of LLM performance on hardware code completion, MHRC-Bench-Eval introduces two annotation levels: code-structure and hardware semantics. At the code-structural level, source files are parsed into CSTs using tree-sitter, and completion targets are assigned to depth-based buckets according to the enclosing CST node, capturing structural complexity. At the hardware-semantic level, code is categorized into nine subcategories, enabling detailed evaluation across diverse hardware semantics.

The benchmark and related code are publicly available\footnote{Dataset: \url{https://huggingface.co/datasets/Zanehpn/MHRC-Bench}. Code: \url{https://github.com/zanehpn/MHRC-Bench}.}.
Our contributions are: (1) \textbf{MHRC-Bench-Eval}, a multilingual repository-level HDL completion benchmark with structural/semantic labels and repository-level functional evaluation; (2) \textbf{MHRC-Bench-Train}, a training split that improves open-weight HDL completion models; and (3) an empirical analysis showing language-dependent difficulty, non-monotonic structural sensitivity, context-length saturation, and HDL-specific retrieval failures.

\begin{table}[t]
\centering
\scriptsize
\captionsetup{font=small, labelfont=small}
\caption{Comparison with hardware code benchmark. FG indicates fine-grained annotation. Train means if it contains training dataset. NL means natural language.}
\label{tab:dataset_comparison}
\renewcommand{\arraystretch}{1.1}
\setlength{\tabcolsep}{2pt}
\begin{tabular}{l c c c c c c c c r}
\toprule
\textbf{Benchmark}
& \textbf{Level}
& \textbf{Task Type}
& \textbf{Input}
& \textbf{Languages}
& \textbf{FG}
& \textbf{Train}
& \textbf{Output}
& \textbf{Data types}
& \textbf{Samples} \\
\midrule
VerilogEval
& Module
& Code Gen
& NL Spec
& RTL
& \texttimes
& \checkmark
& Module Code
& \makecell[c]{Synthetic and\\Real-world}
&  8,502 \\
RTLLM
& Module
& Code Gen
& NL Spec
& RTL
& \texttimes
& \texttimes
& Module Code
& Real-world
& 30 \\
HLS-Eval
& Module
& Code Gen
& \makecell[c]{NL Spec\\or C Code}
& HLS
& \texttimes
& \texttimes
& Module Code
& Real-world
& 94 \\
HDLEval
& Module
& Code Gen
& NL Spec
& RTL, DSL
& \texttimes
& \texttimes
& Module Code
& Real-world
& 461 \\
\midrule
RTL-Repo
& Repository
& Code Completion
& \makecell[c]{Repository\\Code}
& RTL
& \texttimes
& \checkmark
& Single-line
& Real-world
& 4098 \\
\textbf{Ours}
& \textbf{Repository}
& \textbf{Code Completion}
& \textbf{\makecell[c]{Repository\\Code}}
& \textbf{RTL,HLS,DSL}
& \textbf{\checkmark}
& \textbf{\checkmark}
& \textbf{Single \& Multi-line}
& \textbf{Real-world}
& \textbf{47,200} \\
\bottomrule
\end{tabular}
\end{table}

\section{Related Work}

Repository-level code completion has been studied extensively for software, with retrieval-aug\-mented generation (RAG) methods such as RepoCoder~\citep{zhang2023repocoder}, GraphCoder~\citep{liu2024graphcoder}, and RLCoder~\citep{wang2024rlcoder}, and structural-sensitivity analyses such as M2RC~\citep{liu2025m2rc}; however, this line of work has not transferred to hardware code, which differs fundamentally in execution semantics, structural parallelism, and cross-file dependency patterns.

Hardware code-generation benchmarks cover several tasks: VerilogEval~\citep{liu2023verilogeval} and RTLLM~\citep{lu2024rtllm} evaluate module-level Verilog generation, while HDLEval~\citep{kashanaki2024hdleval} covers multiple HDL dialects. HLS-Eval~\citep{abi2025hls} evaluates HLS generation, HLStrans~\citep{zou2025hlstrans} provides paired C-to-HLS programs, and HLS-Seek~\citep{zou2026hlsseek} optimizes HLS code generation for latency and resource usage. Reward-weighted on-policy distillation~\citep{zou2026rwopd} uses property-equivalence verification for natural-language-to-SystemVerilog-assertion generation. Phoenix-bench~\citep{zou2026phoenix} evaluates agents on repository-level hardware issue resolution. For repository-level completion, RTL-Repo~\citep{allam2024rtl} remains limited to one HDL and primarily single-line targets, leaving multilingual completion and fine-grained performance analysis underexplored.

\section{Dataset}
\subsection{Data Collection}
We collect publicly accessible hardware design repositories from GitHub, created after 2000 and with more than five GitHub stars. Source code retains its original license terms, including permissive and copyleft licenses; the released repository inventory records license notices and unresolved cases. To avoid data leakage, we remove files located in common vendor or dependency directories and deduplicate exact repository copies. The resulting corpus contains 47,200 source files written in Chisel, SystemVerilog, Verilog, HLS C/C++, and VHDL, drawn from 583 distinct repositories.

\subsection{Dataset Comparison}

\textbf{Language scope.} As shown in Table \ref{tab:dataset_comparison} and Figure \ref{fig:three_types}, we select Verilog/SystemVerilog, VHDL, High Level Synthesis, and Chisel to construct our dataset, as they cover three complementary categories:

\begin{itemize}[itemsep=2pt, topsep=2pt, parsep=0pt, partopsep=0pt, leftmargin=1.5em]
\item \textit{RTL-based HDLs.} They provide cycle-accurate and explicitly timed hardware descriptions including Verilog/SystemVerilog and VHDL. SystemVerilog and Verilog share similar grammatical structures, whereas VHDL differs significantly. Therefore, we group Verilog and SystemVerilog together (V/SV);
\item \textit{HLS languages.} These describe hardware behavior using C/C++-like abstractions, often augmented with synthesis directives. Among them, Xilinx HLS C/C++ is widely used, so we include it in our benchmark.
\item \textit{DSLs.} DSLs enable parameterized and programmable generation of RTL designs. Chisel is a representative example, and we include it in our benchmark for its growing adoption.
\end{itemize}

As shown in Table~\ref{tab:dataset_comparison}, MHRC-Bench contains the largest number of data samples, supports multilingual repository-level hardware code, and provides fine-grained annotations, enabling comprehensive evaluation of LLMs in hardware code completion tasks compared to existing hardware benchmarks.

\begin{figure}[!t]
  \centering
  \begin{minipage}[c]{0.52\linewidth}
    \centering
    \includegraphics[width=\linewidth]{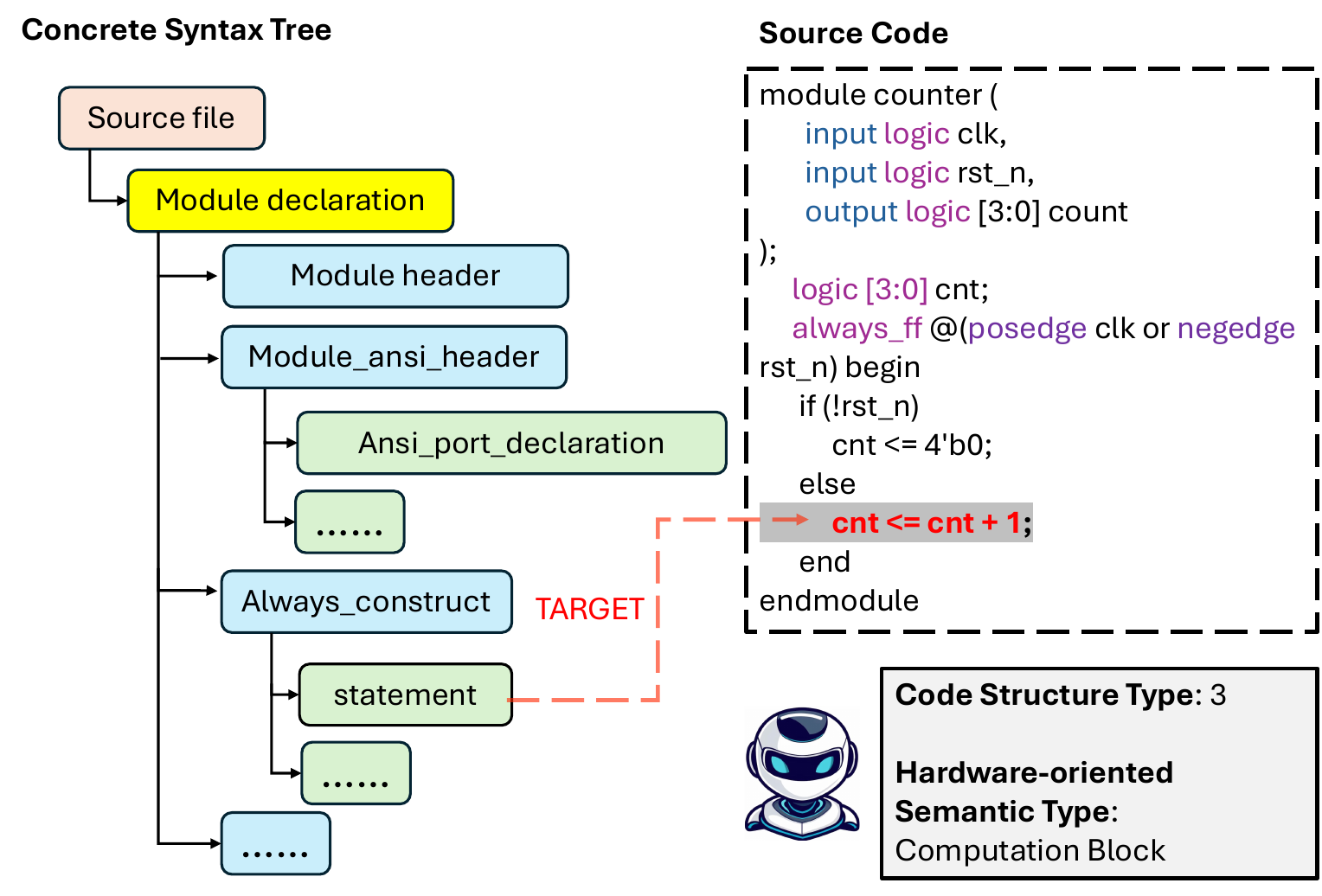}
    \caption{Illustration of fine-grained completion target annotation in our benchmark. Completion targets are identified by aligning source code positions with Tree-sitter CST nodes and labeled with structure depth and hardware-oriented semantic categories.}
    \label{fig:code_structre}
  \end{minipage}\hfill
  \begin{minipage}[c]{0.46\linewidth}
    \centering
    \captionof{table}{Dataset statistics of MHRC-Bench. Val denotes validation.}
    \label{tab:dataset_statistics}
    \small
    \setlength{\tabcolsep}{3pt}
    \renewcommand{\arraystretch}{1.5}
    \begin{tabular}{lcccc}
      \toprule
      \textbf{Metric} & \textbf{Chisel} & \textbf{V/SV} & \textbf{VHDL} & \textbf{HLS} \\
      \midrule
      Train Repos      & 24    & 198    & 172    & 22    \\
      Val Repos        & 4     & 6      & 7      & 3     \\
      Test Repos       & 20    & 43     & 59     & 25    \\
      Train Files      & 2,573 & 25,586 & 13,039 & 3,766 \\
      Val Files        & 119   & 107    & 104    & 123   \\
      Test Files       & 323   & 508    & 496    & 456   \\
      Cross-file Dep.  & 3.12  & 4.02   & 4.34   & 5.85  \\
      \bottomrule
    \end{tabular}
  \end{minipage}
\end{figure}

\subsection{Dataset Quality Control}
The selection of completion targets is critical to the quality and reliability of a code completion benchmark. Accordingly, we adopt the following rules to identify and select target code locations.

\noindent\textbf{Syntactically complete.} Prior work \cite{ding2023crosscodeeval} often selects random character spans as completion targets, which may break syntactic integrity. In contrast, recent studies \citep{liu2025m2rc} show that developers expect LLMs to generate syntactically complete code units (e.g., full lines or blocks). MHRC-Bench-Eval uses tree-sitter\footnote{\url{https://github.com/tree-sitter/tree-sitter}} to parse each file into a concrete syntax tree and randomly selects a syntactic node as the completion target, using its exact source span as the ground-truth output. \noindent\textbf{Data independence and inference difficulty.} To ensure data independence and nontrivial inference, we enforce four rules: (1) repositories and files in the training, validation, and test sets are mutually exclusive; (2) at least 40\% of completion targets contain two to five lines, and test files are drawn from repositories updated after 2022; (3) targets cannot consist only of whitespace, debug messages, or comments; and (4) each file contributes exactly one target position. These rules produce 47,200 samples from 47,200 source files.
Detailed split and leakage-control procedures are provided in Appendix~\ref{app:data_splits}.

\begin{table*}[t]
\centering
\scriptsize
\caption{EM \% and ES \% scores across four hardware description languages. Full context refers to the use of both preceding code and the succeeding code surrounding the target code segment, whereas preceding context-only uses only the preceding code as context.}
\setlength{\tabcolsep}{3pt}
\resizebox{\textwidth}{!}{%
\begin{tabular}{llcccccccccccccccc}
\toprule 
\multirow{2}{*}{\textbf{Model}} & \multirow{2}{*}{\makecell{\textbf{Retrieval}\\\textbf{Method}}} 
& \multicolumn{8}{c}{\textbf{Preceding context}} & \multicolumn{8}{c}{\textbf{Full context}} \\
\cmidrule(lr){3-10} \cmidrule(lr){11-18} 
& & \multicolumn{2}{c}{Chisel} & \multicolumn{2}{c}{V/SV} & \multicolumn{2}{c}{VHDL} & \multicolumn{2}{c}{HLS} 
  & \multicolumn{2}{c}{Chisel} & \multicolumn{2}{c}{V/SV} & \multicolumn{2}{c}{VHDL} & \multicolumn{2}{c}{HLS} \\
\cmidrule(lr){3-4} \cmidrule(lr){5-6} \cmidrule(lr){7-8} \cmidrule(lr){9-10} 
\cmidrule(lr){11-12} \cmidrule(lr){13-14} \cmidrule(lr){15-16} \cmidrule(lr){17-18} 
& & EM & ES & EM & ES & EM & ES & EM & ES & EM & ES & EM & ES & EM & ES & EM & ES \\
\midrule
\multirow{4}{*}{GPT-5} 
& No RAG      & 15.2 & 51.6 & 21.9 & 55.4 & 20.8 & 58.5 & 12.1 & 46.7 & 24.8 & 56.7 & 34.4 & 60.9 & 15.3 & 52.0 & 31.4 & 63.7 \\
& RepoCoder   & \textbf{18.3} & 54.1 & 22.4 & \textbf{56.1} & 22.4 & 60.3 & 14.7 & 47.2 & 27.6 & 59.6 & 35.4 & 61.3 & 16.7 & 54.1 & 34.4 & 63.9 \\
& GraphCoder  & 11.1 & 17.3 & 14.4 & 35.4 & 7.9 & 33.1 & 16.2 & 25.2 & 13.9 & 44.2 & 15.7 & 40.1 & 7.5 & 35.7 & 25.0 & 49.5 \\
& RLCoder     & 18.0 & \textbf{55.3} & \textbf{23.8} & 55.9 & \textbf{26.6} & \textbf{61.4} & \textbf{17.3} & \textbf{56.2} & \textbf{27.9} & \textbf{60.2} & \textbf{36.2} & \textbf{61.7} & \textbf{17.3} & \textbf{55.3} & \textbf{36.8} & \textbf{72.9} \\
\midrule
\multirow{4}{*}{Gemini 2.5 Pro} 
& No RAG      & 11.1 & 56.2 & 28.9 & 61.4 & 23.4 & 61.8 & 18.0 & 52.0 & 32.2 & 60.4 & 31.7 & \textbf{60.4} & 14.9 & 51.3 & 33.3 & 64.3 \\
& RepoCoder   & 11.5 & 58.3 & \textbf{30.9} & 60.8 & 24.8 & 63.6 & 23.0 & 53.2 & 33.7 & 61.7 & 34.8 & 59.6 & 17.9 & 52.3 & 39.5 & 65.3 \\
& GraphCoder  & 2.5 & 17.2 & 2.6 & 26.5 & 3.0 & 25.8 & 5.3 & 17.7 & 13.3 & 50.8 & 22.2 & 50.5 & 9.1 & 45.0 & 29.2 & 57.8 \\
& RLCoder     & \textbf{12.7} & \textbf{59.2} & 30.7 & \textbf{62.0} & \textbf{25.8} & \textbf{64.3} & \textbf{24.1} & \textbf{55.4} & \textbf{34.7} & \textbf{63.2} & \textbf{35.2} & 60.2 & \textbf{19.2} & \textbf{53.7} & \textbf{41.2} & \textbf{66.9} \\
\midrule
\multirow{4}{*}{Grok4} 
& No RAG      & 10.2 & 44.3 & 15.6 & 49.0 & 10.5 & 46.1 & 9.0 & 40.3 & 22.3 & 53.7 & 31.5 & 56.8 & 14.3 & 47.8 & 31.6 & 60.5 \\
& RepoCoder   & 11.1 & 45.6 & 16.5 & 49.0 & 11.7 & 47.9 & 12.1 & 40.8 & 23.8 & 54.7 & 32.1 & 56.1 & 15.7 & 49.3 & 32.2 & 61.1 \\
& GraphCoder  & 2.2 & 24.1 & 6.5 & 23.4 & 0.6 & 24.3 & 6.4 & 23.4 & 12.7 & 35.4 & 16.5 & 38.1 & 8.1 & 35.3 & 18.0 & 38.8 \\
& RLCoder     & \textbf{13.6} & \textbf{46.5} & \textbf{17.3} & \textbf{50.7} & \textbf{13.1} & \textbf{48.4} & \textbf{12.5} & \textbf{42.4} & \textbf{25.7} & \textbf{55.2} & \textbf{33.5} & \textbf{57.2} & \textbf{16.7} & \textbf{51.2} & \textbf{33.3} & \textbf{62.2} \\
\midrule
\multirow{4}{*}{Deepseek V3.2} 
& No RAG      & 10.2 & 47.0 & 22.8 & 56.3 & 16.9 & 53.8 & 9.9 & 42.4 & 22.6 & 51.2 & 28.7 & \textbf{54.1} & 13.3 & 43.0 & 30.3 & 57.3 \\
& RepoCoder   & 10.5 & 48.3 & 23.4 & 57.2 & 17.5 & 54.6 & 11.2 & 43.3 & 23.2 & 52.1 & 28.3 & 52.6 & \textbf{14.1} & 42.8 & 33.3 & 57.9 \\
& GraphCoder  & 0.9 & 24.5 &  0.2 & 39.7 & 0.6 & 32.8 & 0.7 & 28.6 & 9.3 & 39.5 & 15.2 & 39.1 & 9.9 & 39.1 & 22.8 & 47.9 \\
& RLCoder     & \textbf{12.4} & \textbf{48.9} & \textbf{24.2} & \textbf{59.2} & \textbf{19.4} & \textbf{55.3} & \textbf{13.2} & \textbf{44.0} & \textbf{24.1} & \textbf{53.1} & \textbf{29.3} & 45.1 & \textbf{14.1} & \textbf{43.6} & \textbf{34.2} & \textbf{59.2} \\
\bottomrule
\end{tabular}}
\label{tab:retrieval_methods_results}
\end{table*}

\subsection{Dataset Statistics}
\textbf{Training, Validation, and Test Dataset Splits.} To prevent data leakage, the training, validation, and test sets are constructed from disjoint repositories, ensuring that no repository appears in more than one split. This repository-level separation avoids sharing context across splits and enables a fair evaluation of generalization. The sizes of the training, validation, and test sets are shown in Table \ref{tab:dataset_statistics}. \noindent\textbf{Cross-file Dependencies.} A key distinction between repository-level code generation and single-file code generation lies in the presence of cross-file dependencies. To quantify this structural complexity, we measure the average number of cross-file dependencies per task across different hardware description languages, as shown in Table \ref{tab:dataset_statistics}. The results indicate that the average dependency count is consistently greater than one, demonstrating that resolving references across multiple files is common and essential. This highlights the importance of modeling cross-file context for realistic hardware code completion tasks. The procedure for computing cross-file dependencies is described in Appendix~\ref{app:cross-file dependency}.

\FloatBarrier

\subsection{Dataset Categories}
\label{text:dataset_types}
To support fine-grained analysis of LLM behavior, we annotate each completion target at two levels: code structure and hardware semantics, as shown in Figure~\ref{fig:code_structre}.

\noindent\textbf{(1) Code structural level.} We first parse each source file into a CST and select syntactically complete CST nodes as candidate completion targets. For each selected node, we record its tree depth, which reflects the target's structural granularity within the program. We then divide targets into $M$ depth buckets; MHRC-Bench-Eval uses $M=5$. \noindent\textbf{(2) Hardware semantic level.} We categorize target regions according to hardware design and verification semantics. Unlike software programs, where execution is typically sequential, hardware code describes parallel circuits with explicit structural, temporal, and interface semantics. This motivates a hardware-specific categorization of completion targets.
For design code, we categorize targets into six classes: \textit{Design Structure, Declaration and Definition, Storage Block, Computation Block, Control Flow Block, and Interface}, which capture hardware hierarchy, state storage, data-path computation, control logic, and communication interfaces. We also model verification code, which is critical in hardware development. Verification targets are divided into \textit{Property and Assertion Specification, Testbench Stimulus and Environment Modeling, and Monitoring and Checking Logic}, characterizing how correctness properties are specified and how design behavior is observed through simulation or formal verification. The nine semantic categories are assigned by expert-defined, deterministic pattern-matching rules applied to each code snippet (Table~\ref{tab:hardware_semantic_categories}). For each HDL, a priority-ordered decision tree maps syntactic patterns (keywords, declared types, structural roles) to a single category. The distribution of hardware semantic categories is reported in Appendix~\ref{app:hw_semantic_dis}.

\section{Experiments}

\subsection{Experimental Setup}

\textbf{Code models.} We evaluate code LLMs of different scale: DeepSeek-Coder-MQA-5.7B~\cite{deepseek-coder}, DeepSeek-Coder-base-6.7B~\cite{deepseek-coder}, Qwen2.5-Coder-3B-base~\cite{hui2024qwen2}, Qwen2.5-Coder-7B-base~\cite{hui2024qwen2}, and Qwen2.5-Coder-14B-base~\cite{hui2024qwen2}. \textbf{General models.} We also evaluate commercial models including GPT-5~\cite{singh2025openai}, Gemini~2.5~Pro~\cite{comanici2025gemini}, Grok-4~\cite{xai_grok4_model_card_2025}, and DeepSeek-V3.2~\cite{liu2025deepseek}. \textbf{Metrics.} We compare generated code with the reference using \textbf{EM} and \textbf{ES} following~\cite{liu2025m2rc, zhang2023repocoder}; detailed definitions are given in Appendix~\ref{app:eval-config}. We additionally report functional pass rates on evaluated subsets of MHRC-Bench-Eval, using available testbenches or canonical I/O vectors. Code LLMs use greedy decoding (temperature $0$, top-$p$ $1$); commercial API settings are given in Appendix~\ref{app:eval-config}. Statistical uncertainty is discussed in Appendix~\ref{app:seed_var}. \textbf{Baseline and tuning.} The baseline provides the model only with the file containing the completion cursor, so the model must rely on its pretrained ability without explicit inter-file context; we additionally fine-tune the code LLMs on the training set via supervised fine-tuning (SFT), with detailed configuration in Appendix~\ref{app:eval-config}. \textbf{Retrieval framework and prompt.} We evaluate three repository-level methods--RepoCoder~\citep{zhang2023repocoder}, GraphCoder~\citep{liu2024graphcoder}, and RLCoder~\cite{wang2024rlcoder}--on MHRC-Bench-Eval, with retriever-specific hyperparameters tuned per the original papers and a fixed 2{,}048-token input window for all models. We use two prompt formats (Appendix~\ref{app:eval-config}): \emph{preceding-context} (preceding code plus the target line only) and \emph{full-context} (preceding and succeeding code), under the same window budget. The full-context format uses the native FIM mode of code LLMs such as DeepSeek-Coder and Qwen2.5-Coder, where left and right context are wrapped in the model's FIM control tokens (e.g., \texttt{<|fim\_begin|>}, \texttt{<|fim\_hole|>}, \texttt{<|fim\_end|>}) so that the model is invoked exactly as it is during pretraining, rather than via plain concatenation; for proprietary models without FIM tokens, we emulate the same protocol by instructing the model to fill the masked target span between the provided left and right context.

\begin{table*}[t]
\centering
\scriptsize
\caption{EM (\%) and ES (\%) across different languages. Full context refers to the use of both preceding code and the succeeding code surrounding the target code segment, whereas preceding context-only uses only the preceding code as context.}
\setlength{\tabcolsep}{3.2pt}
\begin{tabular}{lcccccccccccccccc}
\toprule
\multirow{3}{*}{\textbf{Model}} 
& \multicolumn{8}{c}{\textbf{Preceding context}}
& \multicolumn{8}{c}{\textbf{Full context}} \\
\cmidrule(lr){2-9}\cmidrule(lr){10-17}
& \multicolumn{2}{c}{\textbf{Chisel}}
& \multicolumn{2}{c}{\textbf{V/SV}}
& \multicolumn{2}{c}{\textbf{VHDL}}
& \multicolumn{2}{c}{\textbf{HLS}}
& \multicolumn{2}{c}{\textbf{Chisel}}
& \multicolumn{2}{c}{\textbf{V/SV}}
& \multicolumn{2}{c}{\textbf{VHDL}}
& \multicolumn{2}{c}{\textbf{HLS}} \\
\cmidrule(lr){2-3}\cmidrule(lr){4-5}\cmidrule(lr){6-7}\cmidrule(lr){8-9}
\cmidrule(lr){10-11}\cmidrule(lr){12-13}\cmidrule(lr){14-15}\cmidrule(lr){16-17}
& EM & ES & EM & ES & EM & ES & EM & ES
& EM & ES & EM & ES & EM & ES & EM & ES \\
\midrule

Qwen2.5-Coder-3B 
& 0.0 & 15.5 & 0.0 & 19.0 & 0.0 & 22.8 & 0.2 & 15.7
& 12.38 & 46.55 & 17.32 & 48.66 & 9.68 & 48.45 & 17.11 & 48.91 \\
+Tuning 
& 17.3 & 56.1 & 28.9 & 61.5 & 26.6 & 64.7 & 17.1 & 54.0
& 24.77 & 54.30 & 34.84 & 65.50 & 28.23 & 59.10 & 33.11 & 55.00 \\
+Retrieval \& Tuning
& 18.9 & 57.6 & 30.1 & 63.4 & 28.2 & 66.2 & 18.2 & 56.0
& 27.55 & 63.95 & 38.78 & 69.94 & 31.45 & 71.88 & 40.35 & 72.63 \\
\midrule

Qwen2.5-Coder-7B 
& 0.0 & 9.8 & 0.0 & 12.8 & 0.2 & 14.0 & 0.0 & 9.7
& 10.22 & 45.79 & 15.55 & 47.00 & 10.48 & 49.26 & 12.72 & 39.51 \\
+Tuning 
& 19.2 & 59.3 & 30.1 & 63.8 & 29.2 & 66.3 & 18.2 & 56.2
& 26.63 & 66.70 & 38.98 & 70.80 & 34.48 & 71.90 & 35.53 & 69.50 \\
+Retrieval \& Tuning
& 20.7 & \underline{61.4} & \textbf{31.9} & 65.2 & 30.6 & 68.2 & 20.2 & \underline{57.8}
& 30.65 & 67.83 & 42.32 & 73.45 & \underline{36.49} & \underline{75.24} & \textbf{44.30} & \textbf{75.91} \\
\midrule

Qwen2.5-Coder-14B  
& 0.0 & 15.5 & 0.0 & 16.1 & 0.0 & 23.3 & 0.2 & 10.4
& 17.96 & 57.29 & 22.44 & 55.99 & 11.09 & 51.44 & 24.34 & 55.66 \\
+Tuning 
& \underline{22.0} & 61.1 & 30.1 & 65.3 & \underline{31.5} & \underline{68.8} & 20.8 & \underline{57.8}
& \underline{31.27} & \underline{68.90}
& 41.93 & 74.40
& 35.28 & 73.80
& 38.16 & 71.50 \\
+Retrieval \& Tuning
& \textbf{24.1} & \textbf{62.7} & \underline{31.7} & \textbf{67.4} & \textbf{33.1} & \textbf{70.0} & \textbf{22.4} & \textbf{59.8}
& \textbf{33.13} & \textbf{70.60}
& \textbf{43.90} & \textbf{76.10}
& \textbf{37.10} & \textbf{75.40}
& 39.91 & 73.20 \\
\midrule

DS-Coder-5.7B 
& 0.0 & 22.9 & 0.4 & 24.5 & 0.4 & 30.1 & 0.2 & 16.9
& 15.48 & 50.28 & 21.06 & 55.57 & 11.90 & 56.18 & 22.81 & 56.36 \\
+Tuning 
& 16.7 & 54.6 & 28.7 & 63.2 & 27.6 & 65.9 & 18.0 & 54.9
& 26.63 & 65.04 & 38.58 & 73.03 & 33.06 & 72.14 & 36.40 & 70.58 \\
+Retrieval \& Tuning
& 18.6 & 56.2 & 29.9 & 65.3 & 29.4 & 67.3 & 19.5 & 56.9
& 28.17 & 63.21 & 40.55 & 73.18 & 33.47 & 73.82 & 42.11 & 73.05 \\
\midrule

DS-Coder-6.7B  
& 0.31 & 20.7 & 0.0 & 21.5 & 0.2 & 28.3 & 0.0 & 15.5
& 10.53 & 43.95 & 20.28 & 52.67 & 9.27 & 52.00 & 17.32 & 49.79 \\
+Tuning 
& 14.6 & 52.4 & 28.5 & 63.7 & 26.6 & 65.2 & 20.0 & 55.7
& 25.70 & 56.50 & 37.80 & 67.40 & 31.45 & 61.10 & 36.84 & 60.40 \\
+Retrieval \& Tuning
& 16.4 & 53.9 & 29.9 & \underline{65.6} & 28.6 & 67.0 & \underline{21.5} & 57.6
& 29.72 & 66.40 & \underline{43.31} & \underline{74.61} & 34.88 & 74.73 & \underline{43.64} & \underline{75.84} \\
\bottomrule
\end{tabular}
\label{tab:model_tuning_results}
\end{table*}
\begin{figure}[!t]
  \centering
  \captionsetup[subfigure]{font=small, labelfont=small}
  \begin{subfigure}[t]{0.49\linewidth}
    \centering
    \begin{tikzpicture}
      \begin{axis}[
        ybar,
        width=0.87\linewidth, height=4.4cm, scale only axis=true,
        bar width=1.2pt,
        enlarge x limits=0.13,
        symbolic x coords={Avg, HLS, SystemVerilog, VHDL, Chisel},
        xtick=data,
        xticklabel style={font=\tiny},
        yticklabel style={font=\tiny},
        ylabel={Pass rate (\%)},
        ylabel style={font=\tiny, yshift=-3pt},
        ymin=0, ymax=80,
        ymajorgrids,
        major grid style={dashed, gray!25},
      ]
        \addplot[fill=orange!70!black, draw=orange!70!black] coordinates {(Avg,38.7) (HLS,48.0) (SystemVerilog,40.0) (VHDL,10.3) (Chisel,56.7)};
        \addplot[fill=red!75!black,    draw=red!75!black]    coordinates {(Avg,35.2) (HLS,68.0) (SystemVerilog,33.3) (VHDL,6.0) (Chisel,33.4)};
        \addplot[fill=brown!70!black,  draw=brown!70!black]  coordinates {(Avg,31.4) (HLS,66.0) (SystemVerilog,31.1) (VHDL,8.3) (Chisel,20.1)};
        \addplot[fill=red!45!black,    draw=red!45!black]    coordinates {(Avg,28.8) (HLS,60.1) (SystemVerilog,23.8) (VHDL,8.3) (Chisel,23.2)};
        \addplot[fill=blue!85!black,   draw=blue!85!black]   coordinates {(Avg,40.0) (HLS,71.9) (SystemVerilog,40.0) (VHDL,14.5) (Chisel,33.4)};
        \addplot[fill=blue!55!black,   draw=blue!55!black]   coordinates {(Avg,35.6) (HLS,64.0) (SystemVerilog,33.3) (VHDL,18.3) (Chisel,26.6)};
        \addplot[fill=cyan!75!black,   draw=cyan!75!black]   coordinates {(Avg,31.3) (HLS,55.9) (SystemVerilog,31.1) (VHDL,8.3) (Chisel,30.0)};
        \addplot[fill=teal!85!black,   draw=teal!85!black]   coordinates {(Avg,36.5) (HLS,66.0) (SystemVerilog,35.6) (VHDL,14.3) (Chisel,30.0)};
        \addplot[fill=teal!50!black,   draw=teal!50!black]   coordinates {(Avg,37.6) (HLS,62.1) (SystemVerilog,40.0) (VHDL,18.3) (Chisel,30.0)};
      \end{axis}
    \end{tikzpicture}
    \subcaption{FIM mode.}
    \label{fig:functional_pass_fim}
  \end{subfigure}\hspace{4pt}%
  \begin{subfigure}[t]{0.49\linewidth}
    \centering
    \begin{tikzpicture}
      \begin{axis}[
        ybar,
        width=0.87\linewidth, height=4.4cm, scale only axis=true,
        bar width=1.2pt,
        enlarge x limits=0.13,
        symbolic x coords={Avg, HLS, SystemVerilog, VHDL, Chisel},
        xtick=data,
        xticklabel style={font=\tiny},
        yticklabel style={font=\tiny},
        ymin=0, ymax=80,
        ymajorgrids,
        major grid style={dashed, gray!25},
      ]
        \addplot[fill=orange!70!black, draw=orange!70!black] coordinates {(Avg,19.7) (HLS,12.1) (SystemVerilog,40.0) (VHDL,0.0) (Chisel,26.6)};
        \addplot[fill=red!75!black,    draw=red!75!black]    coordinates {(Avg,19.7) (HLS,18.0) (SystemVerilog,37.8) (VHDL,0.0) (Chisel,23.2)};
        \addplot[fill=brown!70!black,  draw=brown!70!black]  coordinates {(Avg,15.0) (HLS,20.0) (SystemVerilog,33.3) (VHDL,0.0) (Chisel,6.8)};
        \addplot[fill=red!45!black,    draw=red!45!black]    coordinates {(Avg,20.9) (HLS,23.9) (SystemVerilog,33.3) (VHDL,0.0) (Chisel,26.6)};
        \addplot[fill=blue!85!black,   draw=blue!85!black]   coordinates {(Avg,28.3) (HLS,32.0) (SystemVerilog,35.6) (VHDL,22.4) (Chisel,23.2)};
        \addplot[fill=blue!55!black,   draw=blue!55!black]   coordinates {(Avg,26.4) (HLS,28.1) (SystemVerilog,33.3) (VHDL,14.3) (Chisel,30.0)};
        \addplot[fill=cyan!75!black,   draw=cyan!75!black]   coordinates {(Avg,21.8) (HLS,30.0) (SystemVerilog,28.9) (VHDL,8.3) (Chisel,20.1)};
        \addplot[fill=teal!85!black,   draw=teal!85!black]   coordinates {(Avg,17.0) (HLS,0.0) (SystemVerilog,31.1) (VHDL,20.4) (Chisel,16.7)};
        \addplot[fill=teal!50!black,   draw=teal!50!black]   coordinates {(Avg,18.2) (HLS,0.0) (SystemVerilog,37.8) (VHDL,18.3) (Chisel,16.7)};
      \end{axis}
    \end{tikzpicture}
    \subcaption{Left-only context (preceding-only).}
    \label{fig:functional_pass_l2r}
  \end{subfigure}

  \vspace{1pt}
  \makebox[\linewidth][c]{%
  \begin{tabular}{@{}*{9}{@{\hspace{2pt}}l}@{}}
    \tikz\fill[orange!70!black] (0,0) rectangle (1.6mm,1.2mm); \scalebox{0.55}{Gemini} &
    \tikz\fill[red!75!black]    (0,0) rectangle (1.6mm,1.2mm); \scalebox{0.55}{GPT-5} &
    \tikz\fill[brown!70!black]  (0,0) rectangle (1.6mm,1.2mm); \scalebox{0.55}{Grok-4} &
    \tikz\fill[red!45!black]    (0,0) rectangle (1.6mm,1.2mm); \scalebox{0.55}{DSv3.2} &
    \tikz\fill[blue!85!black]   (0,0) rectangle (1.6mm,1.2mm); \scalebox{0.55}{PT-Qwen-14B} &
    \tikz\fill[blue!55!black]   (0,0) rectangle (1.6mm,1.2mm); \scalebox{0.55}{PT-Qwen-7B} &
    \tikz\fill[cyan!75!black]   (0,0) rectangle (1.6mm,1.2mm); \scalebox{0.55}{PT-Qwen-3B} &
    \tikz\fill[teal!85!black]   (0,0) rectangle (1.6mm,1.2mm); \scalebox{0.55}{PT-DSC-6.7B} &
    \tikz\fill[teal!50!black]   (0,0) rectangle (1.6mm,1.2mm); \scalebox{0.55}{PT-DS-5.7B}
  \end{tabular}}

  \captionsetup{font=small, labelfont=small}
  \caption{Functional pass rate (\%) on evaluated subsets per HDL under (a) FIM and (b) left-only context. Warm bars = proprietary, cool = post-trained (\emph{PT-} prefix indicates low-rank adaptation (LoRA) fine-tuning on MHRC-Bench-Train). Zero-height bars in the left-only panel include settings that were not evaluated and are shown only as placeholders.}
  \label{fig:functional_pass}
\end{figure}
\subsection{Main Results}

Table~\ref{tab:retrieval_methods_results} compares repository-level retrieval methods under the default settings from their original papers. With the exception of GraphCoder, retrieval generally improves performance across multiple hardware languages. RLCoder achieves the strongest overall retrieval result, while GraphCoder, \emph{as released}, often underperforms the no-retrieval baseline because its CST traversal helpers and dependency-graph constructor are tuned for Python/Java and do not natively parse HDL constructs; Appendix~\ref{subsec:graph_coder} details the HDL adaptations we made and the residual mismatch (e.g., systematic dropping of function declarations and HLS pragmas during retrieval).

Table~\ref{tab:model_tuning_results} summarizes base code models, fine-tuned models, and fine-tuned models augmented with RLCoder across the four HDLs. Before SFT, open-weight code models have near-zero EM in preceding-context-only tasks and weak full-context performance, underscoring the difficulty of repository-level hardware code completion for pretrained models. Fine-tuning yields substantial gains across all model families, and retrieval augmentation further improves many settings. Overall, post-trained Qwen2.5-Coder-14B achieves the strongest performance, followed by Qwen2.5-Coder-7B; both outperform the smaller DeepSeek-Coder variants in most settings. Compared with larger general-purpose models, fine-tuned open-weight models are competitive under matched context windows. Qwen2.5-Coder-14B after fine-tuning outperforms GPT-5, Grok-4, and DeepSeek V3.2 across the four HDLs in Table~\ref{tab:model_tuning_results}, and it surpasses Gemini 2.5 Pro in all cases except Chisel.
\paragraph{Functional pass rate.}
We evaluate functional pass rate on subsets of MHRC-Bench-Eval spanning HLS, SystemVerilog, VHDL, and Chisel. Evaluated completions are checked using available testbenches or canonical I/O vectors; the released testbench inventory documents partial coverage. We report \emph{pass@1 under model-specific decoding settings} for four proprietary LLMs and five LoRA fine-tuned models in Figure~\ref{fig:functional_pass}; coverage and per-language toolchains are discussed in Appendix~\ref{app:functional_eval}. Pass rates vary sharply by language and the relative ranking of model families also flips: on HLS the LoRA-tuned Qwen-14B reaches 71.9\%, beating the strongest proprietary model GPT-5 (68.0\%) and trailing only itself; on VHDL every proprietary model stays below 11\%, while the fine-tuned DS-5.7B-LoRA and Qwen-7B-LoRA reach 18.3\%; on SystemVerilog the fine-tuned DS-5.7B-LoRA and Qwen-14B-LoRA tie Gemini-2.5 at 40.0\%; on Chisel, however, Gemini-2.5 retains a clear lead (56.7\%) and only Qwen-14B-LoRA (33.4\%) approaches it. These language-dependent gaps motivate keeping a multi-paradigm benchmark rather than reporting a single average.

\subsection{Evaluation of Other Benchmarks and Models}

We compare Verilog-specific models \citep{wu2025rtlrepocoder,liu2024rtlcoder} across a prior hardware repository benchmark and ours. To ensure a fair comparison in Figure~\ref{fig:other_models_bench}, the DeepSeek-based baselines use the same base model (DeepSeek-Coder-6.7B-base) and the preceding-context-only setting. We remove 11 repositories that overlap between MHRC-Bench-Train and RTL-repo to avoid data leakage, and we additionally include CodeV-All-QC~\citep{zhao2024codev} under a 2{,}048-token context window on the V/SV subset. Despite the increased difficulty of MHRC-Bench-Eval due to multi-line completion tasks, MHRC-DeepSeek achieves strong performance on this benchmark and surpasses RTLCoder-DeepSeek, indicating that our training dataset generalizes without overfitting to RTL-repo.
\FloatBarrier
\enlargethispage{1\baselineskip}
\enlargethispage{1\baselineskip}
\subsection{Experimental Analysis}

\noindent\textbf{Context format.}
As shown in Tables \ref{tab:retrieval_methods_results} and \ref{tab:model_tuning_results}, the full-context outperforms the preceding-context-only in terms of EM and ES across all models and languages except VHDL. This suggests that bidirectional context is generally beneficial for code generation, while the rigid and highly structured nature of VHDL reduces the marginal benefit of right-context information. Hence, we focus on full-context in the following analysis.

\noindent\textbf{Prompt robustness.}
Table~\ref{tab:prompt_strategy} shows that the main conclusion is robust across three prompt formats: few-shot prompting, chain-of-thought prompting, and our default prompt. Under all three settings, the fine-tuned Qwen-14B-LoRA model outperforms the strongest general-purpose models, including GPT-5 and Gemini-2.5 Pro, on average EM and ES across languages. This indicates that the performance gains primarily come from MHRC-Bench-Train and hardware-specific post-training rather than from a prompt-template artifact.

\begin{figure*}[!t]
  \centering
  \providecommand{\panelheight}{3.5cm}%
  \captionsetup{justification=centering,singlelinecheck=false}
  \newcommand{\alignanalysisbounds}{%
    \pgfresetboundingbox
    \path[use as bounding box]
      ([xshift=-0.65cm,yshift=-0.60cm]analysisaxis.south west)
      rectangle ([xshift=0.10cm,yshift=0.10cm]analysisaxis.north east);%
  }
  \begin{minipage}[t]{0.33\linewidth}
    \vspace{0pt}
    \centering
    \begin{tikzpicture}
      \begin{axis}[
        name=analysisaxis, anchor=south west, at={(0,0)},
        width=0.76\linewidth, height=\panelheight, scale only axis=true,
        xtick={25,50,75,100},
        xticklabel style={font=\tiny},
        yticklabel style={font=\tiny},
        xlabel={Training set size (\%)},
        xlabel style={font=\tiny, yshift=2pt},
        ylabel={EM (\%)},
        ylabel style={font=\tiny, yshift=-3pt},
        ymin=20, ymax=42,
        grid=both,
        major grid style={dashed, gray!25},
        legend style={font=\fontsize{4.2pt}{5pt}\selectfont,
                      at={(0.98,0.02)}, anchor=south east,
                      legend columns=2, draw=none,
                      fill=white, fill opacity=0.7, text opacity=1,
                      inner xsep=2pt, inner ysep=1pt, row sep=-2pt,
                      /tikz/every even column/.append style={column sep=3pt}},
        mark size=1.3pt,
        line width=0.7pt,
      ]
        \addplot[solid, color=blue!70!black,    mark=*]         coordinates {(25,24.5) (50,26.0) (75,24.5) (100,26.6)};
        \addplot[solid, color=orange!80!black,  mark=square*]   coordinates {(25,34.5) (50,35.5) (75,36.4) (100,35.5)};
        \addplot[solid, color=green!55!black,   mark=triangle*] coordinates {(25,35.6) (50,37.4) (75,36.6) (100,39.0)};
        \addplot[solid, color=red!75!black,     mark=diamond*]  coordinates {(25,26.8) (50,31.5) (75,31.5) (100,34.5)};
        \legend{Chisel, HLS, V/SV, VHDL}
      \end{axis}
      \alignanalysisbounds
    \end{tikzpicture}
    \caption{EM (\%) vs.\newline training-set size.}
    \label{fig:datasetsize_em}
  \end{minipage}\hfill
  \begin{minipage}[t]{0.33\linewidth}
    \vspace{0pt}
    \centering
    \begin{tikzpicture}
      \begin{axis}[
        name=analysisaxis, anchor=south west, at={(0,0)},
        width=0.76\linewidth, height=\panelheight, scale only axis=true,
        xtick={1,2,3,4,5},
        xticklabel style={font=\tiny},
        yticklabel style={font=\tiny},
        xlabel={Line count},
        xlabel style={font=\tiny, yshift=2pt},
        ylabel={Score (\%)},
        ylabel style={font=\tiny, yshift=-3pt},
        ymin=0, ymax=100,
        grid=both,
        major grid style={dashed, gray!25},
        legend style={font=\fontsize{4.2pt}{5pt}\selectfont,
                      at={(0.98,0.98)}, anchor=north east,
                      legend columns=1, draw=none,
                      fill=white, fill opacity=0.7, text opacity=1,
                      inner xsep=2pt, inner ysep=1pt, row sep=-2pt},
        mark size=1.2pt,
        line width=0.65pt,
      ]
        \addplot[solid,  color=blue!70!black,    mark=*]         coordinates {(1,51) (2,35) (3,18) (4,13) (5,10)};
        \addplot[dashed, color=blue!70!black,    mark=o]         coordinates {(1,79) (2,69) (3,62) (4,60) (5,58)};
        \addplot[solid,  color=red!75!black,     mark=square*]   coordinates {(1,55) (2, 4) (3, 1) (4, 0) (5, 0)};
        \addplot[dashed, color=red!75!black,     mark=square]    coordinates {(1,78) (2,50) (3,43) (4,37) (5,35)};
        \addplot[solid,  color=green!55!black,   mark=triangle*] coordinates {(1,57) (2, 5) (3, 2) (4, 1) (5, 1)};
        \addplot[dashed, color=green!55!black,   mark=triangle]  coordinates {(1,80) (2,52) (3,45) (4,38) (5,36)};
        \legend{Qwen-7B EM, Qwen-7B ES, GPT-5 EM, GPT-5 ES, Gemini-2.5 EM, Gemini-2.5 ES}
      \end{axis}
      \alignanalysisbounds
    \end{tikzpicture}
    \caption{EM/ES (\%) vs.\newline line count.}
    \label{fig:line_performance}
  \end{minipage}\hfill
  \begin{minipage}[t]{0.33\linewidth}
    \vspace{0pt}
    \centering
    \begin{tikzpicture}
      \begin{axis}[
        name=analysisaxis, anchor=south west, at={(0,0)},
        width=0.76\linewidth, height=\panelheight, scale only axis=true,
        xmode=log, log basis x={2},
        log ticks with fixed point,
        xtick={512,1024,2048,4096,8192},
        xticklabels={512,1k,2k,4k,8k},
        xticklabel style={font=\tiny},
        yticklabel style={font=\tiny},
        xlabel={Context length (tokens)},
        xlabel style={font=\tiny, yshift=2pt},
        ylabel={Avg.\ EM (\%)},
        ylabel style={font=\tiny, yshift=-3pt},
        ymin=18, ymax=42,
        grid=both,
        major grid style={dashed, gray!25},
        legend style={font=\fontsize{4.2pt}{5pt}\selectfont,
                      at={(0.02,0.98)}, anchor=north west,
                      legend columns=1, draw=none, fill=white, fill opacity=0.7, text opacity=1,
                      inner xsep=2pt, inner ysep=1pt, row sep=-2pt},
        mark size=1.2pt,
        line width=0.65pt,
      ]
        \addplot[dashed, color=blue!85!black,   mark=*]         coordinates {(512,29.13) (1024,31.94) (2048,36.66) (4096,31.78) (8192,32.55)};
        \addplot[dashed, color=blue!55!black,   mark=square*]   coordinates {(512,28.52) (1024,30.89) (2048,33.90) (4096,27.69) (8192,27.27)};
        \addplot[dashed, color=cyan!75!black,   mark=triangle*] coordinates {(512,22.91) (1024,28.40) (2048,30.24) (4096,26.10) (8192,23.65)};
        \addplot[dashed, color=teal!75!black,   mark=diamond*]  coordinates {(512,20.66) (1024,28.74) (2048,32.95) (4096,25.50) (8192,25.31)};
        \addplot[solid,  color=orange!85!black, mark=o]         coordinates {(512,28.50) (1024,27.08) (2048,31.48) (4096,27.38) (8192,31.94)};
        \addplot[solid,  color=red!75!black,    mark=square]    coordinates {(512,25.58) (1024,26.34) (2048,26.48) (4096,25.77) (8192,28.29)};
        \legend{FT Qwen-14B, FT Qwen-7B, FT Qwen-3B, FT DS-Coder-6.7B, Gemini-2.5, GPT-5}
      \end{axis}
      \alignanalysisbounds
    \end{tikzpicture}
    \caption{Avg.\ EM (\%) vs.\newline context length.}
    \label{fig:context_length_em}
  \end{minipage}
\end{figure*}

\noindent\textbf{Training data size.} In Figure~\ref{fig:datasetsize_em}, we study how training-set size affects the full-context EM of Qwen2.5-Coder-7B-base after fine-tuning on 25\%, 50\%, 75\%, and 100\% of MHRC-Bench-Train. The effect is language dependent: V/SV and VHDL improve most clearly as more data is used, Chisel changes only mildly, and HLS peaks at 75\% before slightly decreasing at 100\%. This suggests that more HDL data generally helps, but HLS remains partly governed by the model's pre-existing C/C++ prior and can be more sensitive to distribution shifts.

\noindent\textbf{Prompt length.} Figure~\ref{fig:context_length_em} compares average EM across 512, 1024, 2048, 4096, and 8192 input tokens. Fine-tuned models consistently improve up to 2048 tokens but degrade at longer windows, suggesting that additional context can introduce noise for smaller post-trained code models. Proprietary models are flatter and benefit from the longest window, but their gains are limited, fluctuating and smaller than the 2048-token peak observed for fine-tuned models. We therefore use 2048 tokens as the default context budget for the main evaluation. 

\noindent\textbf{Extremely large repository input bundles.} We also evaluate proprietary models by filling their much larger input windows with related repository files for each query, as shown in Table~\ref{tab:large_input_windows}. Larger repository input bundles mainly improve HLS and Chisel, while VHDL remains difficult; overall, these proprietary results still trail our fine-tuned 7B- and 14B-scale models in most settings.

\begin{table*}[!t]
\centering
\scriptsize
\caption{EM/ES (\%) of proprietary models when related repository files are provided as large input bundles.}
\label{tab:large_input_windows}
\setlength{\tabcolsep}{5pt}
\renewcommand{\arraystretch}{1.1}
\begin{tabular}{lccccccccc}
\toprule
\textbf{Model} & \textbf{Bundle budget} & \multicolumn{2}{c}{\textbf{Chisel (EM/ES)}} & \multicolumn{2}{c}{\textbf{V/SV (EM/ES)}} & \multicolumn{2}{c}{\textbf{VHDL (EM/ES)}} & \multicolumn{2}{c}{\textbf{HLS (EM/ES)}} \\
\midrule
GPT-5 & 400{,}000 & 30.96 & 63.57 & 38.58 & 62.95 & 14.95 & 50.57 & 44.30 & 71.50 \\
Gemini-2.5 & 1{,}000{,}000 & 32.20 & 64.11 & 36.81 & 61.17 & 11.90 & 43.63 & 46.37 & 71.12 \\
\bottomrule
\end{tabular}
\end{table*}

\FloatBarrier
\noindent\textbf{Target line count.} As shown in Figure~\ref{fig:line_performance}, all evaluated models degrade as the target span grows from one to five lines. EM drops especially sharply: GPT-5 and Gemini-2.5 are strong on single-line targets but nearly lose exact matches on longer spans, while post-trained Qwen2.5-Coder-7B retains nonzero EM through five lines. ES decreases more smoothly, indicating that models often preserve partial structure even when exact multi-line reconstruction fails.

\setlength{\textfloatsep}{6pt}

\begin{figure*}[!t]
  \centering
  \providecommand{\panelheight}{2.9cm}%
  \begin{minipage}[b]{0.31\linewidth}
    \centering
    \begin{tikzpicture}
      \begin{axis}[
        width=0.76\linewidth, height=\panelheight, scale only axis=true,
        xtick={16,32,48,64,76},
        xticklabel style={font=\tiny},
        yticklabel style={font=\tiny},
        xlabel={Node depth},
        xlabel style={font=\tiny, yshift=2pt},
        ylabel={Score (\%)},
        ylabel style={font=\tiny, yshift=-3pt},
        xmin=0, xmax=80,
        ymin=0, ymax=100,
        grid=both,
        major grid style={dashed, gray!25},
        legend style={font=\fontsize{4.2pt}{5pt}\selectfont,
                      at={(0.02,0.98)}, anchor=north west,
                      legend columns=1, draw=none,
                      fill=white, fill opacity=0.7, text opacity=1,
                      inner xsep=2pt, inner ysep=1pt, row sep=-2pt},
        mark size=1.2pt,
        line width=0.65pt,
      ]
        \addplot[solid,  color=blue!70!black, mark=*]       coordinates {(16,34) (32,39) (48,35) (64,10) (76,88)};
        \addplot[dashed, color=blue!70!black, mark=o]       coordinates {(16,70) (32,70) (48,74) (64,36) (76,89)};
        \addplot[solid,  color=red!75!black,  mark=square*] coordinates {(16,26) (32,29) (48,60) (64,88) (76,88)};
        \addplot[dashed, color=red!75!black,  mark=square]  coordinates {(16,58) (32,59) (48,68) (64,90) (76,90)};
        \legend{Qwen-7B EM, Qwen-7B ES, GPT-5 EM, GPT-5 ES}
      \end{axis}
    \end{tikzpicture}
    \caption{EM/ES (\%) vs.\ node depth.}
    \label{fig:node_depth_performance}
  \end{minipage}\hfill
  \begin{minipage}[b]{0.32\linewidth}
    \centering
    \includegraphics[height=\panelheight, width=\linewidth, keepaspectratio]{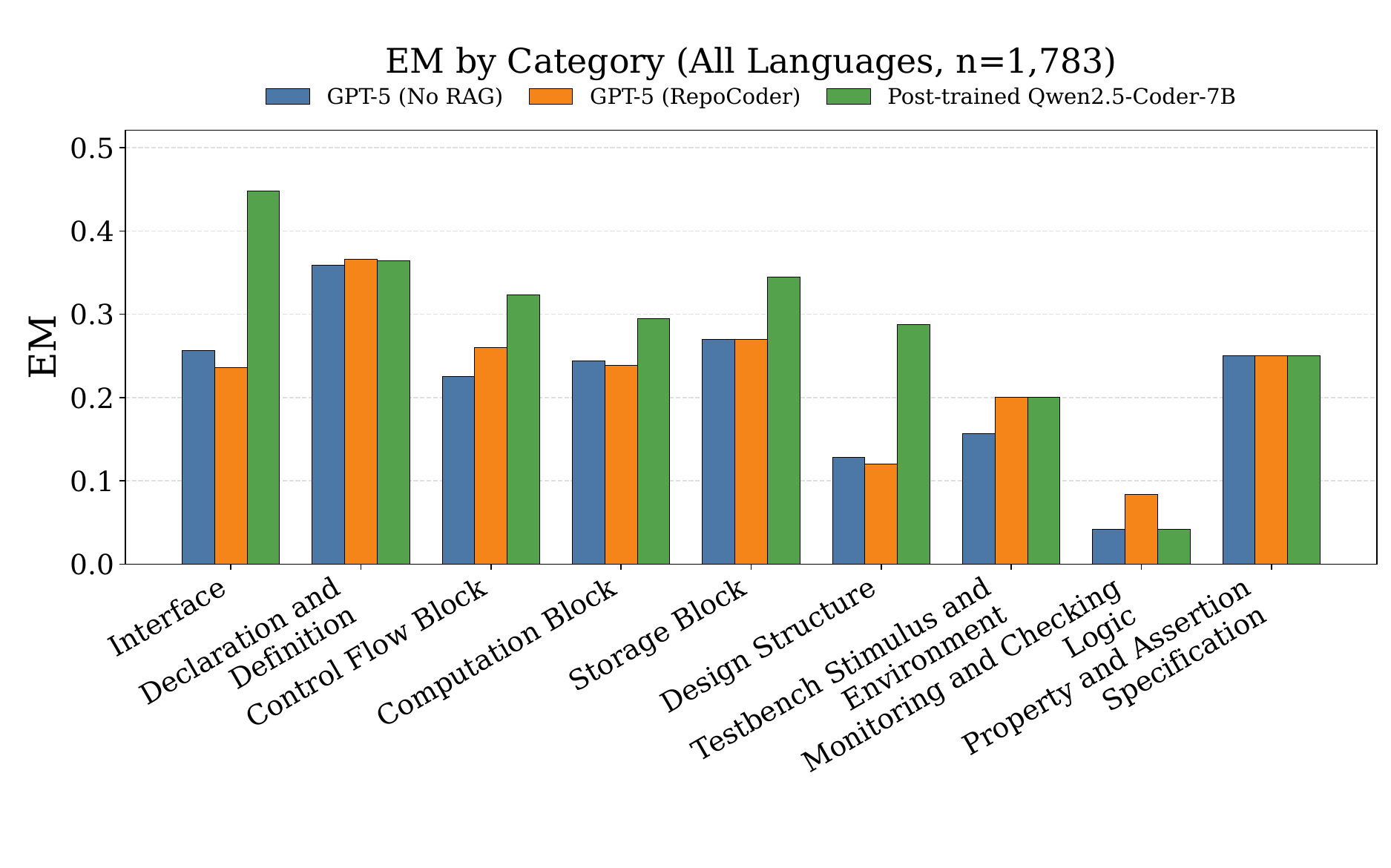}
    \caption{EM of GPT-5 vs. post-trained Qwen2.5-Coder-7B across hardware semantic categories.}
    \label{fig:by_category_EM}
  \end{minipage}\hfill
  \begin{minipage}[b]{0.31\linewidth}
    \centering
    \begin{tikzpicture}
      \begin{axis}[
        ybar,
        width=\linewidth, height=\panelheight,
        bar width=1.9pt,
        enlarge x limits=0.16,
        ylabel={Score (\%)},
        ylabel style={font=\scriptsize, yshift=-3pt},
        xmin=0.45, xmax=4.55,
        xtick={1,2,3,4},
        xticklabels={RTLCoder, MHRC (Ours), CodeV-All-QC, RTLRepo},
        xticklabel style={font=\tiny, rotate=30, anchor=east, yshift=2pt},
        yticklabel style={font=\tiny},
        ymin=0, ymax=100,
        ymajorgrids,
        major grid style={dashed, gray!30},
        legend style={font=\fontsize{4.2pt}{5pt}\selectfont, at={(0.5,1.02)}, anchor=south, legend columns=3,
                      /tikz/every even column/.append style={column sep=2pt},
                      draw=none, fill=none},
      ]
        \addplot[fill=blue!60, draw=blue!60, bar shift=-3.15pt] coordinates {(1,15.5) (2,63.7) (4,13.0)};
        \addplot[fill=blue!25, draw=blue!25, bar shift=-1.05pt] coordinates {(1, 0.0) (2,28.5) (4, 6.3)};
        \addplot[fill=orange!70, draw=orange!70, bar shift=1.05pt] coordinates {(1,48.1) (2,51.7) (4,81.2)};
        \addplot[fill=orange!30, draw=orange!30, bar shift=3.15pt] coordinates {(1,16.2) (2,18.4) (4,50.3)};
        \addplot[fill=violet!70, draw=violet!70, bar shift=-3.15pt] coordinates {(3,51.36)};
        \addplot[fill=violet!30, draw=violet!30, bar shift=-1.05pt] coordinates {(3,18.11)};
        \legend{HDL ES, HDL EM, RTL ES, RTL EM, CodeV ES, CodeV EM}
      \end{axis}
    \end{tikzpicture}
    \caption{ES/EM (\%) of DeepSeek-based models and CodeV-All-QC on MHRC-Bench-Eval (V/SV) and RTL-repo Test.}
    \label{fig:other_models_bench}
  \end{minipage}
\end{figure*}

\noindent\textbf{Code structural levels.} As mentioned in Section~\ref{text:dataset_types}, we bucket test cases by CST node depth. Figure~\ref{fig:node_depth_performance} shows that performance is not monotonic in structural depth. GPT-5 improves on deeper buckets, whereas post-trained Qwen2.5-Coder-7B fluctuates, with a sharp drop at one intermediate depth and strong performance on the deepest bucket. This differs from software-code trends reported by \citet{liu2025m2rc}, where structural depth is associated with a more consistent performance pattern. Detailed per-node results are provided in Appendix~\ref{app:perf_structure}.

\noindent\textbf{Hardware semantic levels.} We evaluate all 1{,}783 test samples using the nine hardware-semantic categories defined in Section~\ref{text:dataset_types}, with the same rule-based labels and category counts for every model. Figure~\ref{fig:by_category_EM} shows that post-trained Qwen2.5-Coder-7B achieves higher EM than GPT-5 on most categories, with its largest advantages over RepoCoder appearing in interface and design-structure snippets. Monitoring/checking remains difficult for both models. The property/assertion category contains only four targets, so its rate is sensitive to individual predictions. Appendix~\ref{app:perf_semantic} reports the per-language semantic breakdown.

\paragraph{Agentic Evaluation.} We also evaluate the Aider code agent~\cite{aider2026} on MHRC-Bench-Eval. The detailed setup and results are in Appendix \ref{app:agentic}. Although Aider was able to attain better performance than the strongest local LLMs fine-tuned on MHRC-Bench-Train, the improvement is 6.0 percentage points in EM. Furthermore, it costs ten times the token usage, suggesting that local, smaller models fine-tuned on our dataset are a more cost-efficient alternative for repository-level HDL code completion.

\subsection{Key Observations}

\textbf{Performance varies significantly across hardware languages.}
In most settings, the four HDLs follow the same ordering across functional pass rate and EM/ES: HLS is easiest, reaching 48.0--71.9\% pass rate across nine models; VHDL is hardest, staying below 23\% even for the best post-trained model; and V/SV and Chisel fall between. We attribute this to three language-specific factors: HLS benefits from C/C++ pretraining priors, whereas VHDL has a smaller public-code footprint; VHDL's verbosity and strict typing make small cast, width, or library-type errors fatal; and VHDL is the only language where full-context FIM does not beat preceding-context-only completion, suggesting preceding declarations are more informative than trailing context. Thus, HDL performance should be reported per language rather than only as an aggregate.

\noindent\textbf{Similarities between hardware and software code.}
(1) Increasing the target length consistently degrades performance for both hardware and software code completion tasks.
(2) LLMs perform well on identifier-level generation in both domains, indicating strong lexical and naming capabilities.

\noindent\textbf{Differences between hardware and software code.}
(1) Structural sensitivity:  
    For software code, deeper CST depths often correlate with improved LLM performance~\cite{liu2025m2rc}. However, for hardware code, increasing structural complexity can either improve or degrade performance.
(2) Prompt length sensitivity.
    Increasing prompt length generally benefits software code completion, whereas for hardware code it yields limited or inconsistent gains (from 2048 to 4096 tokens), reflecting the greater difficulty of reasoning about hardware semantics.

\noindent\textbf{Inappropriate retrieval can degrade performance.} Retrieval-augmented methods are not always beneficial. We observe that retrieving irrelevant or noisy information can harm model performance, as shown by GraphCoder. The drop is not a critique of the GraphCoder algorithm itself; rather, it documents the cost of porting a software-tuned retriever to HDL, where its CST helpers and dependency-graph constructor (released for Python/Java) systematically miss function declarations and HLS-specific directives during retrieval (Appendix~\ref{subsec:graph_coder}). This highlights the importance of retrieval quality and motivates HDL-native retrievers as future work.
\section{Conclusion and Limitations}
We present MHRC-Bench, the first multilingual, repository-level benchmark for hardware code completion, covering RTL, HLS, and generator-based hardware languages across design and verification stages. Fine-grained structural and semantic annotations enable detailed analysis of LLM behavior in hardware-oriented settings, while repository-level functional checks complement EM/ES by measuring behavioral correctness. Post-training on MHRC-Bench-Train substantially improves performance, allowing smaller open-weight models to outperform strong general baselines under matched context windows.
\textbf{Limitations.}
MHRC-Bench is built from open-source repositories, which may not fully reflect proprietary HDL development practices. Its heuristic semantic labels may miss ambiguous or project-specific design intent, and the benchmark currently excludes analog/mixed-signal design and physical-design scripts. Although we strictly enforce repository-level train/test separation and try to avoid internet data leakage, we cannot fully rule out prior exposure of public test repositories in closed-source pretraining data.



\begin{ack}
This research is supported by the Ministry of Education, Singapore, under its Academic Research Fund Tier 3 grant MOE-MOET32024-0003 and Tier 2 grant T2EP20224-0038. Any opinions, findings, and conclusions or recommendations expressed in this material are those of the authors and do not reflect the views of the Ministry of Education, Singapore. We also thank AMD for the generous donation of the Heterogeneous Accelerated Compute Cluster (HACC) at the National University of Singapore. Apart from the support disclosed above, the authors declare no competing interests.
\end{ack}

\begingroup
\raggedright
\bibliographystyle{plainnat}
\bibliography{example_paper}

@misc{aider2026,
  title = {{Aider: AI Pair Programming in Your Terminal}},
  author = {{Aider-AI}},
  year = {2026},
  howpublished = {\url{https://github.com/Aider-AI/aider}},
  note = {Version 0.86.2. Accessed: 2026-05-06}
}

@inproceedings{liu2025m2rc,
  title = {{{M2RC-EVAL}: Massively Multilingual Repository-level Code Completion Evaluation}},
  author = {Liu, Jiaheng and Deng, Ken and Liu, Congnan and Yang, Jian and Liu, Shukai and Zhu, He and Zhao, Peng and Chai, Linzheng and Wu, Yanan and Ke, Jin and others},
  booktitle = {Proceedings of the 63rd Annual Meeting of the Association for Computational Linguistics (Volume 1: Long Papers)},
  pages = {15661--15684},
  year = {2025},
  publisher = {Association for Computational Linguistics},
  doi = {10.18653/v1/2025.acl-long.763},
  url = {https://aclanthology.org/2025.acl-long.763/}
}

@article{liu2024graphcoder,
  title = {{GraphCoder: Enhancing Repository-Level Code Completion via Code Context Graph-based Retrieval and Language Model}},
  author = {Liu, Wei and Yu, Ailun and Zan, Daoguang and Shen, Bo and Zhang, Wei and Zhao, Haiyan and Jin, Zhi and Wang, Qianxiang},
  journal = {arXiv preprint arXiv:2406.07003},
  year = {2024},
  url = {https://arxiv.org/abs/2406.07003}
}

@article{liang2024repofuse,
  title = {{REPOFUSE: Repository-Level Code Completion with Fused Dual Context}},
  author = {Liang, Ming and Xie, Xiaoheng and Zhang, Gehao and Zheng, Xunjin and Di, Peng and Jiang, Wei and Chen, Hongwei and Wang, Chengpeng and Fan, Gang},
  journal = {arXiv preprint arXiv:2402.14323},
  year = {2024},
  url = {https://arxiv.org/abs/2402.14323}
}

@article{wang2024rlcoder,
  title = {{RLCoder: Reinforcement Learning for Repository-Level Code Completion}},
  author = {Wang, Yanlin and Wang, Yanli and Guo, Daya and Chen, Jiachi and Zhang, Ruikai and Ma, Yuchi and Zheng, Zibin},
  journal = {arXiv preprint arXiv:2407.19487},
  year = {2024},
  url = {https://arxiv.org/abs/2407.19487}
}

@article{deng2025enhancing,
  title = {{Enhancing Project-Specific Code Completion by Inferring Internal API Information}},
  author = {Deng, Le and Ren, Xiaoxia and Ni, Chao and Liang, Ming and Lo, David and Liu, Zhongxin},
  journal = {IEEE Transactions on Software Engineering},
  volume = {51},
  number = {9},
  pages = {2566--2582},
  year = {2025},
  publisher = {Institute of Electrical and Electronics Engineers},
  doi = {10.1109/TSE.2025.3592823}
}

@article{li2025coderag,
  title = {{CodeRAG: Supportive Code Retrieval on Bigraph for Real-World Code Generation}},
  author = {Li, Jia and Shi, Xianjie and Zhang, Kechi and Li, Lei and Li, Ge and Tao, Zhengwei and Li, Jia and Liu, Fang and Tao, Chongyang and Jin, Zhi},
  journal = {arXiv preprint arXiv:2504.10046v1},
  year = {2025},
  url = {https://arxiv.org/abs/2504.10046v1}
}

@inproceedings{zhang2023repocoder,
  title = {{RepoCoder: Repository-Level Code Completion Through Iterative Retrieval and Generation}},
  author = {Zhang, Fengji and Chen, Bei and Zhang, Yue and Keung, Jacky and Liu, Jin and Zan, Daoguang and Mao, Yi and Lou, Jian-Guang and Chen, Weizhu},
  booktitle = {Proceedings of the 2023 Conference on Empirical Methods in Natural Language Processing},
  pages = {2471--2484},
  year = {2023},
  publisher = {Association for Computational Linguistics},
  doi = {10.18653/v1/2023.emnlp-main.151},
  url = {https://aclanthology.org/2023.emnlp-main.151/}
}

@article{deepseek-coder,
  title = {{DeepSeek-Coder: When the Large Language Model Meets Programming -- The Rise of Code Intelligence}},
  author = {Guo, Daya and Zhu, Qihao and Yang, Dejian and Xie, Zhenda and Dong, Kai and Zhang, Wentao and Chen, Guanting and Bi, Xiao and Wu, Y. and Li, Y. K. and others},
  journal = {arXiv preprint arXiv:2401.14196},
  year = {2024},
  url = {https://arxiv.org/abs/2401.14196}
}

@article{hui2024qwen2,
  title = {{Qwen2.5-Coder Technical Report}},
  author = {Hui, Binyuan and Yang, Jian and Cui, Zeyu and Yang, Jiaxi and Liu, Dayiheng and Zhang, Lei and Liu, Tianyu and Zhang, Jiajun and Yu, Bowen and Lu, Keming and others},
  journal = {arXiv preprint arXiv:2409.12186},
  year = {2024},
  url = {https://arxiv.org/abs/2409.12186}
}

@article{liu2023repobench,
  title = {{RepoBench: Benchmarking Repository-Level Code Auto-Completion Systems}},
  author = {Liu, Tianyang and Xu, Canwen and McAuley, Julian},
  journal = {arXiv preprint arXiv:2306.03091},
  year = {2023},
  url = {https://arxiv.org/abs/2306.03091}
}

@inproceedings{nam2024using,
  title = {{Using an LLM to Help With Code Understanding}},
  author = {Nam, Daye and Macvean, Andrew and Hellendoorn, Vincent and Vasilescu, Bogdan and Myers, Brad},
  booktitle = {Proceedings of the 46th International Conference on Software Engineering},
  pages = {1--13},
  year = {2024},
  publisher = {Association for Computing Machinery},
  doi = {10.1145/3597503.3639187}
}

@article{liu2024exploring,
  title = {{Exploring and Evaluating Hallucinations in LLM-Powered Code Generation}},
  author = {Liu, Fang and Liu, Yang and Shi, Lin and Huang, Houkun and Wang, Ruifeng and Yang, Zhen and Zhang, Li},
  journal = {arXiv preprint arXiv:2404.00971v1},
  year = {2024},
  url = {https://arxiv.org/abs/2404.00971v1}
}

@inproceedings{pan2024lost,
  title = {{Lost in Translation: A Study of Bugs Introduced by Large Language Models while Translating Code}},
  author = {Pan, Rangeet and Ibrahimizada, Ali Reza and Krishna, Rahul and Sankar, Divya and Wassi, Lambert Pougem and Merler, Michele and Sobolev, Boris and Pavuluri, Raju and Sinha, Saurabh and Jabbarvand, Reyhaneh},
  booktitle = {Proceedings of the 46th International Conference on Software Engineering},
  pages = {1--13},
  year = {2024},
  publisher = {Association for Computing Machinery},
  doi = {10.1145/3597503.3639226}
}

@inproceedings{abedu2025repochat,
  title = {{RepoChat: An LLM-Powered Chatbot for GitHub Repository Question-Answering}},
  author = {Abedu, Samuel and Menneron, Laurine and Khatoonabadi, SayedHassan and Shihab, Emad},
  booktitle = {Proceedings of the 22nd International Conference on Mining Software Repositories},
  pages = {255--259},
  year = {2025},
  organization = {Institute of Electrical and Electronics Engineers},
  url = {https://das.encs.concordia.ca/pdf/abedu2025repochat.pdf}
}

@inproceedings{allam2024rtl,
  title = {{RTL-Repo: A Benchmark for Evaluating LLMs on Large-Scale RTL Design Projects}},
  author = {Allam, Ahmed and Shalan, Mohamed},
  booktitle = {Proceedings of the 2024 International Workshop on Large Language Model-Aided Design},
  pages = {1--5},
  year = {2024},
  organization = {Institute of Electrical and Electronics Engineers},
  doi = {10.1109/LAD62341.2024.10691810}
}

@inproceedings{ding2023crosscodeeval,
  title = {{CrossCodeEval: A Diverse and Multilingual Benchmark for Cross-File Code Completion}},
  author = {Ding, Yangruibo and Wang, Zijian and Ahmad, Wasi Uddin and Ding, Hantian and Tan, Ming and Jain, Nihal and Ramanathan, Murali Krishna and Nallapati, Ramesh and Bhatia, Parminder and Roth, Dan and others},
  booktitle = {Advances in Neural Information Processing Systems},
  volume = {36},
  pages = {46701--46723},
  year = {2023},
  url = {https://proceedings.neurips.cc/paper_files/paper/2023/hash/920f2dced7d32ab2ba2f1970bc306af6-Abstract-Datasets_and_Benchmarks.html}
}

@article{comanici2025gemini,
  title = {{Gemini 2.5: Pushing the Frontier with Advanced Reasoning, Multimodality, Long Context, and Next Generation Agentic Capabilities}},
  author = {Comanici, Gheorghe and Bieber, Eric and Schaekermann, Mike and Pasupat, Ice and Sachdeva, Noveen and Dhillon, Inderjit and Blistein, Marcel and Ram, Ori and Zhang, Dan and Rosen, Evan and others},
  journal = {arXiv preprint arXiv:2507.06261},
  year = {2025},
  url = {https://arxiv.org/abs/2507.06261}
}

@article{liu2025deepseek,
  title = {{DeepSeek-V3.2: Pushing the Frontier of Open Large Language Models}},
  author = {Liu, Aixin and Mei, Aoxue and Lin, Bangcai and Xue, Bing and Wang, Bingxuan and Xu, Bingzheng and Wu, Bochao and Zhang, Bowei and Lin, Chaofan and Dong, Chen and others},
  journal = {arXiv preprint arXiv:2512.02556},
  year = {2025},
  url = {https://arxiv.org/abs/2512.02556}
}

@article{singh2025openai,
  title = {{OpenAI GPT-5 System Card}},
  author = {Singh, Aaditya and Fry, Adam and Perelman, Adam and Tart, Adam and Ganesh, Adi and El-Kishky, Ahmed and McLaughlin, Aidan and Low, Aiden and Ostrow, AJ and Ananthram, Akhila and others},
  journal = {arXiv preprint arXiv:2601.03267},
  year = {2025},
  url = {https://arxiv.org/abs/2601.03267}
}

@article{hua2025researchcodebench,
  title = {{ResearchCodeBench: Benchmarking LLMs on Implementing Novel Machine Learning Research Code}},
  author = {Hua, Tianyu and Hua, Harper and Xiang, Violet and Klieger, Benjamin and Truong, Sang T. and Liang, Weixin and Sun, Fan-Yun and Haber, Nick},
  journal = {arXiv preprint arXiv:2506.02314},
  year = {2025},
  url = {https://arxiv.org/abs/2506.02314}
}

@inproceedings{raihan2025mhumaneval,
  title = {{mHumanEval -- A Multilingual Benchmark to Evaluate Large Language Models for Code Generation}},
  author = {Raihan, Nishat and Anastasopoulos, Antonios and Zampieri, Marcos},
  booktitle = {Proceedings of the 2025 Conference of the Nations of the Americas Chapter of the Association for Computational Linguistics: Human Language Technologies (Volume 1: Long Papers)},
  pages = {11432--11461},
  year = {2025},
  publisher = {Association for Computational Linguistics},
  doi = {10.18653/v1/2025.naacl-long.570},
  url = {https://aclanthology.org/2025.naacl-long.570/}
}

@article{yang2024execrepobench,
  title = {{ExecRepoBench: Multi-level Executable Code Completion Evaluation}},
  author = {Yang, Jian and Zhang, Jiajun and Yang, Jiaxi and Jin, Ke and Zhang, Lei and Peng, Qiyao and Deng, Ken and Miao, Yibo and Liu, Tianyu and Cui, Zeyu and others},
  journal = {arXiv preprint arXiv:2412.11990},
  year = {2024},
  url = {https://arxiv.org/abs/2412.11990}
}

@article{abi2025hls,
  title = {{HLS-Eval: A Benchmark and Framework for Evaluating LLMs on High-Level Synthesis Design Tasks}},
  author = {Abi-Karam, Stefan and Hao, Cong},
  journal = {arXiv preprint arXiv:2504.12268},
  year = {2025},
  url = {https://arxiv.org/abs/2504.12268}
}

@inproceedings{liu2023verilogeval,
  title = {{VerilogEval: Evaluating Large Language Models for Verilog Code Generation}},
  author = {Liu, Mingjie and Pinckney, Nathaniel and Khailany, Brucek and Ren, Haoxing},
  booktitle = {Proceedings of the 2023 International Conference on Computer-Aided Design},
  pages = {1--8},
  year = {2023},
  organization = {Institute of Electrical and Electronics Engineers},
  url = {https://arxiv.org/abs/2309.07544}
}

@inproceedings{lu2024rtllm,
  title = {{RTLLM: An Open-Source Benchmark for Design RTL Generation with Large Language Model}},
  author = {Lu, Yao and Liu, Shang and Zhang, Qijun and Xie, Zhiyao},
  booktitle = {Proceedings of the 29th Asia and South Pacific Design Automation Conference},
  pages = {722--727},
  year = {2024},
  organization = {Institute of Electrical and Electronics Engineers},
  doi = {10.1109/ASP-DAC58780.2024.10473904}
}

@inproceedings{kashanaki2024hdleval,
  title = {{HDLEval Benchmarking LLMs for Multiple HDLs}},
  author = {Kashanaki, Farzaneh Rabiei and Zakharov, Mark and Renau, Jose},
  booktitle = {Proceedings of the 2024 International Workshop on Large Language Model-Aided Design},
  pages = {1--5},
  year = {2024},
  organization = {Institute of Electrical and Electronics Engineers},
  doi = {10.1109/LAD62341.2024.10691770}
}

@misc{xai_grok4_model_card_2025,
  title = {{{Grok 4 Model Card}}},
  author = {{xAI}},
  year = {2025},
  howpublished = {\url{https://data.x.ai/2025-08-20-grok-4-model-card.pdf}}
}

@article{liu2024rtlcoder,
  title = {{RTLCoder: Fully Open-Source and Efficient LLM-Assisted RTL Code Generation Technique}},
  author = {Liu, Shang and Fang, Wenji and Lu, Yao and Wang, Jing and Zhang, Qijun and Zhang, Hongce and Xie, Zhiyao},
  journal = {IEEE Transactions on Computer-Aided Design of Integrated Circuits and Systems},
  volume = {44},
  number = {4},
  pages = {1448--1461},
  year = {2025},
  publisher = {Institute of Electrical and Electronics Engineers},
  doi = {10.1109/TCAD.2024.3483089}
}

@article{wu2025rtlrepocoder,
  title = {{RTLRepoCoder: Repository-Level RTL Code Completion through the Combination of Fine-Tuning and Retrieval Augmentation}},
  author = {Wu, Peiyang and Guo, Nan and Lv, Junliang and Xiao, Xiao and Ye, Xiaochun},
  journal = {arXiv preprint arXiv:2504.08862},
  year = {2025},
  url = {https://arxiv.org/abs/2504.08862}
}

@article{zou2025hlstrans,
  title = {{HLStrans: Dataset for C-to-HLS Hardware Code Synthesis}},
  author = {Zou, Qingyun and Chen, Nuo and Chen, Yao and He, Bingsheng and Wong, Weng-Fai},
  journal = {arXiv preprint arXiv:2507.04315},
  year = {2025},
  url = {https://arxiv.org/abs/2507.04315}
}

@article{zou2026hlsseek,
  title = {{HLS-Seek: QoR-Aware Code Generation for High-Level Synthesis via Proxy Comparative Reward Reinforcement Learning}},
  author = {Zou, Qingyun and Yu, Feng and Tan, Hongshi and Chen, Yao and He, Bingsheng and Wong, Weng-Fai},
  journal = {arXiv preprint arXiv:2605.13536},
  year = {2026},
  url = {https://arxiv.org/abs/2605.13536}
}

@article{zou2026phoenix,
  title = {{Is Agentic AI Ready for Real-World Hardware Engineering? A Deep Dive with Phoenix-bench}},
  author = {Zou, Qingyun and Yu, Feng and Tan, Hongshi and He, Bingsheng and Wong, Weng-Fai},
  journal = {arXiv preprint arXiv:2605.15226},
  year = {2026},
  url = {https://arxiv.org/abs/2605.15226}
}

@article{zou2026rwopd,
  title = {{Reward-Weighted On-Policy Distillation with an Open Property-Equivalence Verifier for NL-to-SVA Generation}},
  author = {Zou, Qingyun and Li, Yingze and Liu, Tianen and He, Bingsheng and Wong, Weng-Fai},
  journal = {arXiv preprint arXiv:2605.13501},
  year = {2026},
  url = {https://arxiv.org/abs/2605.13501}
}

@article{zhao2024codev,
  title = {{CodeV: Empowering LLMs with HDL Generation through Multi-Level Summarization}},
  author = {Zhao, Yang and Huang, Di and Li, Chongxiao and Jin, Pengwei and Song, Muxin and Xu, Yinan and Nan, Ziyuan and Gao, Mingju and Ma, Tianyun and Qi, Lei and others},
  journal = {arXiv preprint arXiv:2407.10424},
  year = {2024},
  url = {https://arxiv.org/abs/2407.10424}
}
\endgroup

\clearpage

\raggedbottom
\appendix
\section{Appendix}
\newcommand{\appendixtablefont}{\scriptsize\renewcommand{\arraystretch}{1.06}}
\providecommand{\rebuttaltablefont}{\appendixtablefont}
\makeatletter
\setlength{\@fptop}{0pt}
\makeatother

\subsection{Dataset Splits}
\label{app:data_splits}
To ensure that no repositories or files overlap between evaluation and training, we perform the leakage check shown in Algorithm~\ref{alg:leakage_check}. Specifically, for each example we concatenate the left context, ground-truth span, and right context into a single string, normalize text (line endings and optional whitespace normalization), and compute a SHA-1 digest. We build a hash index over all training examples and flag any evaluation example whose digest appears in the training index; such overlaps are removed to prevent train--test leakage.

\begin{algorithm}[ht]
\caption{Leakage-Free Test Set Construction}
\label{alg:leakage_check}
\begin{algorithmic}[1]

\STATE \textbf{Input:} Training set $\mathcal{D}_{train}$, candidate test files $\mathcal{C}$
\STATE \textbf{Output:} Leakage-free test set $\mathcal{D}_{test}$

\STATE Build hash index $\mathcal{H}_{train}$ from $\mathcal{D}_{train}$ by hashing
left-context $+$ ground-truth $+$ right-context
\STATE $\mathcal{D}_{test} \leftarrow \emptyset$

\FOR{each candidate file $c \in \mathcal{C}$}
    \STATE Construct full-context string $s_c$
    \STATE Normalize $s_c$ (line endings, optional whitespace)
    \STATE Compute hash $h_c \leftarrow \mathrm{SHA1}(s_c)$
    \IF{$h_c \notin \mathcal{H}_{train}$}
        \STATE $\mathcal{D}_{test} \leftarrow \mathcal{D}_{test} \cup \{c\}$
    \ELSE
        \STATE Discard $c$ \textit{(potential train--test leakage)}
    \ENDIF
\ENDFOR

\STATE \textbf{return} $\mathcal{D}_{test}$
\end{algorithmic}
\end{algorithm}

\subsection{Cross-file dependency}
\label{app:cross-file dependency}
The target file often depends on functions, definitions, or modules defined in other files to be completed correctly. Since different hardware description languages follow distinct grammatical and structural rules, we adopt language-specific strategies to identify and extract cross-file dependencies.

For each task file, we compute cross-file dependencies as the number of distinct other files within the same repository that the file depends on, using deterministic, language-specific resolution rules. For HLS C/C++, dependencies are inferred from resolved include directives, macro references, and user-defined type usages. For VHDL, dependencies are determined by references to design units such as entities, packages, and contexts. For Chisel/Scala, dependencies are identified through resolved imports and statically resolvable symbol usages. For SystemVerilog/Verilog, dependencies are computed from includes, package references, module or interface instantiations, macro usage, and typedef references. In all cases, the final dependency count equals the number of uniquely resolved dependent files, excluding the file itself, and results are reported as averages across task items (or aggregated per file or repository, depending on the evaluation setting).

\subsection{Distribution Across Hardware Semantic Categories}
\label{app:hw_semantic_dis}
\begin{figure}[H]
  \centering
  \setlength{\abovecaptionskip}{4pt}
  \setlength{\belowcaptionskip}{-2pt}

  \begin{subfigure}[t]{0.495\textwidth}
    \centering
    \includegraphics[width=\linewidth,keepaspectratio]{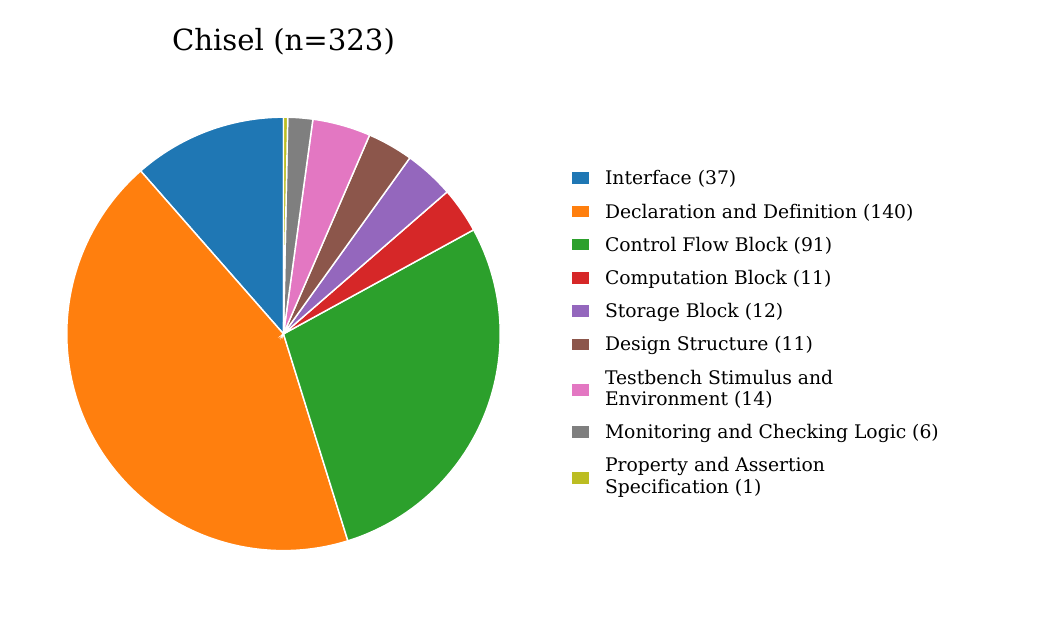}
    \caption{Chisel}
    \label{fig:cat_chisel}
  \end{subfigure}\hfill
  \begin{subfigure}[t]{0.495\textwidth}
    \centering
    \includegraphics[width=\linewidth,keepaspectratio]{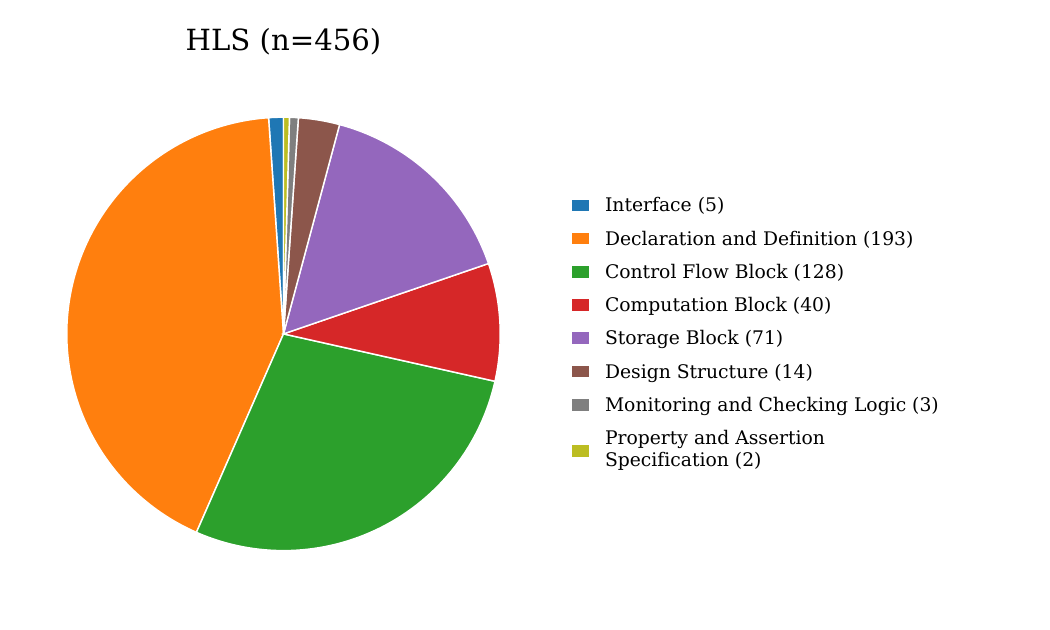}
    \caption{HLS}
    \label{fig:cat_hls}
  \end{subfigure}

  \vspace{0.35em}

  \begin{subfigure}[t]{0.495\textwidth}
    \centering
    \includegraphics[width=\linewidth,keepaspectratio]{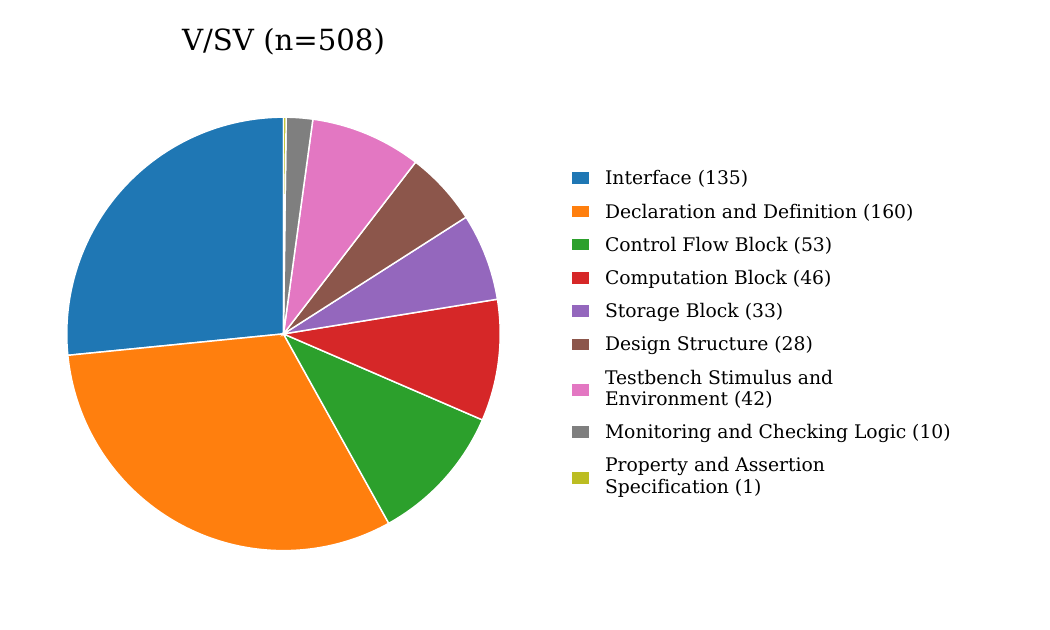}
    \caption{SystemVerilog}
    \label{fig:cat_sv}
  \end{subfigure}\hfill
  \begin{subfigure}[t]{0.495\textwidth}
    \centering
    \includegraphics[width=\linewidth,keepaspectratio]{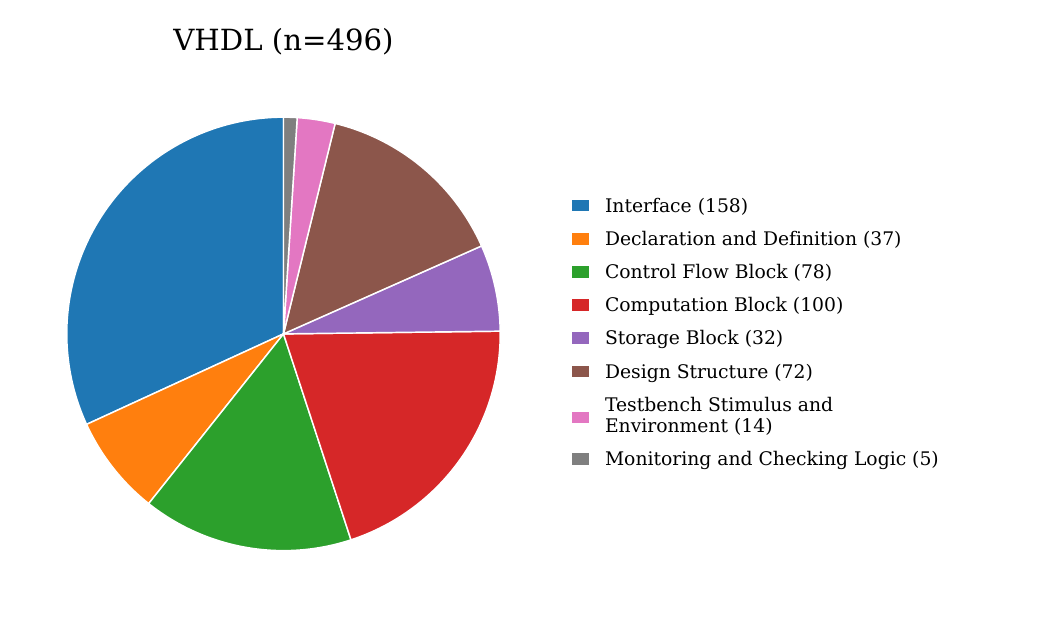}
    \caption{VHDL}
    \label{fig:cat_vhdl}
  \end{subfigure}

  \caption{\textbf{Category distribution of completion targets in MHRC-Bench-Eval.}
  The distributions reveal clear language-specific patterns, reflecting differences in hardware design abstractions, coding styles, and verification practices across Chisel, HLS, SystemVerilog/Verilog, and VHDL.}
  \label{fig:category_distribution}
\end{figure}

Based on the hardware semantic categories defined in Table~\ref{tab:hardware_semantic_categories}, we categorize completion targets in each hardware description language and present their distributions in Figure~\ref{fig:category_distribution}.

\begin{table}[!t]
\centering
\caption{Hardware-oriented semantic categories with language-specific representative constructs. AXI denotes Advanced eXtensible Interface; PSL denotes Property Specification Language.}
\label{tab:hardware_semantic_categories}
\renewcommand{\arraystretch}{1.25}
\setlength{\tabcolsep}{4pt}
\scriptsize
\begin{tabularx}{\textwidth}{|>{\raggedright\arraybackslash}p{3.8cm}|Y|Y|Y|Y|}
\hline
\textbf{Category}
& \textbf{SystemVerilog / Verilog}
& \textbf{VHDL}
& \textbf{Chisel}
& \textbf{HLS C/C++} \\
\hline

Design Structure
& ``module'', ``package''
& ``entity'', ``architecture''
& ``Module'', ``Bundle''
& ``Top Function'', ``Kernel'' \\
\hline

Declaration and Definition
& ``wire'', ``reg''
& ``signal'', ``generic''
& ``IO'', ``Wire''
& ``Function Signature''\\
\hline

Storage Block
& ``reg'', ``always\_ff''
& ``clocked signal'', ``process''
& ``Reg'', ``Mem''
& ``Array'', ``hls::stream'' \\
\hline

Computation Block
& ``Arithmetic Op'', ``Bit Slice''
& ``Arithmetic Op'', ``Type Conversion''
& ``Arithmetic Op'', ``Mux''
& ``Arithmetic Expression'', ``Array Access'' \\
\hline

Control Flow Block
& ``always'', ``if''
& ``process'', ``if''
& ``when'', ``switch''
& ``for loop'', ``if'' \\
\hline

Interface
& ``Port List'', ``Ready/Valid''
& ``Port Map'', ``Component Instantiation''
& ``Decoupled'', ``Valid''
& ``AXI Interface'', ``Stream Interface'' \\
\hline

Property and Assertion Specification
& ``assert'', ``property''
& ``assert'', ``PSL''
& ``assert''
& ``assert'', ``check'' \\
\hline

Testbench Stimulus and Environment
& ``initial'', ``always''
& ``process'', ``wait''
& ``poke'', ``step''
& ``testbench loop'' \\
\hline

Monitoring and Checking Logic
& ``display'', ``error''
& ``report'', ``assert''
& ``expect''
& ``golden check'' \\
\hline

\end{tabularx}
\end{table}

\subsection{Expert-Defined Rule-Based Classification}
\label{app:label_rules}

This section documents the deterministic, expert-defined rules that map each HDL code snippet into one of the nine semantic categories of Table~\ref{tab:hardware_semantic_categories}. Each rule inspects the \emph{ground truth} (the completion target) and its \emph{nearby context} (the last 400 characters of the left context plus the first 400 characters of the right context) using regex pattern matching. Classification is performed per language with a priority-ordered rule chain: the \emph{first matching rule} determines the category. Each rule also produces a subcategory, a predefined confidence score in $[0.50, 0.90]$, and a human-readable reason. The language-specific fallback rules assign snippets without a specific pattern match to one of these nine categories.

\paragraph{Rule design principles.}
\begin{itemize}[itemsep=2pt, topsep=2pt, parsep=0pt, partopsep=0pt, leftmargin=1.5em]
\item \emph{Verification-first priority.} Testbench, assertion, and monitoring rules are checked before design rules, because file-level context (test vs.\ design) is the strongest classification signal.
\item \emph{Keyword + context.} Most rules combine a keyword match in the ground truth with a context check (e.g., \texttt{<=} alone is ambiguous, but \texttt{<=} inside an \texttt{always\_ff} block $\to$ Storage Block).
\item \emph{Confidence scoring.} Each rule carries a confidence score (0.50--0.90): 0.85--0.90 for unambiguous keywords (e.g., \texttt{always\_ff}, \texttt{entity\,...\,is}, \texttt{RegInit}); 0.70--0.80 for strong context-dependent signals (e.g., \texttt{if} inside \texttt{always\_comb}); 0.50--0.65 for fallback rules using weak signals (e.g., a bare \texttt{<=} assignment).
\item \emph{Language-specific adaptation.} Each language has its own classifier with rules tailored to its syntax and idioms, while mapping to the same nine unified categories.
\item \emph{First-match wins.} Rules are ordered by priority; the first matching rule determines the classification, eliminating ambiguity when multiple patterns match.
\end{itemize}

\paragraph{Per-language label-confidence distribution.}
All 1{,}783 evaluation snippets are classified with confidence~$\geq\!0.50$, using the predefined score of the first matching rule. Table~\ref{tab:label_confidence} reports the resulting per-language distribution across the high ($\geq\!0.80$), medium ($[0.60, 0.80)$), and lower ($[0.50, 0.60)$) bands. The majority of snippets in every language fall in the high or medium band; the lower band peaks at $25.7\%$ for Chisel, where the Scala-side syntax has more structural overloading.

\begin{table}[H]
\centering
\small
\captionsetup{font=small, labelfont=small}
\caption{Per-language confidence distribution of the deterministic semantic labels (\% of snippets per band).}
\label{tab:label_confidence}
\setlength{\tabcolsep}{6pt}
\renewcommand{\arraystretch}{1.15}
\begin{tabular}{lccc}
\toprule
\textbf{Language} & \textbf{High ($\geq 0.80$)} & \textbf{Medium ($[0.60, 0.80)$)} & \textbf{Lower ($[0.50, 0.60)$)} \\
\midrule
SystemVerilog & 38.6 & 41.9 & 19.5 \\
VHDL          & 55.6 & 38.1 & 6.2  \\
Chisel        & 55.4 & 18.9 & 25.7 \\
HLS C/C++     & 41.4 & 35.5 & 23.0 \\
\bottomrule
\end{tabular}
\end{table}

\subsubsection{SystemVerilog rules}

\begin{table}[H]
\centering
\scriptsize
\captionsetup{font=small, labelfont=small}
\caption{Priority-ordered classification rules for SystemVerilog.}
\label{tab:rules_sv}
\setlength{\tabcolsep}{4pt}
\renewcommand{\arraystretch}{1.06}
\begin{tabularx}{\linewidth}{lY lc}
\toprule
\textbf{Rule} & \textbf{Pattern (and context)} & \textbf{Subcategory} & \textbf{Conf.} \\
\midrule
\multicolumn{4}{l}{\emph{Priority 1 -- Property and Assertion Specification}} \\
SV-R1 & \texttt{assert property}, \texttt{cover property}, \texttt{assume property} & Assertion Definition & 0.90 \\
SV-R2 & \texttt{property <name>} + \texttt{endproperty} & Assertion Definition & 0.85 \\
SV-R3 & \texttt{property <name>}; nearby context contains \texttt{assert property} & Assertion Definition & 0.80 \\
\midrule
\multicolumn{4}{l}{\emph{Priority 2 -- Monitoring and Checking Logic}} \\
SV-R4 & \texttt{`define}/\allowbreak \texttt{`ASSERT} + \texttt{\$display}/\allowbreak \texttt{\$error}/\allowbreak \texttt{\$warning}/\allowbreak \texttt{\$fatal} & Error Reporting and Logging & 0.85 \\
SV-R5 & \texttt{\$display}/\allowbreak \texttt{\$error}/\allowbreak \texttt{\$warning}/\allowbreak \texttt{\$fatal}/\allowbreak \texttt{\$info}/\allowbreak \texttt{\$write}/\allowbreak \texttt{\$monitor}/\allowbreak \texttt{\$fwrite} ($\geq\!2$ matches or not in \texttt{initial}) & Error Reporting and Logging & 0.75 \\
\midrule
\multicolumn{4}{l}{\emph{Priority 3 -- Testbench Stimulus and Environment}} \\
SV-R6 & \texttt{extends uvm\_test/\allowbreak env/\allowbreak agent/\allowbreak driver/\allowbreak monitor/\allowbreak scoreboard/\allowbreak subscriber} & Environment and Harness Construction & 0.90 \\
SV-R7 & \texttt{extends uvm\_sequence} & Test Scenario and Sequencing & 0.90 \\
SV-R8 & \texttt{uvm\_config\_db} & Environment and Harness Construction & 0.85 \\
SV-R9 & \texttt{initial} + (\texttt{clk}/\allowbreak \texttt{clock} assignment, \texttt{forever}, \texttt{\#delay}, \texttt{\$finish}, \texttt{\$stop}) & Stimulus Generation & 0.85 \\
SV-R10 & \texttt{initial} (standalone) & Stimulus Generation & 0.70 \\
SV-R11 & \texttt{\#<delay>} appears $\geq\!2$ times; not in \texttt{always} block & Stimulus Generation & 0.75 \\
SV-R12 & \texttt{\$finish} or \texttt{\$stop} & Stimulus Generation & 0.70 \\
SV-R13 & \texttt{dut}/\allowbreak \texttt{uut} + \texttt{.<port>(<signal>)} & Environment and Harness Construction & 0.80 \\
SV-R14 & \texttt{wait(...)}; nearby context contains \texttt{initial}/\allowbreak \texttt{\$finish} & Test Scenario and Sequencing & 0.75 \\
SV-R15 & \texttt{clocking} + \texttt{endclocking} & Environment and Harness Construction & 0.80 \\
SV-R16 & \texttt{forever} & Stimulus Generation & 0.75 \\
\midrule
\multicolumn{4}{l}{\emph{Priority 4 -- Design Structure}} \\
SV-R17 & \texttt{module <name>} & Top-level Structure & 0.85 \\
SV-R18 & \texttt{endmodule} & Top-level Structure & 0.80 \\
SV-R19 & \texttt{package <name>} or \texttt{endpackage} & Package and Namespace Definition & 0.85 \\
SV-R20 & \texttt{import <pkg>::} or \texttt{`include}; no \texttt{always}/\allowbreak \texttt{initial}/\allowbreak \texttt{assign} & Package and Namespace Definition & 0.75 \\
\midrule
\multicolumn{4}{l}{\emph{Priority 5 -- Interface}} \\
SV-R21 & \texttt{input}/\allowbreak \texttt{output}/\allowbreak \texttt{inout} declarations ($\geq\!1$); no \texttt{always}/\allowbreak \texttt{initial}/\allowbreak \texttt{if}/\allowbreak \texttt{case} & Port Definition & 0.85 \\
SV-R22 & \texttt{uvm\_*\_port}/\allowbreak \texttt{imp}/\allowbreak \texttt{export} & Port Definition & 0.85 \\
SV-R23 & \texttt{.<port>(<signal>)} ($\geq\!2$); no \texttt{always}/\allowbreak \texttt{initial}/\allowbreak \texttt{input}/\allowbreak \texttt{output} & Component and Module Connection & 0.80 \\
SV-R24 & \texttt{valid} + \texttt{ready} signal assignments & Handshake and Flow Control & 0.80 \\
SV-R25 & AXI handshake signals (\texttt{tvalid}/\allowbreak \texttt{tready}/\allowbreak \texttt{awvalid}/\allowbreak \dots) with assignment & Handshake and Flow Control & 0.80 \\
\midrule
\multicolumn{4}{l}{\emph{Priority 6 -- Storage Block}} \\
SV-R26 & \texttt{always\_ff @(...)} + \texttt{memory[}/\allowbreak \texttt{mem[}/\allowbreak \texttt{ram[} & Memory and Array Storage & 0.85 \\
SV-R27 & \texttt{always\_ff @(...)} & Register-based Storage & 0.85 \\
SV-R28 & \texttt{always @(posedge ...)} + \texttt{<=} + \texttt{memory}/\allowbreak \texttt{mem}/\allowbreak \texttt{ram[} & Memory and Array Storage & 0.80 \\
SV-R29 & \texttt{always @(posedge ...)} + \texttt{<=} & Register-based Storage & 0.80 \\
SV-R30 & \texttt{memory[}/\allowbreak \texttt{mem[}/\allowbreak \texttt{ram[} + \texttt{<=} & Memory and Array Storage & 0.75 \\
\midrule
\multicolumn{4}{l}{\emph{Priority 7 -- Control Flow Block}} \\
SV-R31 & \texttt{case} + \texttt{state}/\allowbreak \texttt{STATE\_} or \texttt{next\_state} in nearby context & State-machine Control & 0.85 \\
SV-R32 & \texttt{case} (general) & State-machine Control & 0.75 \\
SV-R33 & \texttt{next\_*state} assignment & State-machine Control & 0.85 \\
SV-R34 & \texttt{STATE\_*} constants ($\geq\!2$) & State-machine Control & 0.80 \\
SV-R35 & \texttt{if(...)} in \texttt{always\_comb} or \texttt{always @(*)} & Conditional Control & 0.75 \\
\midrule
\multicolumn{4}{l}{\emph{Priority 8 -- Declaration and Definition}} \\
SV-R36 & \texttt{parameter} or \texttt{localparam} & Parameter and Generic Definition & 0.85 \\
SV-R37 & \texttt{`define} or \texttt{`timescale} & Parameter and Generic Definition & 0.85 \\
SV-R38 & \texttt{function}/\allowbreak \texttt{endfunction}/\allowbreak \texttt{task}/\allowbreak \texttt{endtask} & Function and Procedure Signature & 0.80 \\
SV-R39 & \texttt{typedef} & Type Definition & 0.80 \\
SV-R40 & \texttt{wire}/\allowbreak \texttt{logic}/\allowbreak \texttt{reg} declaration; no \texttt{input}/\allowbreak \texttt{output}/\allowbreak \texttt{always}/\allowbreak \texttt{initial} & Signal and Wire Declaration & 0.75 \\
\midrule
\multicolumn{4}{l}{\emph{Priority 9 -- Computation Block}} \\
SV-R41 & \texttt{assign} + bit operators ($\{\}$, $[n\!:\!m]$, \texttt{<<}/\allowbreak \texttt{>>}, \texttt{\textasciicircum}/\allowbreak \texttt{\textasciitilde}/\allowbreak \texttt{|}/\allowbreak \texttt{\&}) & Bit-level Manipulation & 0.80 \\
SV-R42 & \texttt{assign} + arithmetic (\texttt{+}/\allowbreak \texttt{-}/\allowbreak \texttt{*}/\allowbreak \texttt{/}/\allowbreak \texttt{\%}) & Bit-level Manipulation & 0.75 \\
SV-R43 & \texttt{assign} (general) & Bit-level Manipulation & 0.65 \\
SV-R44 & \texttt{\{...,\,...\}} concatenation + \texttt{[n:m]} slice & Bit-level Manipulation & 0.70 \\
\midrule
\multicolumn{4}{l}{\emph{Fallback rules (SV)}} \\
SV-F1 & \texttt{<=} in \texttt{always\_ff}/\allowbreak \texttt{posedge} context & Storage Block & 0.60 \\
SV-F2 & Ternary \texttt{?\,:} & Computation Block & 0.65 \\
SV-F3 & \texttt{begin} + \texttt{if} & Control Flow Block & 0.60 \\
SV-F4 & \texttt{uvm\_*} macro & Testbench Stimulus & 0.65 \\
SV-F5 & \texttt{<=} (non-blocking, no other match) & Storage Block & 0.55 \\
SV-F6 & Near \texttt{initial}/\allowbreak \texttt{\$finish} & Testbench Stimulus & 0.60 \\
SV-F7 & Near \texttt{always} & Control Flow Block & 0.60 \\
SV-F8 & No match & Declaration and Definition & 0.50 \\
\bottomrule
\end{tabularx}
\end{table}

\subsubsection{VHDL rules}

\begin{table}[H]
\centering
\scriptsize
\captionsetup{font=small, labelfont=small}
\caption{Priority-ordered classification rules for VHDL.}
\label{tab:rules_vhdl}
\setlength{\tabcolsep}{4pt}
\renewcommand{\arraystretch}{1.08}
\begin{tabularx}{\linewidth}{lX lc}
\toprule
\textbf{Rule} & \textbf{Pattern (and context)} & \textbf{Subcategory} & \textbf{Conf.} \\
\midrule
\multicolumn{4}{l}{\emph{Priority 1 -- Testbench / Monitoring}} \\
VH-R1 & \texttt{wait for} & Stimulus Generation & 0.85 \\
VH-R2 & \texttt{wait until} & Stimulus Generation & 0.80 \\
VH-R3 & \texttt{assert} + \texttt{report} & Error Reporting and Logging & 0.85 \\
VH-R4 & \texttt{report "..."} & Error Reporting and Logging & 0.75 \\
VH-R5 & \texttt{assert} + \texttt{severity} & Error Reporting and Logging & 0.80 \\
\midrule
\multicolumn{4}{l}{\emph{Priority 2 -- Design Structure}} \\
VH-R6 & \texttt{entity <name> is} & Top-level Structure & 0.90 \\
VH-R7 & \texttt{architecture <name> of} & Top-level Structure & 0.90 \\
VH-R8 & \texttt{end entity}/\texttt{end architecture} & Top-level Structure & 0.80 \\
VH-R9 & \texttt{package <name> is} or \texttt{package body} & Package and Namespace Definition & 0.85 \\
VH-R10 & \texttt{library} + \texttt{use} & Package and Namespace Definition & 0.80 \\
VH-R11 & \texttt{generic(} & Parameterization and Configuration & 0.80 \\
VH-R12 & \texttt{generate} & Hierarchical Composition & 0.80 \\
\midrule
\multicolumn{4}{l}{\emph{Priority 3 -- Interface}} \\
VH-R13 & \texttt{port map} & Component and Module Connection & 0.85 \\
VH-R14 & \texttt{generic map} & Component and Module Connection & 0.80 \\
VH-R15 & \texttt{component <name>} & Component and Module Connection & 0.75 \\
VH-R16 & \texttt{:\,in/out/inout/buffer std\_logic/unsigned/signed/integer/...} & Port Definition & 0.85 \\
VH-R17 & \texttt{port(} & Port Definition & 0.80 \\
VH-R18 & \texttt{axi}/\texttt{wishbone}/\texttt{amba}/\texttt{avalon} + direction or assignment & Protocol-based Interface & 0.75 \\
\midrule
\multicolumn{4}{l}{\emph{Priority 4 -- Storage Block}} \\
VH-R19 & \texttt{rising\_edge()}/\texttt{falling\_edge()}/\texttt{'event and} & Register-based Storage & 0.85 \\
VH-R20 & \texttt{process} + \texttt{clk}/\texttt{clock} + \texttt{<=} & Register-based Storage & 0.80 \\
\midrule
\multicolumn{4}{l}{\emph{Priority 5 -- Control Flow Block}} \\
VH-R21 & \texttt{case ... is} & Conditional Control & 0.80 \\
VH-R22 & \texttt{process} + \texttt{if ... then} & Conditional Control & 0.75 \\
VH-R23 & \texttt{if ... then} in \texttt{process} context & Conditional Control & 0.70 \\
\midrule
\multicolumn{4}{l}{\emph{Priority 6 -- Computation Block}} \\
VH-R24 & \texttt{with ... select} & Mux and Selection Logic & 0.85 \\
VH-R25 & \texttt{<= ... when ... else} & Mux and Selection Logic & 0.80 \\
VH-R26 & \texttt{<= ... when} & Mux and Selection Logic & 0.75 \\
VH-R27 & \texttt{<=} + arithmetic (\texttt{+}/\texttt{-}/\texttt{*}) & Arithmetic Computation & 0.75 \\
VH-R28 & \texttt{to\_unsigned}/\texttt{to\_signed}/\texttt{to\_integer}/\texttt{std\_logic\_vector()} & Type Casting and Conversion & 0.75 \\
VH-R29 & \texttt{sll}/\texttt{srl}/\texttt{sla}/\texttt{sra}/\texttt{rol}/\texttt{ror} & Bit-level Manipulation & 0.75 \\
\midrule
\multicolumn{4}{l}{\emph{Priority 7 -- Declaration and Definition}} \\
VH-R30 & \texttt{signal <name>} & Signal and Wire Declaration & 0.85 \\
VH-R31 & \texttt{variable <name>} & Signal and Wire Declaration & 0.80 \\
VH-R32 & \texttt{constant <name>} & Parameter and Generic Definition & 0.80 \\
VH-R33 & \texttt{type <name> is} & Type Definition & 0.85 \\
VH-R34 & \texttt{subtype <name>} & Type Definition & 0.80 \\
\midrule
\multicolumn{4}{l}{\emph{Fallback rules (VHDL)}} \\
VH-F1 & \texttt{<=} in clocked \texttt{process} context & Storage Block & 0.60 \\
VH-F2 & \texttt{<=} in combinational \texttt{process} context & Control Flow Block & 0.60 \\
VH-F3 & \texttt{elsif}/\texttt{else} in \texttt{process} & Control Flow Block & 0.65 \\
VH-F4 & \texttt{end if}/\texttt{end process}/\texttt{end case} & Control Flow Block & 0.60 \\
VH-F5 & \texttt{when ... =>} in \texttt{case} context & Control Flow Block & 0.65 \\
VH-F6 & \texttt{:=} in \texttt{process} & Control Flow Block & 0.60 \\
VH-F7 & \texttt{:\,in/out} keyword & Interface & 0.55 \\
VH-F8 & \texttt{=>} operator & Interface & 0.55 \\
VH-F9 & No match & Computation Block & 0.50 \\
\bottomrule
\end{tabularx}
\end{table}

\subsubsection{Chisel (Scala) rules}

\begin{table}[H]
\centering
\scriptsize
\captionsetup{font=small, labelfont=small}
\caption{Priority-ordered classification rules for Chisel.}
\label{tab:rules_chisel}
\setlength{\tabcolsep}{4pt}
\renewcommand{\arraystretch}{1.08}
\begin{tabularx}{\linewidth}{lX lc}
\toprule
\textbf{Rule} & \textbf{Pattern (and context)} & \textbf{Subcategory} & \textbf{Conf.} \\
\midrule
\multicolumn{4}{l}{\emph{Priority 1 -- Monitoring / Testbench}} \\
CH-R1 & \texttt{expect(} & Error Reporting and Logging & 0.85 \\
CH-R2 & \texttt{poke(}/\texttt{peek(}/\texttt{step(} & Stimulus Generation & 0.85 \\
\midrule
\multicolumn{4}{l}{\emph{Priority 2 -- Design Structure}} \\
CH-R3 & \texttt{package <name>} & Package and Namespace Definition & 0.85 \\
CH-R4 & \texttt{import <pkg>.} (standalone) & Package and Namespace Definition & 0.75 \\
CH-R5 & \texttt{class <name> extends *Module} & Top-level Structure & 0.85 \\
CH-R6 & \texttt{object <name> extends App} & Top-level Structure & 0.80 \\
CH-R7 & \texttt{case class\,...\,Params}/\texttt{Config} & Parameterization and Configuration & 0.85 \\
\midrule
\multicolumn{4}{l}{\emph{Priority 3 -- Interface}} \\
CH-R8 & \texttt{val io = IO(} & Port Definition & 0.90 \\
CH-R9 & \texttt{class <name> extends Bundle} & Port Definition & 0.90 \\
CH-R10 & \texttt{Input(} + \texttt{Output(} & Port Definition & 0.85 \\
CH-R11 & \texttt{Decoupled(} & Handshake and Flow Control & 0.85 \\
CH-R12 & \texttt{Irrevocable} & Handshake and Flow Control & 0.80 \\
CH-R13 & \texttt{.ready :=} or \texttt{.valid :=} & Handshake and Flow Control & 0.80 \\
CH-R14 & \texttt{Flipped(} & Port Definition & 0.75 \\
\midrule
\multicolumn{4}{l}{\emph{Priority 4 -- Storage Block}} \\
CH-R15 & \texttt{RegInit} & Register-based Storage & 0.85 \\
CH-R16 & \texttt{RegNext} & Register-based Storage & 0.85 \\
CH-R17 & \texttt{RegEnable} & Register-based Storage & 0.85 \\
CH-R18 & \texttt{Reg(} & Register-based Storage & 0.80 \\
CH-R19 & \texttt{ShiftRegister} & Register-based Storage & 0.80 \\
CH-R20 & \texttt{Mem(}/\texttt{SyncReadMem} & Memory and Array Storage & 0.85 \\
\midrule
\multicolumn{4}{l}{\emph{Priority 5 -- Control Flow Block}} \\
CH-R21 & \texttt{when(} & Conditional Control & 0.85 \\
CH-R22 & \texttt{.elsewhen(} & Conditional Control & 0.85 \\
CH-R23 & \texttt{.otherwise} & Conditional Control & 0.80 \\
CH-R24 & \texttt{switch(} & Conditional Control & 0.85 \\
CH-R25 & \texttt{is(} in \texttt{switch} context & Conditional Control & 0.75 \\
CH-R26 & \texttt{for(} & Loop-based Control & 0.80 \\
CH-R27 & \texttt{.foreach}/\texttt{.map}/\texttt{.foldLeft}/\texttt{.reduce} & Loop-based Control & 0.75 \\
\midrule
\multicolumn{4}{l}{\emph{Priority 6 -- Computation Block}} \\
CH-R28 & \texttt{Mux(}/\texttt{Mux1H}/\texttt{MuxLookup}/\texttt{MuxCase}/\texttt{PriorityMux} & Mux and Selection Logic & 0.85 \\
CH-R29 & \texttt{.asUInt}/\texttt{.asSInt}/\texttt{.asBool}/\texttt{.asTypeOf} & Type Casting and Conversion & 0.80 \\
CH-R30 & \texttt{Cat(}/\texttt{Fill(} & Type Casting and Conversion & 0.75 \\
\midrule
\multicolumn{4}{l}{\emph{Priority 7 -- Declaration and Definition}} \\
CH-R31 & \texttt{Wire(}/\texttt{WireInit}/\texttt{WireDefault} & Signal and Wire Declaration & 0.85 \\
CH-R32 & \texttt{def <name>[(:=} & Function and Procedure Signature & 0.80 \\
CH-R33 & \texttt{ChiselEnum} & Type Definition & 0.80 \\
\midrule
\multicolumn{4}{l}{\emph{Fallback rules (Chisel)}} \\
CH-F1 & \texttt{:=} in \texttt{when}/\texttt{switch} context & Control Flow Block & 0.60 \\
CH-F2 & \texttt{:=} in \texttt{valid}/\texttt{ready}/\texttt{Decoupled} context & Interface & 0.55 \\
CH-F3 & \texttt{:=} to \texttt{io.*} & Interface & 0.50 \\
CH-F4 & \texttt{case <name> =>} (Scala match) & Control Flow Block & 0.70 \\
CH-F5 & \texttt{Module(new ...)} & Design Structure & 0.70 \\
CH-F6 & \texttt{require(}/\texttt{assert(} & Property and Assertion & 0.65 \\
CH-F7 & \texttt{val <name> =} & Declaration and Definition & 0.60 \\
CH-F8 & \texttt{:=} (general) & Control Flow Block & 0.55 \\
CH-F9 & No match & Declaration and Definition & 0.50 \\
\bottomrule
\end{tabularx}
\end{table}

\subsubsection{HLS (C/C++) rules}

\begin{table}[H]
\centering
\scriptsize
\captionsetup{font=small, labelfont=small}
\caption{Priority-ordered classification rules for HLS C/C++.}
\label{tab:rules_hls}
\setlength{\tabcolsep}{4pt}
\renewcommand{\arraystretch}{1.08}
\begin{tabularx}{\linewidth}{lX lc}
\toprule
\textbf{Rule} & \textbf{Pattern (and context)} & \textbf{Subcategory} & \textbf{Conf.} \\
\midrule
\multicolumn{4}{l}{\emph{Priority 1 -- Testbench / Monitoring}} \\
HL-R1 & \texttt{testbench} or \texttt{golden}/\texttt{reference}/\texttt{expected} comparison in a loop & Stimulus Generation & 0.75 \\
HL-R2 & \texttt{sc\_main}/\texttt{sc\_start}/\texttt{sc\_trace} & Stimulus Generation & 0.80 \\
HL-R3 & \texttt{int main(} + \texttt{golden}/\texttt{reference}/\texttt{test} in context & Stimulus Generation & 0.70 \\
HL-R4 & \texttt{golden}/\texttt{reference}/\texttt{expected} + \texttt{!=}/\texttt{==} & Error Reporting and Logging & 0.75 \\
HL-R5 & \texttt{printf}/\texttt{fprintf}/\texttt{cout} + \texttt{error}/\texttt{fail}/\texttt{mismatch} & Error Reporting and Logging & 0.75 \\
\midrule
\multicolumn{4}{l}{\emph{Priority 2 -- Interface}} \\
HL-R6 & \texttt{\#pragma HLS INTERFACE} + \texttt{axi}/\texttt{axis}/\texttt{ap\_ctrl}/\texttt{s\_axilite}/\texttt{m\_axi} & Protocol-based Interface & 0.85 \\
HL-R7 & \texttt{\#pragma HLS INTERFACE} (general) & Port Definition & 0.80 \\
HL-R8 & \texttt{hls::stream} + \texttt{read(}/\texttt{write(} & Protocol-based Interface & 0.75 \\
\midrule
\multicolumn{4}{l}{\emph{Priority 3 -- Design Structure}} \\
HL-R9 & \texttt{\#pragma HLS top}/\texttt{interface} or \texttt{\#pragma hls\_design} & Top-level Structure & 0.85 \\
HL-R10 & \texttt{\#include} (standalone) & Package and Namespace Definition & 0.80 \\
HL-R11 & \texttt{\#ifndef}/\texttt{\#endif}/\texttt{\#pragma once} & Package and Namespace Definition & 0.75 \\
HL-R12 & \texttt{namespace}/\texttt{using namespace} & Package and Namespace Definition & 0.75 \\
\midrule
\multicolumn{4}{l}{\emph{Priority 4 -- Storage Block}} \\
HL-R13 & \texttt{hls::stream}/\texttt{ac\_channel} & Memory and Array Storage & 0.85 \\
HL-R14 & \texttt{\#pragma HLS array\_partition}/\texttt{array\_reshape}/\texttt{stream}/\texttt{bind\_storage} & Memory and Array Storage & 0.85 \\
HL-R15 & \texttt{int}/\texttt{float}/\texttt{ap\_uint}/\texttt{...} \texttt{<name>[} (array declaration) & Memory and Array Storage & 0.80 \\
\midrule
\multicolumn{4}{l}{\emph{Priority 5 -- Control Flow Block}} \\
HL-R16 & \texttt{for(} & Conditional Control & 0.80 \\
HL-R17 & \texttt{while(} & Conditional Control & 0.80 \\
HL-R18 & \texttt{\#pragma HLS unroll}/\texttt{pipeline}/\texttt{loop} & Conditional Control & 0.80 \\
HL-R19 & \texttt{if(} & Conditional Control & 0.75 \\
HL-R20 & \texttt{switch(} & Conditional Control & 0.75 \\
\midrule
\multicolumn{4}{l}{\emph{Priority 6 -- Declaration and Definition}} \\
HL-R21 & \texttt{void/int/float/...\,<name>(} (function signature) & Function and Procedure Signature & 0.80 \\
HL-R22 & \texttt{template<} & Function and Procedure Signature & 0.75 \\
HL-R23 & \texttt{class}/\texttt{struct} & Function and Procedure Signature & 0.75 \\
HL-R24 & \texttt{\#define <name>} & Parameter and Generic Definition & 0.80 \\
HL-R25 & \texttt{static const}/\texttt{constexpr} & Parameter and Generic Definition & 0.80 \\
HL-R26 & \texttt{typedef}/\texttt{enum} & Parameter and Generic Definition & 0.75 \\
HL-R27 & \texttt{int/float/ap\_uint/...\,<name>;} or \texttt{= ...;} & Signal and Wire Declaration & 0.70 \\
\midrule
\multicolumn{4}{l}{\emph{Priority 7 -- Computation Block}} \\
HL-R28 & Compound assignment (\texttt{+=}, \texttt{-=}, \texttt{*=}, \texttt{/=}, \texttt{\%=}) & Arithmetic Computation & 0.75 \\
HL-R29 & \texttt{= <var> +/-/*\,<var>} & Arithmetic Computation & 0.70 \\
HL-R30 & Ternary \texttt{?\,:} & Mux and Selection Logic & 0.70 \\
\midrule
\multicolumn{4}{l}{\emph{Fallback rules (HLS)}} \\
HL-F1 & \texttt{stream<} type & Interface & 0.65 \\
HL-F2 & \texttt{new[]}/\texttt{malloc}/\texttt{calloc} & Storage Block & 0.65 \\
HL-F3 & \texttt{<type>* <name>} (pointer) & Declaration and Definition & 0.60 \\
HL-F4 & \texttt{return} & Control Flow Block & 0.60 \\
HL-F5 & \texttt{<var> = <var>} (general assignment) & Computation Block & 0.55 \\
HL-F6 & \texttt{\}} (closing brace) & Design Structure & 0.55 \\
HL-F7 & \texttt{=} (any) & Declaration and Definition & 0.55 \\
HL-F8 & No match & Declaration and Definition & 0.50 \\
\bottomrule
\end{tabularx}
\end{table}

\subsection{Evaluation Configurations} \label{app:eval-config}
\subsubsection{Prompt}
The exact prompt used during model inference for preceding-context prompt is given in Figure~\ref{fig:prompt_norag_preceding}. For full-context tasks on local LLMs, we use their native fill-in-the-middle tokens rather than this natural-language instruction.



\begin{figure}[ht]
    \centering
    \begin{tcolorbox}[colback=gray!5!white, colframe=gray!75!black, title=Task Prompt]
    \small
    \ttfamily
    \raggedright
    You are a precise code-completion model completing a repository blank completion task. Input: a repository concatenation where each file of the repo is wrapped in --- FILE: <repo\_name>/<rel/path/to/file> (BEGIN) ---<file content...>
    --- FILE: <repo name>/<rel/path/to/file> (END) ---. Some lines have been replaced by the exact token <TARGET>. Your task: reconstruct \textbf{only} the replaced lines, using the repository context.\\
    OUTPUT: the reconstructed lines (and nothing else). Target line count: {target\_lines}. Output exactly {target\_lines} line(s). 

    <Left Context>... ...</Left Context>\\
    <TARGET>\par
    \end{tcolorbox}
    \caption{The preceding-context prompt used for repository-level code completion (No RAG). If the context length exceeds the token limit, the head of the left context and the tail of the right context are trimmed.}
    \label{fig:prompt_norag_preceding}
\end{figure}

\subsubsection{Fine-Tuning}
\begin{figure}[ht]
    \centering
    \includegraphics[width=\linewidth]{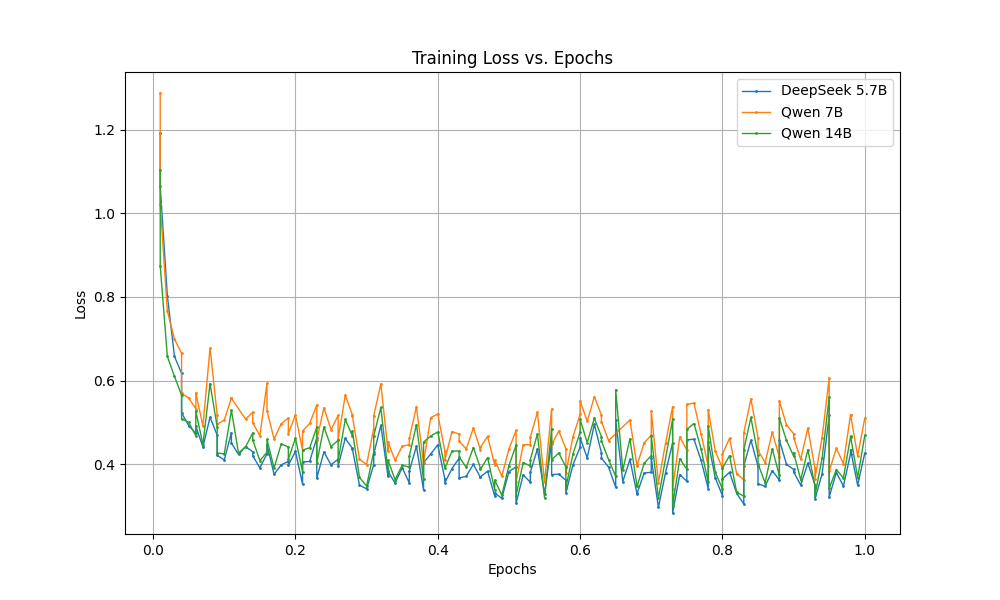} 
    
    \caption{Training loss over epochs for Qwen2.5 and DeepSeek models.}
    \label{fig:loss_plot}
\end{figure}

For fine-tuning local LLMs (Qwen and DeepSeek) under preceding-context-only settings, we model training as a completion instruction task. The prompt contains concatenated repository context with file delimiters and a \texttt{\textless TARGET\textgreater} token representing the missing span. For full-context tasks, where both left and right context are available, we use the model's native FIM tokens (for instance, \texttt{<|fim\_prefix|>}, \texttt{<|fim\_suffix|>}, and \texttt{<|fim\_middle|>}) instead of natural-language task descriptions. To manage context-window constraints, inputs are dynamically truncated to a maximum of 2,048 tokens with a balanced budget that prioritizes the immediate left and right context around the target. Fine-tuning uses LoRA on all linear attention and feed-forward layers, with rank $r=16$, scaling factor $\alpha=32$, and dropout $0.05$. We optimize with AdamW using a peak learning rate of $2 \times 10^{-4}$ and a cosine decay schedule with a 3\% warm-up over one epoch. Cross-entropy loss is computed only on the generated ground-truth span, masking instruction and context tokens. Figure~\ref{fig:loss_plot} plots training loss for Qwen2.5-Coder-7B, Qwen2.5-Coder-14B, and DeepSeek-Coder-5.7B-MQA-base.

\subsubsection{Evaluation Metrics}
During inference, the model is configured with temperature $0.0$, top\_p $=1$, and a maximum generation length of $64$ tokens. We use this maximum length because all ground-truth completions in our dataset contain no more than 64 tokens. Generation stops when the \texttt{eos\_token} is produced or when the maximum length is reached.

Prior to calculating EM and ES, generated code is stripped of comments, trimmed for whitespace, and filtered to remove empty lines. EM is set to $1$ if the cleaned generation exactly matches the ground truth and $0$ otherwise. ES is computed as:
\[
\mathrm{ES} \;=\; 1 - \frac{\operatorname{editdistance}\!\left(\text{target\_str}, \text{prediction\_str}\right)}{\max\!\left(1,\; \lvert \text{target\_str} \rvert,\; \lvert \text{prediction\_str} \rvert\right)}
\]
where $editdistance$ is the Levenshtein edit distance between the generated code and the ground truth.

\subsubsection{Commercial API Setting}
For commercial LLMs, we use API services provided by ChatfireAPI. The model names are \texttt{gpt-5}, \texttt{deepseek-v3.2}, \texttt{gemini-2.5-pro}, and \texttt{grok-4}. We use the API defaults for temperature, top\_p, max\_tokens, and other generation parameters.

\subsection{Functional Evaluation Setup}
\label{app:functional_eval}

\paragraph{Coverage.} The functional evaluation in
Figure~\ref{fig:functional_pass} covers evaluated subsets of MHRC-Bench-Eval.
The released testbench inventory documents partial coverage.
An evaluated completion passes only when its testbench or canonical I/O
checks confirm the expected behaviour.

\paragraph{Per-language toolchains.} Table~\ref{tab:func_eval_toolchain}
lists the simulator/compiler used per language. Reported numbers are
\textbf{pass@1}, using one completion per problem under the
model-specific decoding settings in Appendix~\ref{app:eval-config}.
We do not evaluate multiple sampled completions per problem.

\begin{table}[ht]
\centering
\small
\caption{Functional-evaluation toolchain by language.}
\label{tab:func_eval_toolchain}
\begin{tabular}{ll}
\toprule
Language & Tool \\
\midrule
HLS C/C++       & Vitis HLS / g++ with \texttt{ac\_int}   \\
SystemVerilog   & Icarus Verilog / Verilator              \\
VHDL            & GHDL                                    \\
Chisel          & sbt + ChiselTest                        \\
\bottomrule
\end{tabular}
\end{table}

\paragraph{Semantic categories.} The full MHRC-Bench-Eval test split spans all
nine hardware-semantic categories of Section~\ref{text:dataset_types};
the per-category breakdown is reported in
Appendix~\ref{app:hw_semantic_dis}.

\subsection{Decoding and Statistical Uncertainty}
\label{app:seed_var}

\paragraph{Decoding settings.} Code LLMs use greedy decoding with
fixed model weights, temperature $0$, and top-$p$ $1$. Commercial API
models use the provider defaults described in
Appendix~\ref{app:eval-config}. Reproducing a run also depends on the
model checkpoint and serving implementation.

\paragraph{Statistical uncertainty.} The reported EM, ES, and
functional pass rates are point estimates. We do not report confidence
intervals, significance tests, or means and standard deviations across
independent fine-tuning seeds. The results therefore do not quantify
sampling uncertainty or variation across fine-tuning runs.

\subsection{Case Study of Pretrained-Model Responses}
\label{app:case_study_finetuned}
To illustrate the effect of fine-tuning on local models, we analyze the completions generated by Qwen2.5-Coder-7B on fill-in-the-middle tasks before and after fine-tuning. Specifically, we list for each tested language common errors made by the model that were significantly improved by fine-tuning.

    \subsubsection{Verilog/SystemVerilog} 
    
    \textbf{Syntax Hallucination:} One of the most frequent mistakes made by the pre-trained model is completely losing track of the syntax context. in \texttt{AXI-SDCard-High-Speed-Controller}, for instance, the model hallucinates a module instantiation statement \texttt{.waddr(waddr),.wdata(wdata)...} instead of the expected IO declaration \texttt{input logic[2:0] axilite\_arprot}
    
    \textbf{Structural Hallucination:} The pre-trained model is prone to regurgitating file headers, copyright notices, or artificial file boundary markers in the middle of code blocks. In \texttt{custom\_\allowbreak uvm\_\allowbreak report\_\allowbreak server} for instance, the pre-trained model suddenly outputs a full copyright header \texttt{// Copyright 2007-2010 Mentor Graphics...}inside the \texttt{build\_phase} function instead of the expected \texttt{super.build\_phase(phase);} call.
    
    \textbf{Pattern Oversight:} The pre-trained model fails to follow obvious naming sequences or structural patterns established immediately before the target line. In \texttt{SystemVerilog Assertions}, for instance, the model failed to recognize the naming pattern of \texttt{logic gnt1; logic gnt2;} and even missed the comment \texttt{//3 Requests to Arbiter logic req1;logic req2; logic req3;}. Instead, it regurgitated earlier comments and produced irrelevant code.
    
    \textbf{Logical Confusion:} The pre-trained model sometimes generates code that is syntactically valid SystemVerilog but logically disconnected from the immediate control flow (e.g., jumping to a different state in a Finite State Machine). In \texttt{virtio}, the code is strictly defining an \texttt{always\_ff} block for reset logic. The pre-trained model attempts to insert FSM state logic \texttt{\_EVENT\_IDX: begin...} from a case statement that belongs elsewhere, ignoring the  \texttt{always\_ff} structure.   
        
    \subsubsection{HLS} 
        
        \textbf{Structural Hallucination:} Similar to other languages, the pre-trained model for HLS code often completely loses track of the file position and hallucinates the start of a new file. Instead of generating the next line of code, the model outputs copyright notices, include guards, or library imports. For example, in \texttt{hls-spmv}, the pre-trained model outputs \texttt{\#include "spmv.h"} despite the cursor being positioned in the \texttt{spmv\_kernel} function.
        
        \textbf{C++/HLS Confusion:} HLS often requires specific syntax not included in standard C++ that are used for reporting and optimization directives. The pre-trained model often writes valid standard C++ but misses these hardware-description necessities. In \texttt{hls\_spmv}, for instance, the ground truth required an \texttt{L1:} in the loop \texttt{L1: for (int i = 0; i < NUM\_ROWS; i++)}, which the pre-trained model neglected.
        
        \textbf{Syntax Hallucination:} Similar to other languages, the pre-trained model fails to recognize the contextual logic and syntax. In \texttt{libllsm}, for instance, while the context explicitly requires a memory allocation (\texttt{calloc}), the pre-trained model instead attempts to define a new function \texttt{FP\_TYPE* llsm\_envelope...}.

    \subsubsection{VHDL} 

        \textbf{Structural Confusion:} Similar to other languages, the pre-trained model struggles to comprehend the context and has the tendency to hallucinate file headers, library declarations, or copyright notices in the middle of an active code block. For the \texttt{FPGA-Speech-Recognition} repository, for instance, the pre-trained model outputted \texttt{library IEEE; use IEEE.STD\_LOGIC\_1164.ALL;... entity UART\_PARITY....} despite the target line being deep inside a \texttt{generate} block for odd parity. More severely, the model often confuses code segments with prompt structures and output strings used to mark file boundaries in prompts. In the \texttt{cpu\_for\_nscscc2022\_single} repository, the model outputs \texttt{--- FILE: dcache\_tag.vhd (END) ---} immediately, skipping the required generic map assignments \texttt{C\_MEM\_TYPE => 1}.
        
        \textbf{Syntax Confusion:} VHDL distinguishes heavily between concurrent statements and sequential statements inside processes or subprograms. The pre-trained model often mixes these contexts. In \texttt{FPGA-I2C-Minion}, the pre-trained model hallucinates unrelated procedure calls \texttt{i2c\_wait\_quarter\_clock} or definitions that belong in a testbench or architecture body, not the sequential spike generator logic that the target line is in.

    \subsubsection{Chisel}
        \textbf{Structural Confusion:} Similar to other languages, the pre-trained model often hallucinates copyright headers and library declarations.
        
        \textbf{Language Confusion:} The pre-trained model sometimes fails to generate valid Chisel syntax or uses generic tokens that don't fit the Chisel hardware construction paradigm. For example, in \texttt{rjrouter}, the ground truth is \texttt{c => new GCDUnitTester(c)}, but the pre-trained model outputs \texttt{<|fim\_middle|>}, indicating a complete failure to generate content.

A list of examples referenced above can be found in Table \ref{tab:pretrain_outputs}


\begin{figure*}[t]
  \centering
  \setlength{\tabcolsep}{3pt}  
  \renewcommand{\arraystretch}{0.95} 

  \begin{tabular}{@{}cc@{}}
    \subcaptionbox{\textbf{HLS:} Line Number\label{fig:hls_line}}{%
      \includegraphics[width=0.43\textwidth]{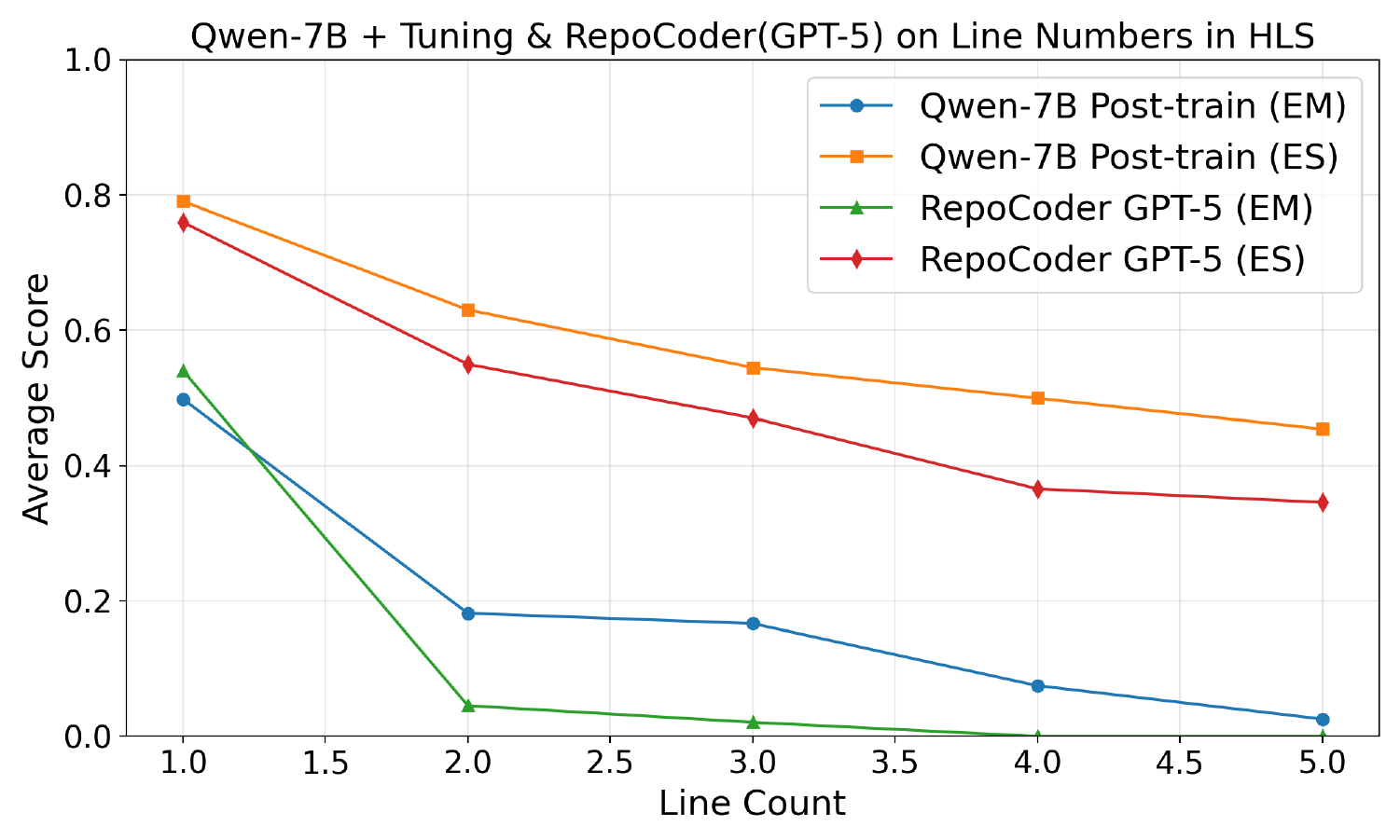}} &
    \subcaptionbox{\textbf{SystemVerilog:} Line Number\label{fig:sv_line}}{%
      \includegraphics[width=0.43\textwidth]{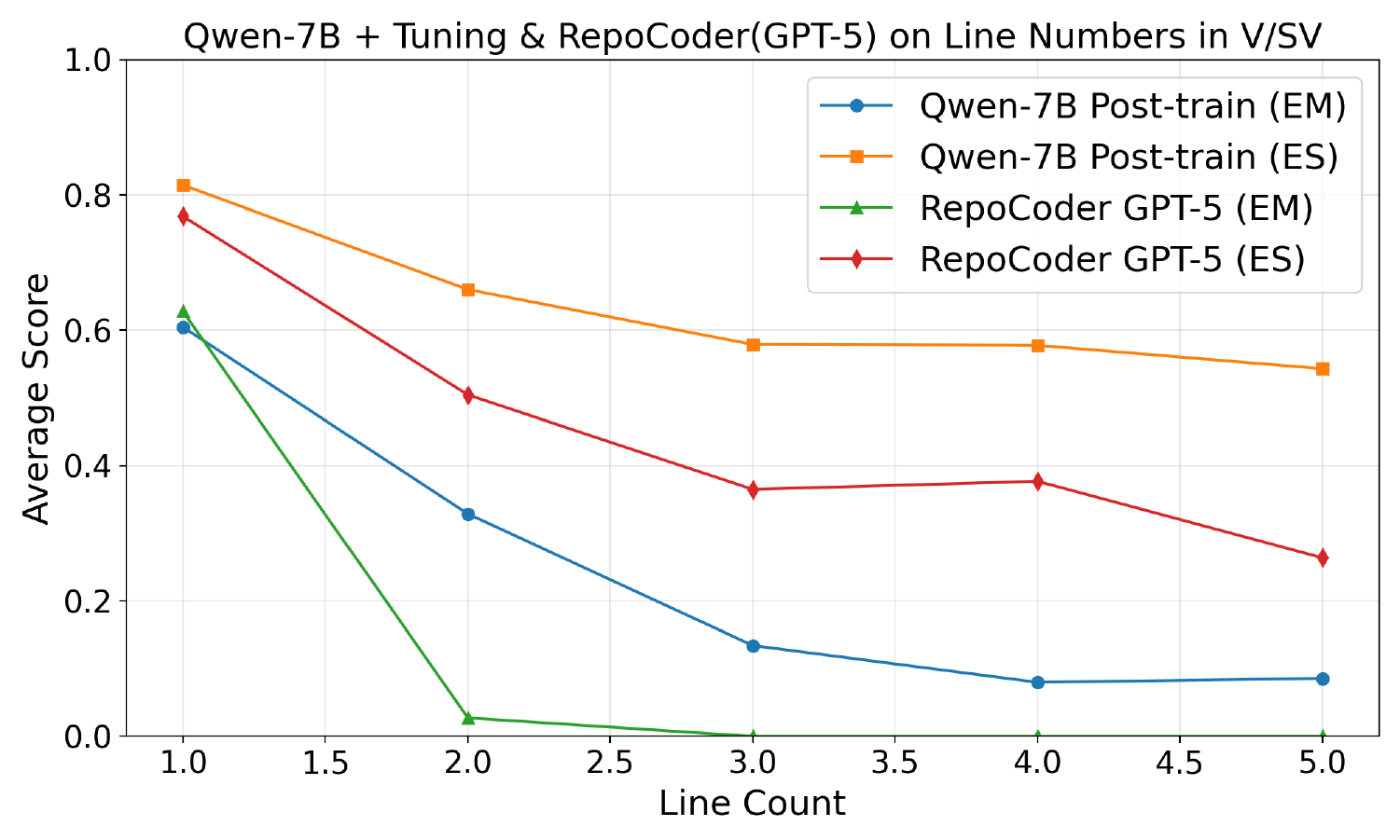}} \\[-2pt]

    \subcaptionbox{\textbf{HLS:} Structural Complexity\label{fig:hls_depth}}{%
      \includegraphics[width=0.43\textwidth]{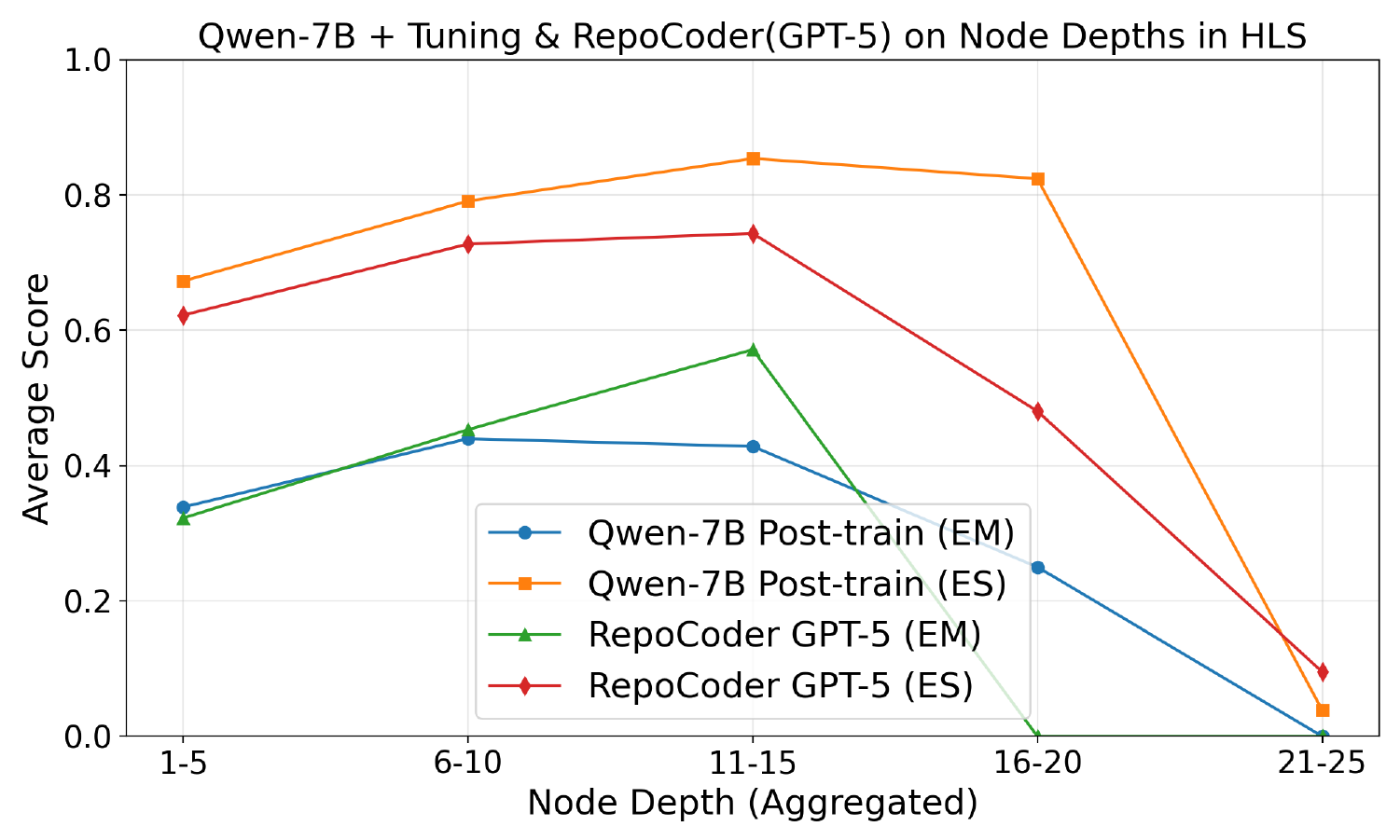}} &
    \subcaptionbox{\textbf{SystemVerilog:} Structural Complexity\label{fig:sv_depth}}{%
      \includegraphics[width=0.43\textwidth]{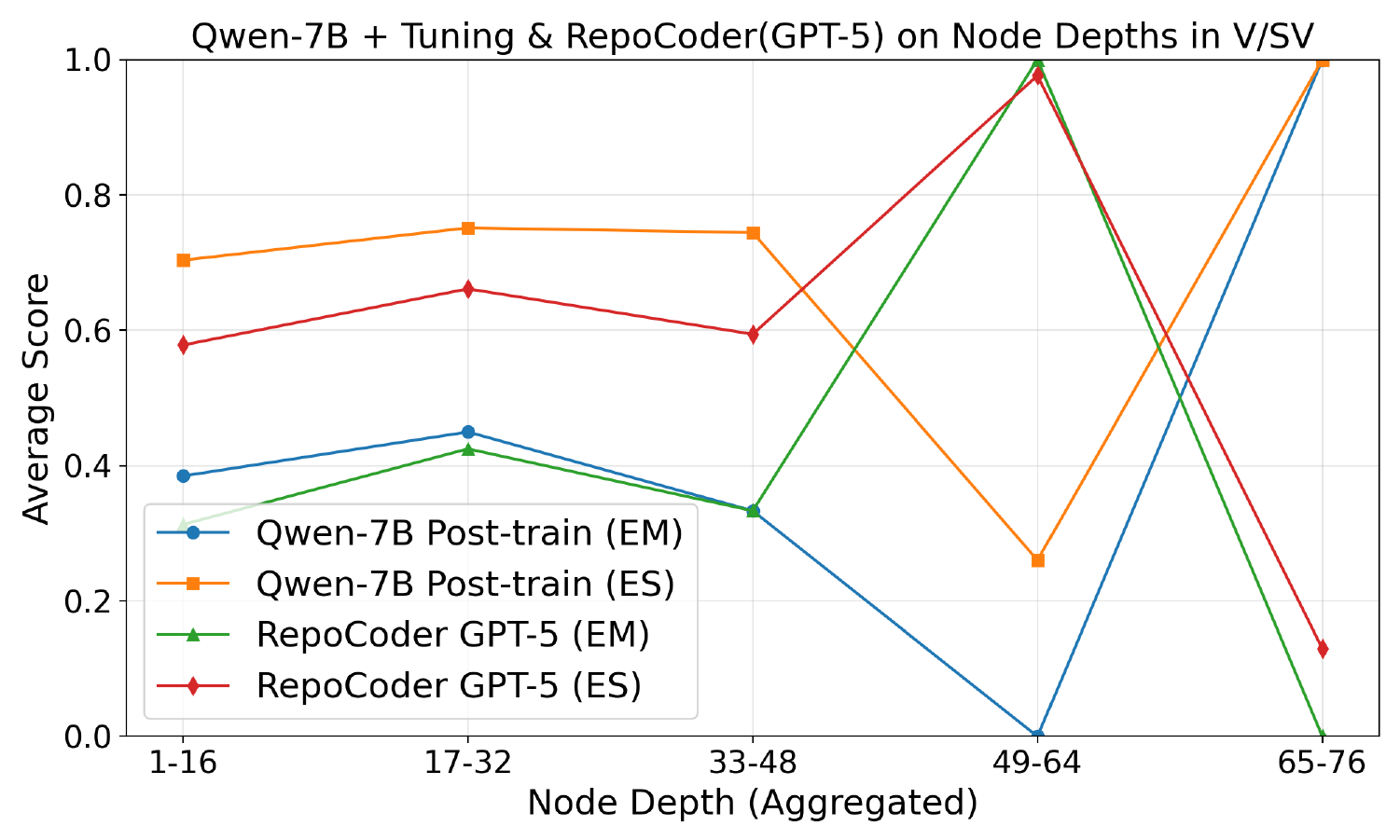}} \\[2pt]

    \subcaptionbox{\textbf{Chisel:} Line Number\label{fig:chisel_line}}{%
      \includegraphics[width=0.43\textwidth]{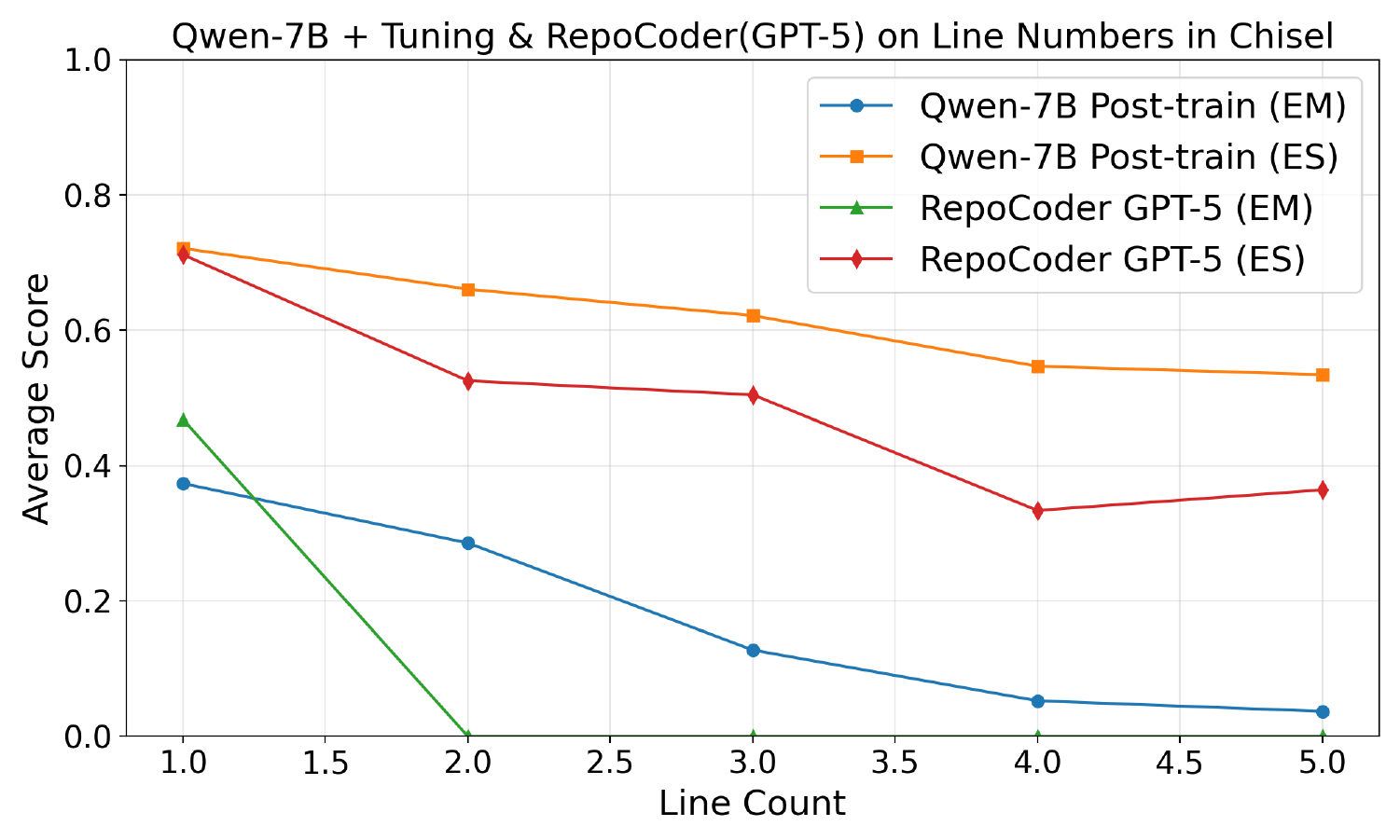}} &
    \subcaptionbox{\textbf{VHDL:} Line Number\label{fig:vhdl_line}}{%
      \includegraphics[width=0.43\textwidth]{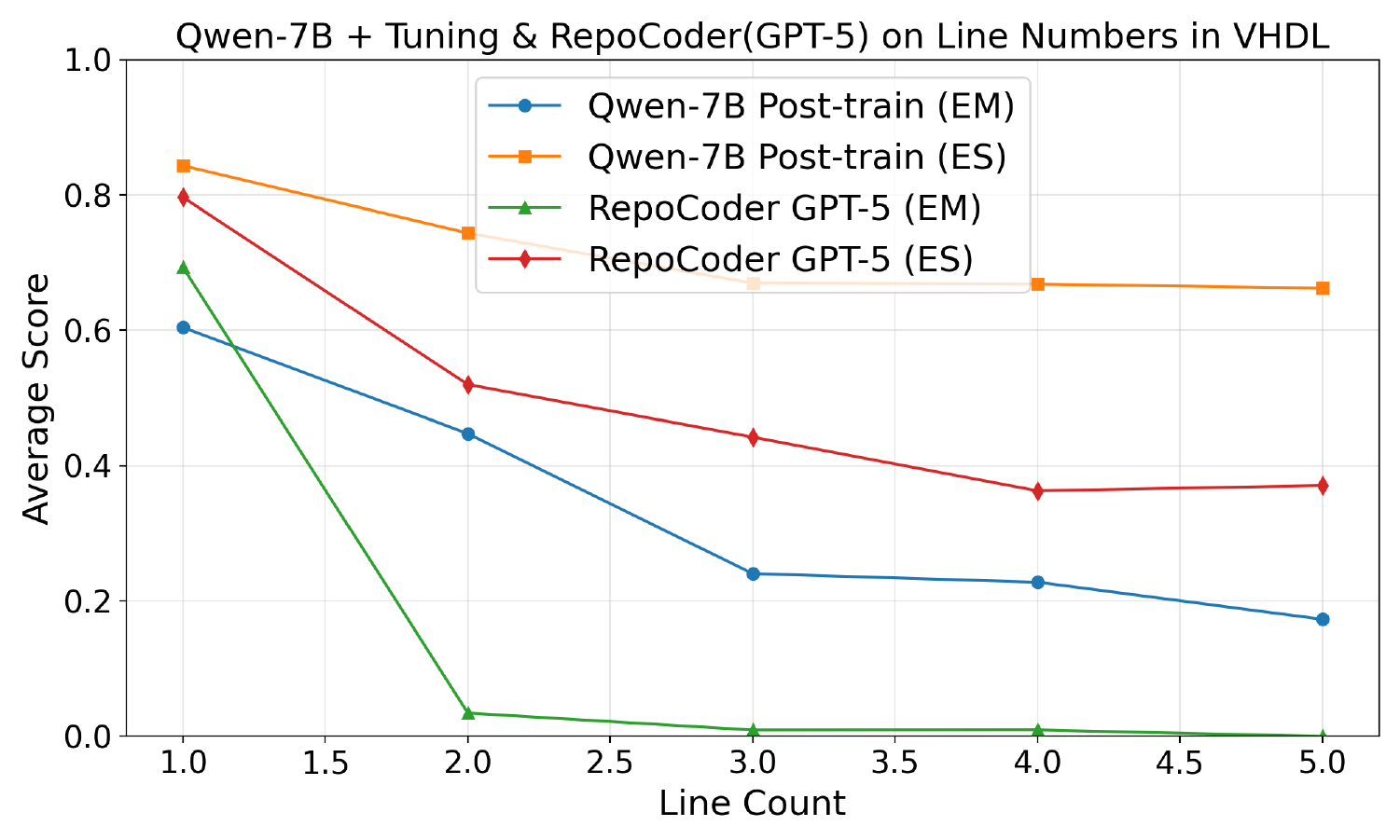}} \\[-2pt]

    \subcaptionbox{\textbf{Chisel:} Structural Complexity\label{fig:chisel_depth}}{%
      \includegraphics[width=0.43\textwidth]{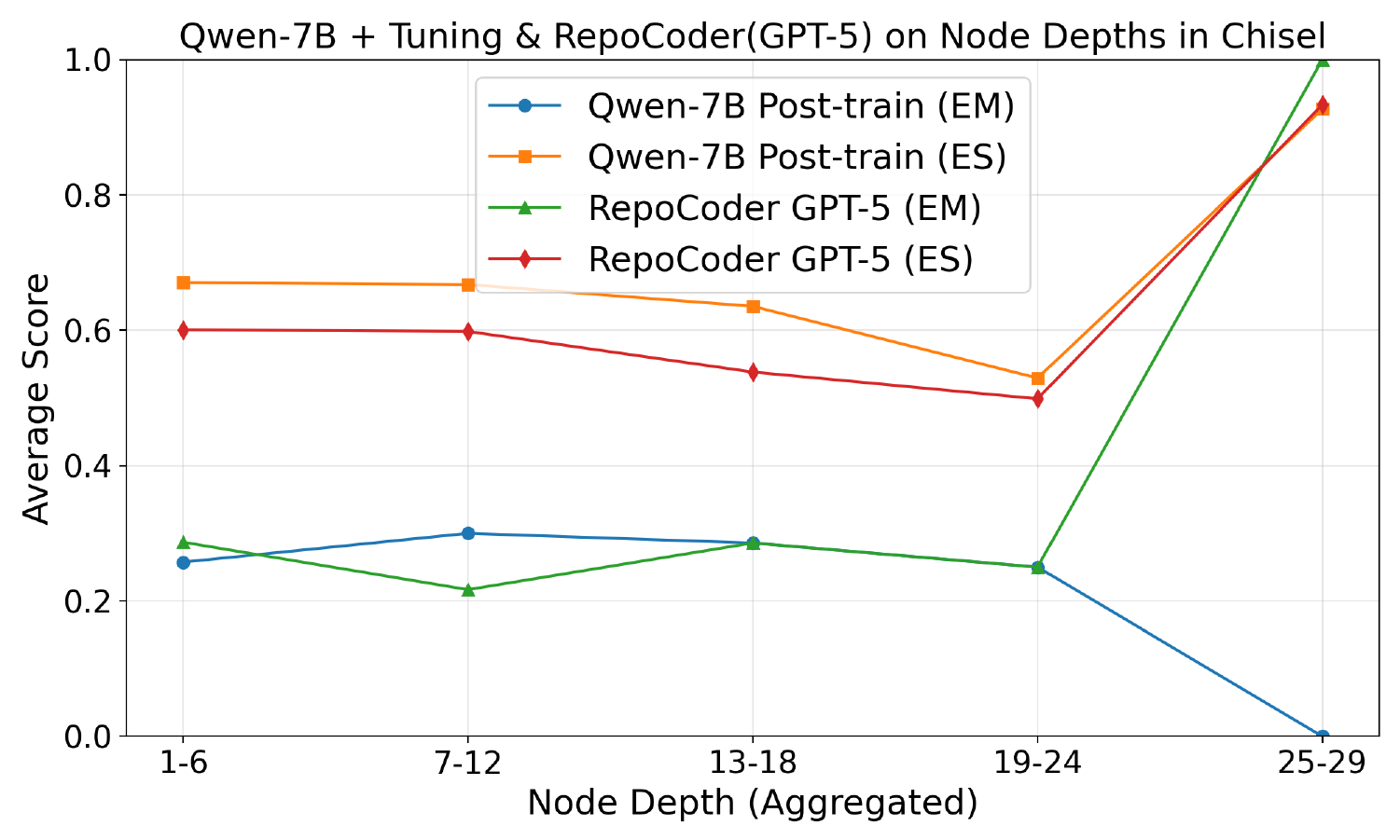}} &
    \subcaptionbox{\textbf{VHDL:} Structural Complexity\label{fig:vhdl_depth}}{%
      \includegraphics[width=0.43\textwidth]{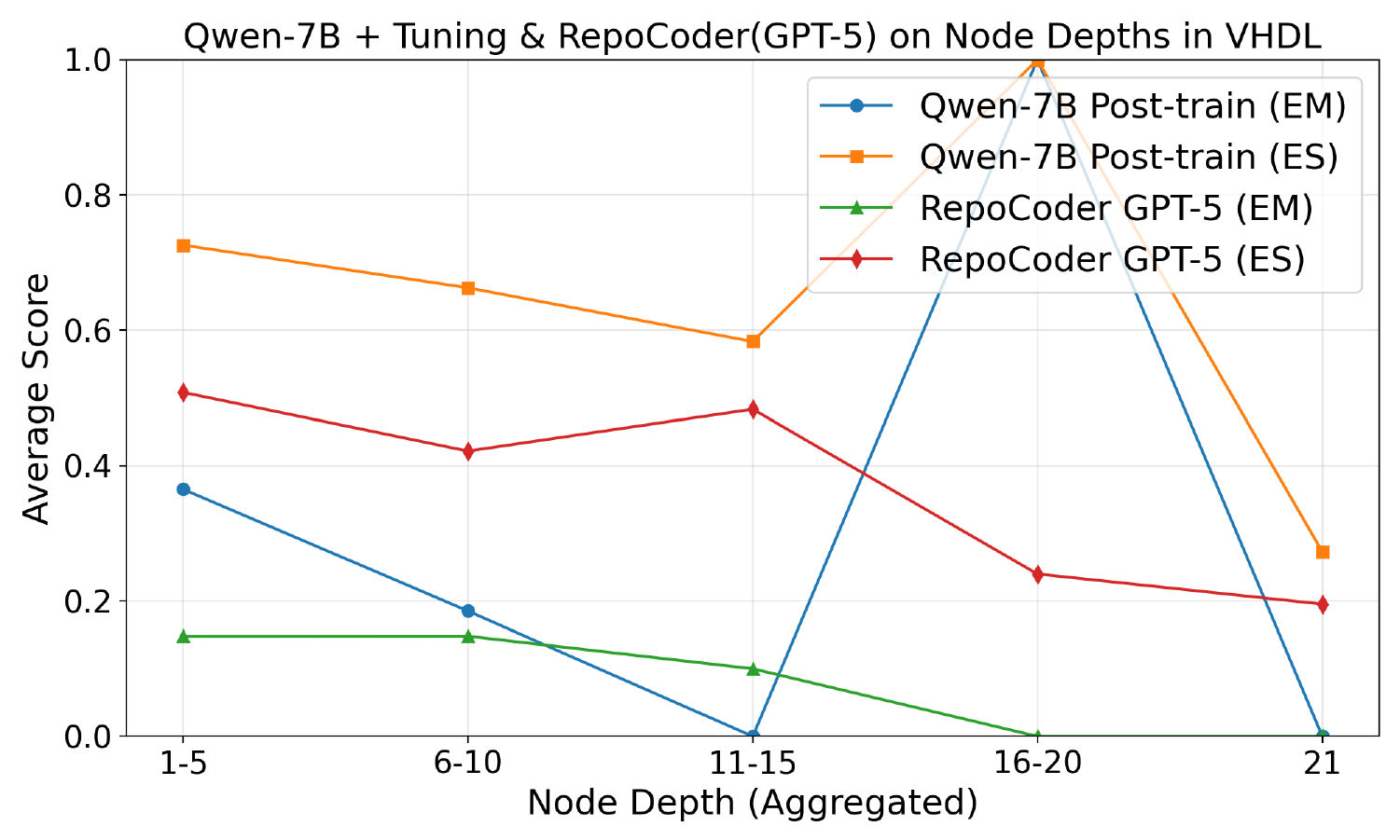}} \\
  \end{tabular}

  \caption{EM performance across languages. Rows correspond to line number (rows 1 and 3) and structural complexity (rows 2 and 4).}
  \label{fig:lang_line_node_performance_all}
\end{figure*}

\subsection{GraphCoder Modifications and Limitations} \label{subsec:graph_coder}
The original implementation of GraphCoder is designed for preceding context tasks. To evaluate its performance on full context tasks, we modified GraphCoder such that the right context is also preserved and included during prompt assembly. As the graph builder and the retriever operate independently of the left and right contexts, this has no other effect on the functionality of GraphCoder.

Furthermore, the original implementation of GraphCoder included helper functions in \texttt{utils/\allowbreak ccg.py} for traversing CSTs, but these functions only supported Java and Python. To extend support to hardware languages, we added helper functions for each language's syntax and for identifying control-flow and data dependencies. We also made minor changes to include HDL languages during \texttt{tree-sitter} parser selection and code-file filtering.

GraphCoder underperforms on HDL repositories compared with other methods. We attribute this to its retriever, which has lower hit@k on HDL languages than on software languages such as Python, as shown in Table~\ref{tab:hitk_comparison}. By GraphCoder's design, ineffective retrieval floods the context window with irrelevant code snippets, causing GraphCoder to perform worse than No RAG.

Further analysis reveals properties of HDL languages that turn subtle GraphCoder design choices into liabilities for effective retrieval. Using HLS as an example, we compare snippets retrieved by GraphCoder and RepoCoder and identify two key differences.

\textbf{Absence of function declarations.} We found that 16.9\% of snippets retrieved by RepoCoder contain HLS function definitions, while the number drops to 1.8\% for GraphCoder. This arises from a design in the official GraphCoder implementation: line 79 of \texttt{utils/slicing.py} explicitly filters out nodes with \texttt{definition} in their type names, likely to focus on executable statements. However, this excludes many function signatures and affects HLS retrieval because key synthesis constraints (types, interfaces, and pragmas that shape parallelism) are often encoded in definitions and headers rather than in nearby data/control-flow statements.

\textbf{Absence of HLS constructs.} Because \texttt{tree-sitter} has no dedicated HLS grammar, we use \texttt{tree-sitter-cpp} to support HLS. However, the C++ grammar does not process HLS-specific constructs, causing them to be left out of retrieval. For instance, 4.2\% of snippets retrieved by RepoCoder contain HLS-specific directives, while the rate for GraphCoder is only 0.7\%. This causes many information-carrying synthesis constraints to be systematically omitted from GraphCoder's retrieved context. In HLS, pragmas and vendor-specific types are not merely stylistic annotations: they encode parallelization, memory partitioning, and interface protocols that directly affect correctness and implementability.

We also observe a cross-language weakness in GraphCoder retrieval: snippet length divergence. Table~\ref{tab:snippet_length_center90} reports the mean and standard deviation of retrieved snippet lengths for RepoCoder and GraphCoder. GraphCoder's snippets have roughly half the mean length but triple the standard deviation of RepoCoder snippets, suggesting instability in GraphCoder's handling of HDL CST structures.

\begin{table}[H]
\centering
\small
\caption{Snippet length statistics (characters) over the center 90\% of retrieved snippets.}
\begin{tabular}{lccc}
\toprule
Retriever & \# Snippets (center 90\%) & Mean Length (chars) & Std. Dev. (chars) \\
\midrule
GraphCoder & 1595 & 2868.10 & 3223.60 \\
RepoCoder  & 1596 & 4536.98 & 1060.49 \\
\bottomrule
\end{tabular}

\label{tab:snippet_length_center90}
\end{table}

\begin{table}[t]
\centering
\small
\caption{Retriever hit@k on HDL languages and Python}
\begin{tabular}{lcc}
\toprule
\textbf{Language} & \textbf{Hit@1} & \textbf{Hit@5} \\
\midrule
Scala                 & 0.03 & 0.05 \\
VHDL                  & 0.11 & 0.16 \\
SystemVerilog               & 0.09 & 0.12 \\
HLS                   & 0.19 & 0.25 \\
\midrule
Python (API-level)  & 0.29 & 0.35 \\
Python (Line-level) & 0.27 & 0.32 \\
\bottomrule
\end{tabular}

\label{tab:hitk_comparison}
\end{table}

\subsection{Context Window Sensitivity}
\label{app:ctx_window}
To assess how sensitive results are to context window size, we evaluate four fine-tuned models and two proprietary LLMs across context windows from 512 to 8{,}192 tokens. Table~\ref{tab:ctx_window} reports EM (\%) and ES (\%) for each setting; the model's best context window is highlighted in bold per row group. All fine-tuned models peak at 2{,}048 tokens with sharp degradation beyond, whereas the two proprietary LLMs (Gemini-2.5 and GPT-5) keep improving up to 8{,}192 tokens on most languages. Longer contexts overwhelm smaller fine-tuned models but benefit larger general-purpose ones; we therefore adopt 2{,}048 tokens in the main paper, where our contribution targets efficient fine-tuned models.

\begin{table}[!h]
\centering
\scriptsize
\captionsetup{font=small, labelfont=small}
\caption{EM (\%) and ES (\%) across context-window sizes for four fine-tuned models and two proprietary LLMs. Bolded rows mark each model's best context window: 2{,}048 for fine-tuned, 8{,}192 for proprietary.}
\label{tab:ctx_window}
\setlength{\tabcolsep}{3pt}
\renewcommand{\arraystretch}{1.05}
\begin{tabular}{llcccccccc}
\toprule
\textbf{Model} & \textbf{Ctx} & \multicolumn{2}{c}{\textbf{Chisel}} & \multicolumn{2}{c}{\textbf{V/SV}} & \multicolumn{2}{c}{\textbf{VHDL}} & \multicolumn{2}{c}{\textbf{HLS}} \\
\cmidrule(lr){3-4}\cmidrule(lr){5-6}\cmidrule(lr){7-8}\cmidrule(lr){9-10}
& & EM & ES & EM & ES & EM & ES & EM & ES \\
\midrule
\multirow{5}{*}{FT DS-Coder-6.7B}
 & 512    & 16.10 & 50.64 & 24.41 & 54.34 & 17.14 & 51.33 & 25.00 & 55.00 \\
 & 1{,}024   & 22.29 & 61.99 & 34.25 & 69.36 & 26.61 & 67.59 & 31.80 & 67.70 \\
 & \textbf{2{,}048}   & \textbf{25.70}  & \textbf{56.50}  & \textbf{37.80}  & \textbf{67.40}  & \textbf{31.45}  & \textbf{61.10}  & \textbf{36.84}  & \textbf{60.40}  \\
 & 4{,}096   & 19.81 & 59.97 & 30.12 & 64.59 & 23.79 & 61.45 & 28.29 & 62.36 \\
 & 8{,}192   & 19.81 & 57.40 & 30.12 & 62.62 & 22.58 & 58.38 & 28.73 & 61.47 \\
\midrule
\multirow{5}{*}{FT Qwen-14B}
 & 512    & 26.93 & 64.83 & 32.48 & 67.39 & 26.41 & 64.92 & 30.70 & 66.50 \\
 & 1{,}024   & 29.10 & 67.42 & 34.45 & 71.29 & 30.65 & 70.46 & 33.55 & 70.59 \\
 & \textbf{2{,}048}   & \textbf{31.27}  & \textbf{68.90}  & \textbf{41.93}  & \textbf{74.40}  & \textbf{35.28}  & \textbf{73.80}  & \textbf{38.16}  & \textbf{71.50}  \\
 & 4{,}096   & 28.17 & 65.59 & 36.42 & 67.98 & 29.84 & 64.54 & 32.68 & 66.12 \\
 & 8{,}192   & 28.17 & 66.68 & 37.60 & 68.63 & 30.65 & 66.97 & 33.77 & 67.76 \\
\midrule
\multirow{5}{*}{FT Qwen-7B}
 & 512    & 22.60 & 61.07 & 33.66 & 66.89 & 25.81 & 65.12 & 32.02 & 65.52 \\
 & 1{,}024   & 25.08 & 64.78 & 34.65 & 69.69 & 30.04 & 70.83 & 33.77 & 69.42 \\
 & \textbf{2{,}048}   & \textbf{26.63}  & \textbf{66.70}  & \textbf{38.98}  & \textbf{70.80}  & \textbf{34.48}  & \textbf{71.90}  & \textbf{35.53}  & \textbf{69.50}  \\
 & 4{,}096   & 21.98 & 60.68 & 31.69 & 63.93 & 26.81 & 63.79 & 30.26 & 62.97 \\
 & 8{,}192   & 21.36 & 61.81 & 31.50 & 62.77 & 26.61 & 60.95 & 29.61 & 63.29 \\
\midrule
\multirow{5}{*}{FT Qwen-3B}
 & 512    & 20.74 & 56.64 & 25.00 & 57.43 & 19.15 & 55.33 & 26.75 & 60.07 \\
 & 1{,}024   & 22.91 & 61.73 & 32.87 & 66.60 & 26.01 & 65.94 & 31.80 & 66.70 \\
 & \textbf{2{,}048}   & \textbf{24.77}  & \textbf{54.30}  & \textbf{34.84}  & \textbf{65.50}  & \textbf{28.23}  & \textbf{59.10}  & \textbf{33.11}  & \textbf{55.00}  \\
 & 4{,}096   & 21.36 & 59.57 & 29.72 & 64.16 & 24.80 & 62.06 & 28.51 & 64.95 \\
 & 8{,}192   & 18.89 & 56.19 & 26.57 & 57.06 & 22.38 & 54.78 & 26.75 & 58.31 \\
\midrule
\multirow{5}{*}{Gemini-2.5}
 & 512    & 27.02 & 59.32 & 34.78 & 61.27 & 13.77 & 41.92 & 38.44 & 70.74 \\
 & 1{,}024   & 29.81 & 64.31 & 34.98 & 61.34 & 16.02 & 48.99 & 27.49 & 52.22 \\
 & 2{,}048   & 33.7  & 61.7  & 34.8  & 59.6  & 17.9  & 52.3  & 39.5  & 65.3  \\
 & 4{,}096   & 26.63 & 60.26 & 33.40 & 59.22 & 12.75 & 43.67 & 36.73 & 69.80 \\
 & \textbf{8{,}192}   & \textbf{32.51} & \textbf{66.93} & \textbf{36.69} & \textbf{64.97} & \textbf{17.41} & \textbf{53.59} & \textbf{41.15} & \textbf{71.73} \\
\midrule
\multirow{5}{*}{GPT-5}
 & 512    & 27.24 & 61.34 & 29.13 & 57.98 & 13.91 & 49.67 & 32.02 & 64.02 \\
 & 1{,}024   & 28.17 & 61.38 & 30.13 & 57.33 & 13.51 & 48.32 & 33.55 & 64.63 \\
 & 2{,}048   & 24.8 & 56.7 & 34.4 & 60.9 & 15.3 & 52.0 & 31.4 & 63.7 \\
 & 4{,}096   & 24.46 & 61.25 & 32.87 & 61.16 & 13.51 & 48.82 & 32.24 & 65.67 \\
 & \textbf{8{,}192}   & \textbf{28.17} & \textbf{61.23} & \textbf{34.71} & \textbf{62.49} & \textbf{13.56} & \textbf{51.02} & \textbf{36.73} & \textbf{69.12} \\
\bottomrule
\end{tabular}
\end{table}

\subsection{Agentic Evaluation Results}
\label{app:agentic}

To better demonstrate that MHRC-Bench is compatible with agentic systems, we additionally evaluate the Aider code agent. Although the strongest Aider configuration (Gemini-2.5-Pro) achieves better performance than the strongest fine-tuned Qwen2.5-Coder-14B baseline with RepoCoder, its advantage is modest, achieving an overall EM rate of 45.0\% compared to Qwen-14B's 38.9\%, resulting in a gain of 6.0 percentage points. This comes at immense computational costs: the agentic runs consume 81.42M total tokens over 3,566 instances, or 22,833 tokens per instance on average, whereas the fine-tuned local models are evaluated with a 2,048-token input budget and a 64-token maximum generation length. It also suffers from longer latency compared with local inference on small models, suggesting that our fine-tuned model is perhaps a more practical solution in practical code completion scenarios that require fast, low-cost inference.

Tables~\ref{tab:rebuttal_agent_usage} and~\ref{tab:rebuttal_aider_full} report the token usage and EM/ES results of agentic evaluation results.

\begin{table}[!t]
\centering
\rebuttaltablefont
\caption{Token usage for the agentic evaluation runs. Sent and received tokens are reported in millions.}
\label{tab:rebuttal_agent_usage}
\setlength{\tabcolsep}{4pt}
\begin{tabular}{llrrrrr}
\toprule
\textbf{Language} & \textbf{Model} & \textbf{Tasks} & \textbf{Sent (M)} & \textbf{Recv. (M)} & \textbf{Avg./Inst.} \\
\midrule
Chisel & Gemini-2.5-Pro & 323 & 4.02 & 2.00 & 18{,}638 \\
Chisel & GPT-5 & 323 & 4.71 & 0.92 & 17{,}426 \\
HLS & Gemini-2.5-Pro & 456 & 7.73 & 2.48 & 22{,}397 \\
HLS & GPT-5 & 456 & 10.17 & 1.45 & 25{,}471 \\
SystemVerilog & Gemini-2.5-Pro & 508 & 6.13 & 1.85 & 15{,}709 \\
SystemVerilog & GPT-5 & 508 & 10.77 & 1.93 & 25{,}005 \\
VHDL & Gemini-2.5-Pro & 496 & 10.63 & 3.66 & 28{,}818 \\
VHDL & GPT-5 & 496 & 11.59 & 1.38 & 26{,}147 \\
\midrule
Total & -- & 3{,}566 & 65.75 & 15.67 & 22{,}833 \\
\bottomrule
\end{tabular}
\end{table}

\begin{table}[!t]
\centering
\rebuttaltablefont
\caption{Aider code-agent evaluation on full test cases.}
\label{tab:rebuttal_aider_full}
\setlength{\tabcolsep}{3pt}
\resizebox{\textwidth}{!}{%
\begin{tabular}{llrrrrrrrr}
\toprule
\textbf{Agent / harness} & \textbf{Model} & \textbf{Chisel EM} & \textbf{Chisel ES} & \textbf{V/SV EM} & \textbf{V/SV ES} & \textbf{VHDL EM} & \textbf{VHDL ES} & \textbf{HLS EM} & \textbf{HLS ES} \\
\midrule
Aider & GPT-5 & \underline{42.41} & \underline{78.40} & \textbf{43.11} & \textbf{72.57} & \underline{43.15} & \underline{81.41} & \underline{44.52} & \underline{75.25} \\
Aider & Gemini-2.5-Pro & \textbf{46.44} & \textbf{80.46} & \underline{39.17} & \underline{67.30} & \textbf{45.77} & \textbf{81.61} & \textbf{48.46} & \textbf{75.59} \\
\bottomrule
\end{tabular}%
}
\end{table}

\subsection{Prompt Strategy Comparison}
\label{app:prompt_strategy}
To check whether our results are sensitive to the specific prompt template, we compare three prompt formats on the full MHRC-Bench-Eval benchmark with a fixed 2{,}048-token input window and the full-context (FIM) variant: (i)~\emph{Few-shot} prepends three repository-context demonstrations of completion before the target; (ii)~\emph{CoT} (chain-of-thought) instructs the model to first produce a brief plan in a hidden scratchpad and then emit the completion; (iii)~\emph{Our Prompt} is the template described in Appendix~\ref{app:eval-config}, which simply specifies the FIM tokens and a target-length hint. Table~\ref{tab:prompt_strategy} reports EM (\%) and ES (\%) averaged across the four HDLs for each combination.

The default prompt achieves the highest average EM and ES for all five fine-tuned models in this comparison. We use this prompt for the main evaluation.

\begin{table}[H]
\centering
\small
\captionsetup{font=small, labelfont=small}
\caption{EM (\%) and ES (\%) averaged across the four HDLs under three prompt strategies (Few-shot, CoT, Our Prompt) at a fixed 2{,}048-token input window. Best per row in bold.}
\label{tab:prompt_strategy}
\setlength{\tabcolsep}{4pt}
\renewcommand{\arraystretch}{1.15}
\begin{tabular}{lcccccc}
\toprule
\multirow{2}{*}{\textbf{Model}}
& \multicolumn{2}{c}{\textbf{Few-shot}}
& \multicolumn{2}{c}{\textbf{CoT}}
& \multicolumn{2}{c}{\textbf{Our Prompt}} \\
\cmidrule(lr){2-3}\cmidrule(lr){4-5}\cmidrule(lr){6-7}
& EM & ES & EM & ES & EM & ES \\
\midrule
\multicolumn{7}{l}{\textit{Proprietary LLMs}} \\
\midrule
Gemini-2.5 Pro    & 27.8 & \textbf{60.5} & \textbf{28.1} & 60.4 & 28.0 & 59.1 \\
GPT-5             & 26.2 & 57.9          & 26.0 & \textbf{59.5} & \textbf{26.5} & 58.3 \\
DeepSeek-V3.2     & 21.6 & 51.3          & 22.9 & \textbf{53.8} & \textbf{23.7} & 51.4 \\
\midrule
\multicolumn{7}{l}{\textit{Fine-tuned Models (LoRA)}} \\
\midrule
Qwen-14B-LoRA     & 35.1 & 70.7 & 34.9 & 70.6 & \textbf{36.7} & \textbf{72.2} \\
Qwen-7B-LoRA      & 32.4 & 68.2 & 32.7 & 68.0 & \textbf{33.9} & \textbf{69.7} \\
DS-6.7B-LoRA      & 31.3 & 59.9 & 31.6 & 60.1 & \textbf{32.9} & \textbf{61.4} \\
DS-5.7B-LoRA      & 29.9 & 66.5 & 30.2 & 66.3 & \textbf{33.7} & \textbf{70.2} \\
Qwen-3B-LoRA      & 28.7 & 57.1 & 28.9 & 57.0 & \textbf{30.2} & \textbf{58.5} \\
\bottomrule
\end{tabular}
\end{table}

\clearpage

\subsubsection{Repository context, retrieval, and long-context evaluation}
\label{app:rebuttal_context}
Tables~\ref{tab:rebuttal_long_context}--\ref{tab:rebuttal_native_fim} compare long-context, context-sensitivity, and native-FIM settings.

\begin{table}[!t]
\centering
\rebuttaltablefont
\caption{Long-context FIM evaluation with relevant files.}
\label{tab:rebuttal_long_context}
\setlength{\tabcolsep}{3pt}
\resizebox{\textwidth}{!}{%
\begin{tabular}{llrrrrrrrrr}
\toprule
\textbf{Model} & \textbf{Setting} & \textbf{Ctx} & \textbf{Chisel EM} & \textbf{Chisel ES} & \textbf{V/SV EM} & \textbf{V/SV ES} & \textbf{VHDL EM} & \textbf{VHDL ES} & \textbf{HLS EM} & \textbf{HLS ES} \\
\midrule
GPT-5 & Long-context FIM with relevant files & 400K & 30.96 & 63.57 & \underline{38.58} & \underline{62.95} & \underline{14.95} & \underline{50.57} & \underline{44.30} & \underline{71.50} \\
Gemini-2.5-Pro & Long-context FIM with relevant files & 1M & \underline{32.20} & \underline{64.11} & 36.81 & 61.17 & 11.90 & 43.63 & \textbf{46.37} & 71.12 \\
Qwen2.5-Coder-14B & Tuning + retrieval & 2,048 & \textbf{33.13} & \textbf{70.60} & \textbf{43.90} & \textbf{76.10} & \textbf{37.10} & \textbf{75.40} & 39.91 & \textbf{73.20} \\
\bottomrule
\end{tabular}%
}
\end{table}

\begin{table}[p]
\centering
\rebuttaltablefont
\caption{Context-window sensitivity from 512 to 8,192 tokens. The Gemini-2.5-Pro + RepoCoder rows correspond to the same run reported in both the context-window sensitivity and retrieval-focused analyses, and are listed only once.}
\label{tab:rebuttal_context_sensitivity}
\setlength{\tabcolsep}{2.5pt}
\resizebox{\textwidth}{!}{%
\begin{tabular}{llrrrrrrrrr}
\toprule
\textbf{Model} & \textbf{Method} & \textbf{Ctx} & \textbf{Chisel EM} & \textbf{Chisel ES} & \textbf{V/SV EM} & \textbf{V/SV ES} & \textbf{VHDL EM} & \textbf{VHDL ES} & \textbf{HLS EM} & \textbf{HLS ES} \\
\midrule
\multirow{5}{*}{FT DS-Coder-6.7B} & \multirow{5}{*}{Fine-tuned} & 512 & 16.10 & 50.64 & 24.41 & 54.34 & 17.14 & 51.33 & 25.00 & 55.00 \\
& & 1,024 & 22.29 & 61.99 & 34.25 & 69.36 & 26.61 & 67.59 & 31.80 & 67.70 \\
& & 2,048 & 25.70 & 56.50 & 37.80 & 67.40 & 31.45 & 61.10 & 36.84 & 60.40 \\
& & 4,096 & 19.81 & 59.97 & 30.12 & 64.59 & 23.79 & 61.45 & 28.29 & 62.36 \\
& & 8,192 & 19.81 & 57.40 & 30.12 & 62.62 & 22.58 & 58.38 & 28.73 & 61.47 \\
\midrule
\multirow{5}{*}{FT Qwen-14B} & \multirow{5}{*}{Fine-tuned} & 512 & 26.93 & 64.83 & 32.48 & 67.39 & 26.41 & 64.92 & 30.70 & 66.50 \\
& & 1,024 & 29.10 & 67.42 & 34.45 & \underline{71.29} & 30.65 & 70.46 & 33.55 & 70.59 \\
& & 2,048 & 31.27 & \textbf{68.90} & \textbf{41.93} & \textbf{74.40} & \textbf{35.28} & \textbf{73.80} & 38.16 & 71.50 \\
& & 4,096 & 28.17 & 65.59 & 36.42 & 67.98 & 29.84 & 64.54 & 32.68 & 66.12 \\
& & 8,192 & 28.17 & 66.68 & 37.60 & 68.63 & 30.65 & 66.97 & 33.77 & 67.76 \\
\midrule
\multirow{5}{*}{FT Qwen-3B} & \multirow{5}{*}{Fine-tuned} & 512 & 20.74 & 56.64 & 25.00 & 57.43 & 19.15 & 55.33 & 26.75 & 60.07 \\
& & 1,024 & 22.91 & 61.73 & 32.87 & 66.60 & 26.01 & 65.94 & 31.80 & 66.70 \\
& & 2,048 & 24.77 & 54.30 & 34.84 & 65.50 & 28.23 & 59.10 & 33.11 & 55.00 \\
& & 4,096 & 21.36 & 59.57 & 29.72 & 64.16 & 24.80 & 62.06 & 28.51 & 64.95 \\
& & 8,192 & 18.89 & 56.19 & 26.57 & 57.06 & 22.38 & 54.78 & 26.75 & 58.31 \\
\midrule
\multirow{5}{*}{FT Qwen-7B} & \multirow{5}{*}{Fine-tuned} & 512 & 22.60 & 61.07 & 33.66 & 66.89 & 25.81 & 65.12 & 32.02 & 65.52 \\
& & 1,024 & 25.08 & 64.78 & 34.65 & 69.69 & 30.04 & 70.83 & 33.77 & 69.42 \\
& & 2,048 & 26.63 & 66.70 & \underline{38.98} & 70.80 & \underline{34.48} & \underline{71.90} & 35.53 & 69.50 \\
& & 4,096 & 21.98 & 60.68 & 31.69 & 63.93 & 26.81 & 63.79 & 30.26 & 62.97 \\
& & 8,192 & 21.36 & 61.81 & 31.50 & 62.77 & 26.61 & 60.95 & 29.61 & 63.29 \\
\midrule
\multirow{5}{*}{GPT-5} & \multirow{5}{*}{Direct} & 512 & 27.24 & 61.34 & 29.13 & 57.98 & 13.91 & 49.67 & 32.02 & 64.02 \\
& & 1,024 & 28.17 & 61.38 & 30.13 & 57.33 & 13.51 & 48.32 & 33.55 & 64.63 \\
& & 2,048 & 24.8 & 56.7 & 34.4 & 60.9 & 15.3 & 52.0 & 31.4 & 63.7 \\
& & 4,096 & 24.46 & 61.25 & 32.87 & 61.16 & 13.51 & 48.82 & 32.24 & 65.67 \\
& & 8,192 & 28.17 & 61.23 & 34.71 & 62.49 & 13.56 & 51.02 & 36.73 & 69.12 \\
\midrule
\multirow{10}{*}{Gemini-2.5-Pro} & \multirow{5}{*}{RepoCoder} & 512 & 27.02 & 59.32 & 34.78 & 61.27 & 13.77 & 41.92 & 38.44 & 70.74 \\
& & 1,024 & 29.81 & 64.31 & 34.98 & 61.34 & 16.02 & 48.99 & 27.49 & 52.22 \\
& & 2,048 & 33.7 & 61.7 & 34.8 & 59.6 & 17.9 & 52.3 & 39.5 & 65.3 \\
& & 4,096 & 26.63 & 60.26 & 33.40 & 59.22 & 12.75 & 43.67 & 36.73 & 69.80 \\
& & 8,192 & 32.51 & 66.93 & 36.69 & 64.97 & 17.41 & 53.59 & 41.15 & 71.73 \\
\cmidrule(lr){2-11}
& \multirow{5}{*}{RLCoder} & 512 & 28.44 & 61.17 & 36.53 & 63.14 & 15.29 & 43.87 & 40.12 & \underline{72.58} \\
& & 1,024 & 31.63 & 66.04 & 36.87 & 63.21 & 17.84 & 50.73 & 29.31 & 54.09 \\
& & 2,048 & \textbf{34.7} & 63.2 & 35.2 & 60.2 & 19.2 & 53.7 & \underline{41.2} & 66.9 \\
& & 4,096 & 28.41 & 62.09 & 35.17 & 61.05 & 14.38 & 45.84 & 38.51 & 71.63 \\
& & 8,192 & \underline{34.28} & \underline{68.71} & 38.44 & 66.83 & 19.14 & 55.47 & \textbf{43.06} & \textbf{73.95} \\
\bottomrule
\end{tabular}%
}
\end{table}

\begin{table}[!t]
\centering
\rebuttaltablefont
\caption{Native FIM-token evaluation at a 2,048-token context window. CodeV was not re-evaluated in this run because its earlier evaluation already used FIM tokens.}
\label{tab:rebuttal_native_fim}
\setlength{\tabcolsep}{2.5pt}
\resizebox{\textwidth}{!}{%
\begin{tabular}{llrrrrrrrr}
\toprule
\textbf{Setting} & \textbf{Model} & \textbf{Chisel EM} & \textbf{Chisel ES} & \textbf{V/SV EM} & \textbf{V/SV ES} & \textbf{VHDL EM} & \textbf{VHDL ES} & \textbf{HLS EM} & \textbf{HLS ES} \\
\midrule
\multirow{7}{*}{Pretrained native FIM} & Deepseek-5.7B & 15.48 & 50.28 & 21.06 & 55.57 & 11.90 & 56.18 & 22.81 & 56.36 \\
& Deepseek-6.7B & 10.53 & 43.95 & 20.28 & 52.67 & 9.27 & 52.00 & 17.32 & 49.79 \\
& Qwen-Coder-3B & 12.38 & 46.55 & 17.32 & 48.66 & 9.68 & 48.45 & 17.11 & 48.91 \\
& Qwen-Coder-7B & 10.22 & 45.79 & 15.55 & 47.00 & 10.48 & 49.26 & 12.72 & 39.51 \\
& Qwen-Coder-14B & 17.96 & 57.29 & 22.44 & 55.99 & 11.09 & 51.44 & 24.34 & 55.66 \\
& GPT-5 & 24.8 & 56.7 & 34.4 & 60.9 & 15.3 & 52.0 & 31.4 & 63.7 \\
& Gemini-2.5-Pro & 32.2 & 60.4 & 31.7 & 60.4 & 14.9 & 51.3 & 33.3 & 64.3 \\
\midrule
\multirow{5}{*}{Tuned native FIM} & Deepseek-5.7B & 26.63 & 65.04 & 38.58 & 73.03 & 33.06 & 72.14 & 36.40 & 70.58 \\
& Deepseek-6.7B & 25.70 & 56.50 & 37.80 & 67.40 & 31.45 & 61.10 & 36.84 & 60.40 \\
& Qwen-Coder-3B & 24.77 & 54.30 & 34.84 & 65.50 & 28.23 & 59.10 & 33.11 & 55.00 \\
& Qwen-Coder-7B & 26.63 & 66.70 & 38.98 & 70.80 & 34.48 & 71.90 & 35.53 & 69.50 \\
& Qwen-Coder-14B & 31.27 & 68.90 & 41.93 & 74.40 & 35.28 & 73.80 & 38.16 & 71.50 \\
\midrule
\multirow{7}{*}{RepoCoder} & Deepseek-5.7B & 26.63 & 61.53 & 38.78 & 71.50 & 31.85 & 71.75 & 40.35 & 71.52 \\
& Deepseek-6.7B & 27.55 & 64.17 & 41.14 & 72.58 & 32.46 & 72.50 & 41.45 & 73.47 \\
& Qwen-Coder-3B & 26.01 & 61.48 & 36.42 & 67.57 & 29.03 & 69.61 & 38.60 & 70.50 \\
& Qwen-Coder-7B & 28.79 & 65.24 & 40.16 & 71.17 & 34.27 & 72.99 & 42.11 & 73.67 \\
& Qwen-Coder-14B & \textbf{34.67} & \textbf{70.68} & 42.13 & 74.22 & 36.09 & 74.76 & 42.76 & 74.49 \\
& GPT-5 & 27.6 & 59.6 & 35.4 & 61.3 & 16.7 & 54.1 & 34.4 & 63.9 \\
& Gemini-2.5-Pro & \underline{33.7} & 61.7 & 34.8 & 59.6 & 17.9 & 52.3 & 39.5 & 65.3 \\
\midrule
\multirow{7}{*}{RLCoder} & Deepseek-5.7B & 28.17 & 63.21 & 40.55 & 73.18 & 33.47 & 73.82 & 42.11 & 73.05 \\
& Deepseek-6.7B & 29.72 & 66.40 & \underline{43.31} & \underline{74.61} & 34.88 & 74.73 & \underline{43.64} & \underline{75.84} \\
& Qwen-Coder-3B & 27.55 & 63.95 & 38.78 & 69.94 & 31.45 & 71.88 & 40.35 & 72.63 \\
& Qwen-Coder-7B & 30.65 & 67.83 & 42.32 & 73.45 & \underline{36.49} & \underline{75.24} & \textbf{44.30} & \textbf{75.91} \\
& Qwen-Coder-14B & 33.13 & \underline{70.60} & \textbf{43.90} & \textbf{76.10} & \textbf{37.10} & \textbf{75.40} & 39.91 & 73.20 \\
& GPT-5 & 27.9 & 60.2 & 36.2 & 61.7 & 17.3 & 55.3 & 36.8 & 72.9 \\
& Gemini-2.5-Pro & \textbf{34.7} & 63.2 & 35.2 & 60.2 & 19.2 & 53.7 & 41.2 & 66.9 \\
\bottomrule
\end{tabular}%
}
\end{table}

\subsubsection{Prompt and output-format experiments}
\label{app:rebuttal_prompt}
The corrected expanded prompt-format run is reported in Table~\ref{tab:rebuttal_prompt}. An earlier few-shot run was superseded because its example was not changed when switching target languages.

\begin{table}[!t]
\centering
\rebuttaltablefont
\caption{Corrected prompt-format comparison.}
\label{tab:rebuttal_prompt}
\setlength{\tabcolsep}{4pt}
\begin{tabular}{lrrrrrr}
\toprule
\textbf{Model} & \textbf{Few-shot EM} & \textbf{Few-shot ES} & \textbf{CoT EM} & \textbf{CoT ES} & \textbf{Task EM} & \textbf{Task ES} \\
\midrule
Gemini-2.5 Pro & 27.8 & 60.5 & 28.1 & 60.4 & 28.0 & 59.1 \\
GPT-5 & 26.2 & 57.9 & 26.0 & 59.5 & 26.5 & 58.3 \\
DeepSeek-V3.2 & 21.6 & 51.3 & 22.9 & 53.8 & 23.7 & 51.4 \\
Qwen-14B-LoRA & \textbf{35.1} & \textbf{70.7} & \textbf{34.9} & \textbf{70.6} & \textbf{36.7} & \textbf{72.2} \\
Qwen-7B-LoRA & \underline{32.4} & \underline{68.2} & \underline{32.7} & \underline{68.0} & \underline{33.9} & 69.7 \\
DS-6.7B-LoRA & 31.3 & 59.9 & 31.6 & 60.1 & 32.9 & 61.4 \\
DS-5.7B-LoRA & 29.9 & 66.5 & 30.2 & 66.3 & 33.7 & \underline{70.2} \\
Qwen-3B-LoRA & 28.7 & 57.1 & 28.9 & 57.0 & 30.2 & 58.5 \\
\bottomrule
\end{tabular}
\end{table}

\FloatBarrier

\subsection{Detailed Performance on Line Counts and Structural Levels} \label{app:perf_structure}
Figure~\ref{fig:lang_line_node_performance_all} shows how the performance of Qwen2.5-Coder-7B and RepoCoder (driven by GPT-5) varies with target line count and structural complexity across hardware languages.

\subsection{Detailed Performance on Hardware Semantic Levels} \label{app:perf_semantic}

We report how Qwen2.5-Coder-7B and RepoCoder (driven by GPT-5) perform across hardware semantic levels for each hardware language using EM, ES, and Bilingual Evaluation Understudy (BLEU).

\begin{figure}[t]
  \centering

  \begin{subfigure}[t]{\linewidth}
    \centering
    \includegraphics[width=0.6\linewidth]{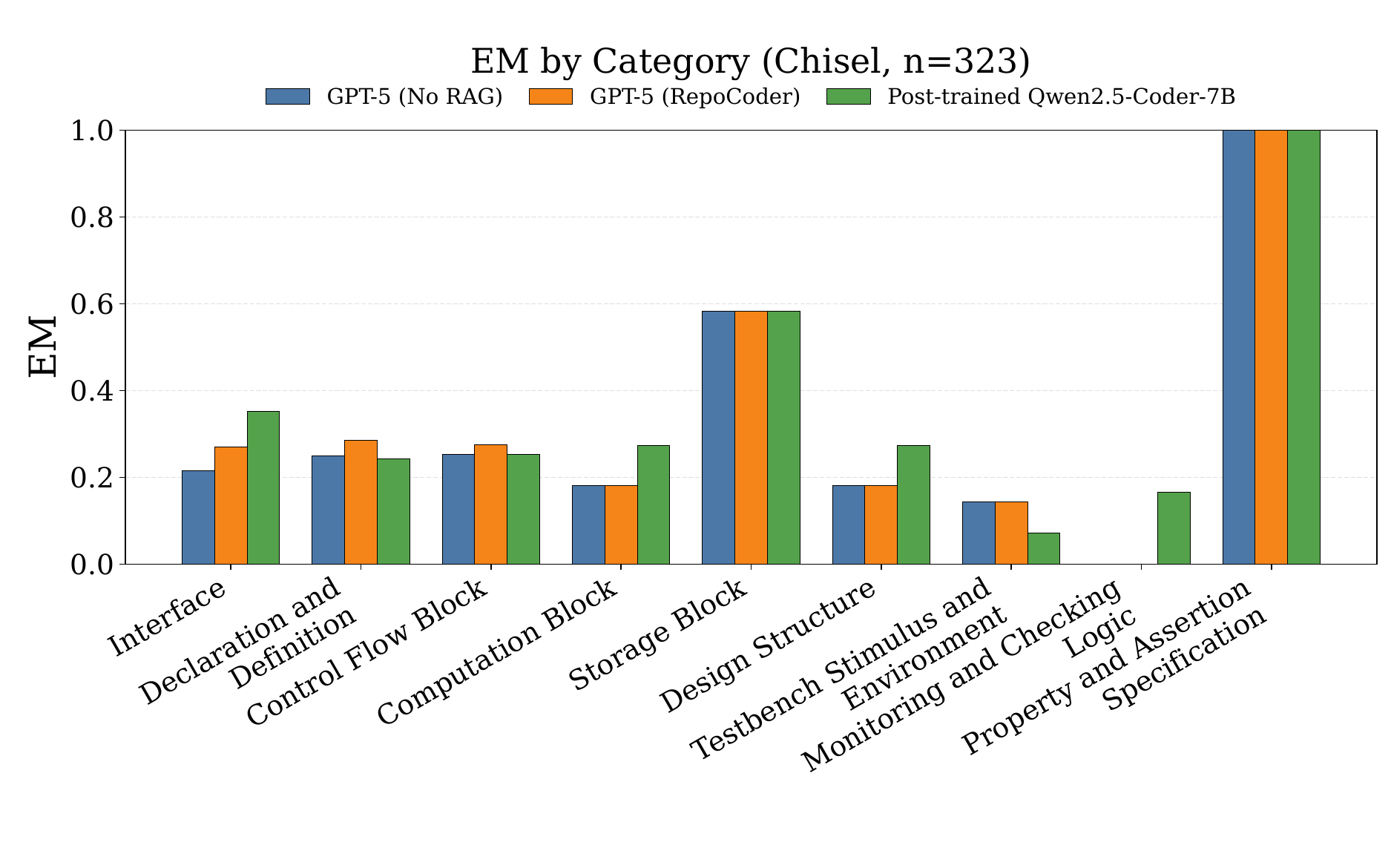}
    \caption{Chisel: EM by Category}
    \label{fig:chisel_cat_em}
  \end{subfigure}

  \begin{subfigure}[t]{\linewidth}
    \centering
    \includegraphics[width=0.6\linewidth]{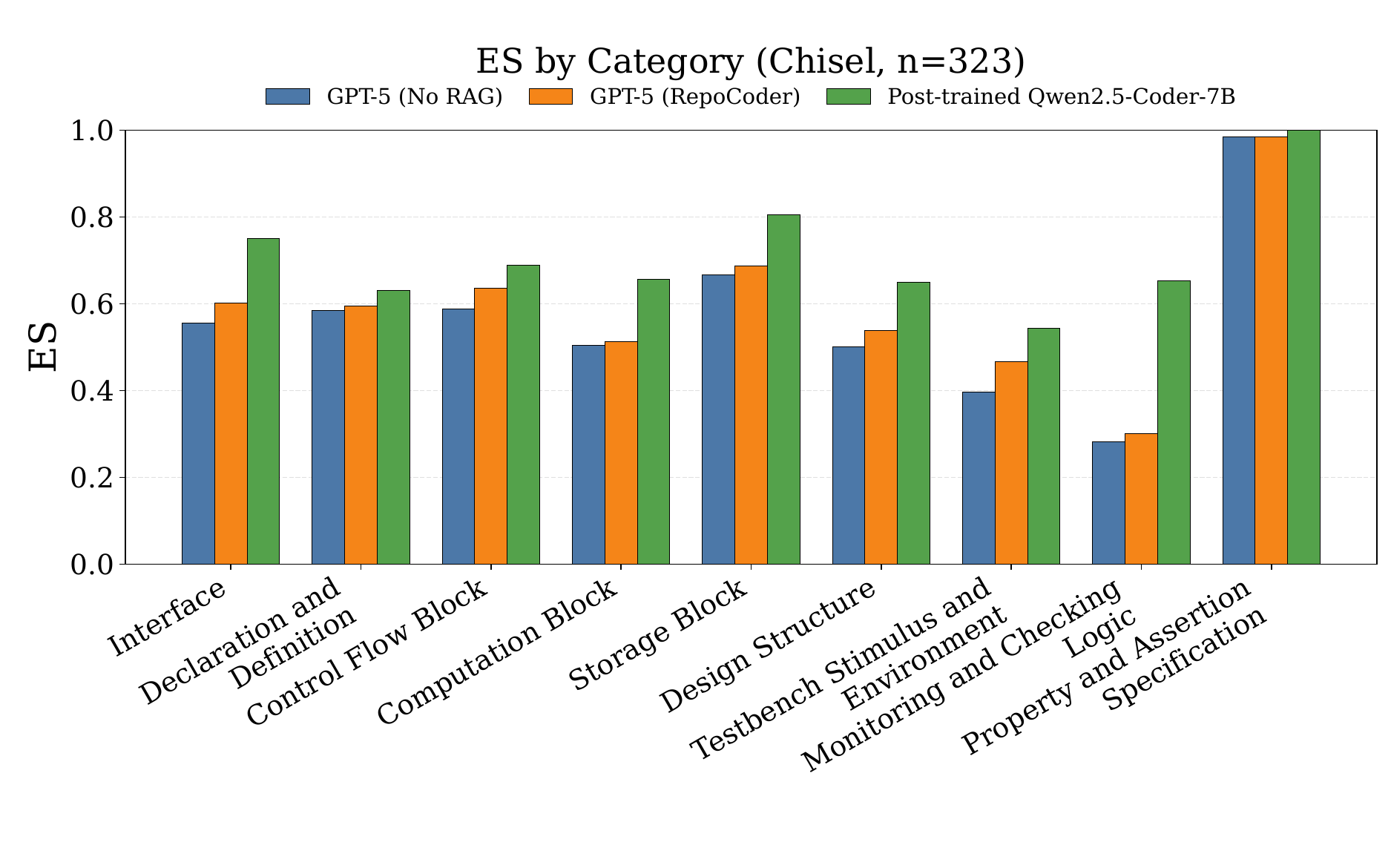}
    \caption{Chisel: ES by Category}
    \label{fig:chisel_cat_es}
  \end{subfigure}

  \begin{subfigure}[t]{\linewidth}
    \centering
    \includegraphics[width=0.6\linewidth]{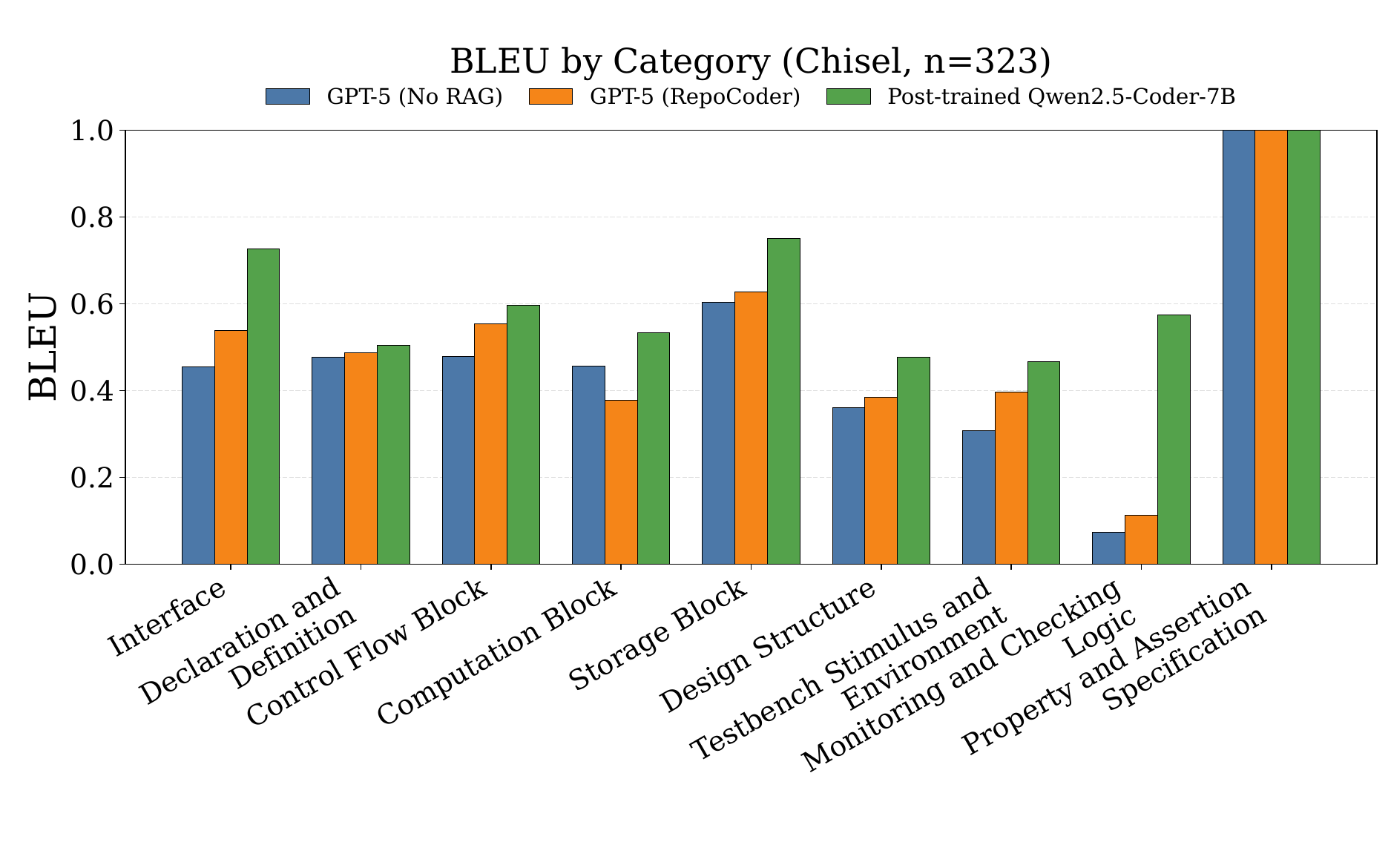}
    \caption{Chisel: BLEU by Category}
    \label{fig:chisel_cat_bleu}
  \end{subfigure}

  \caption{Per-category performance on Chisel across EM, ES, and BLEU.}
  \label{fig:by_category_chisel}
\end{figure}

\begin{figure}[t]
  \centering

  \begin{subfigure}[t]{\linewidth}
    \centering
    \includegraphics[width=0.6\linewidth]{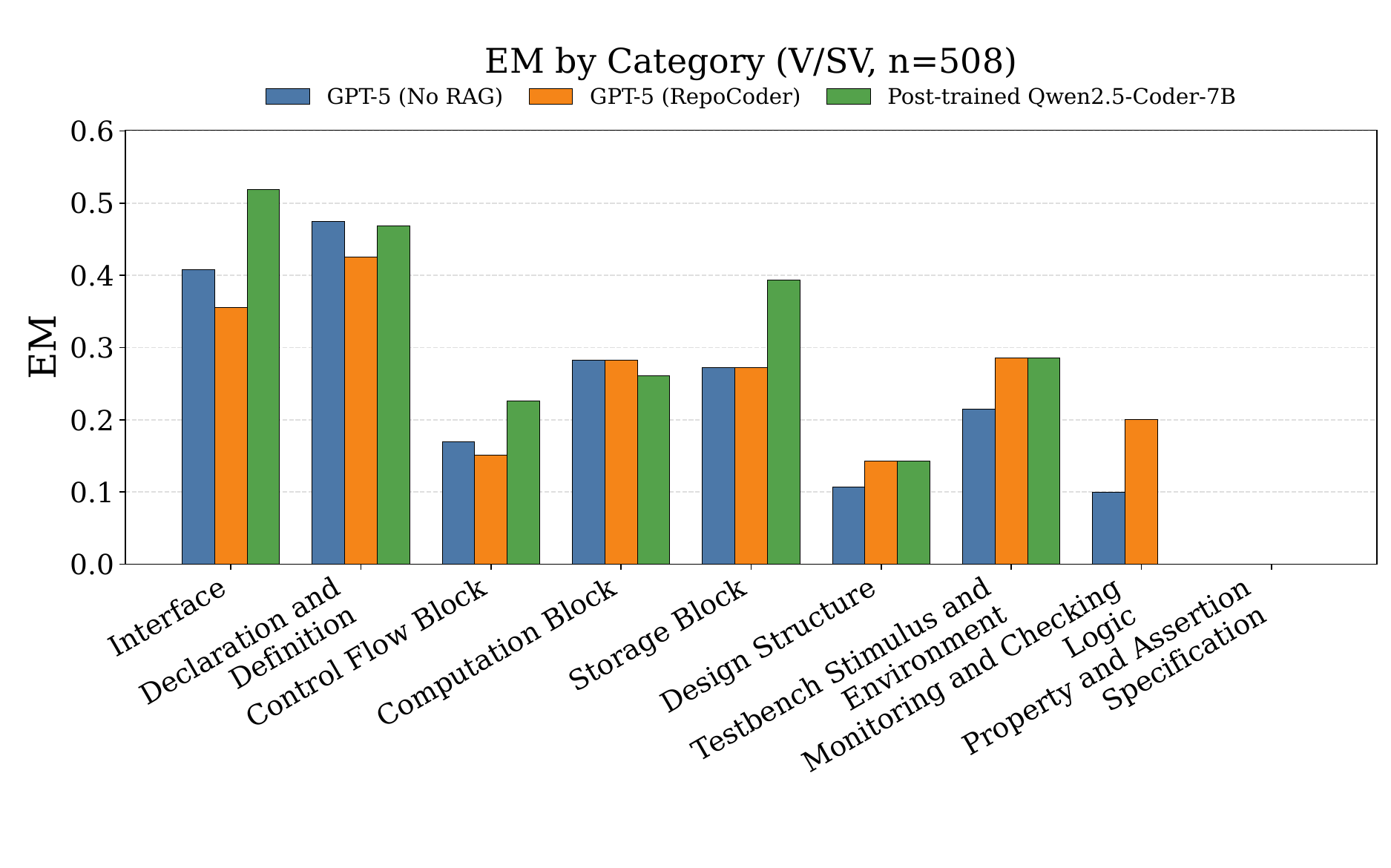}
    \caption{SystemVerilog: EM by Category}
    \label{fig:sv_cat_em}
  \end{subfigure}

  \begin{subfigure}[t]{\linewidth}
    \centering
    \includegraphics[width=0.6\linewidth]{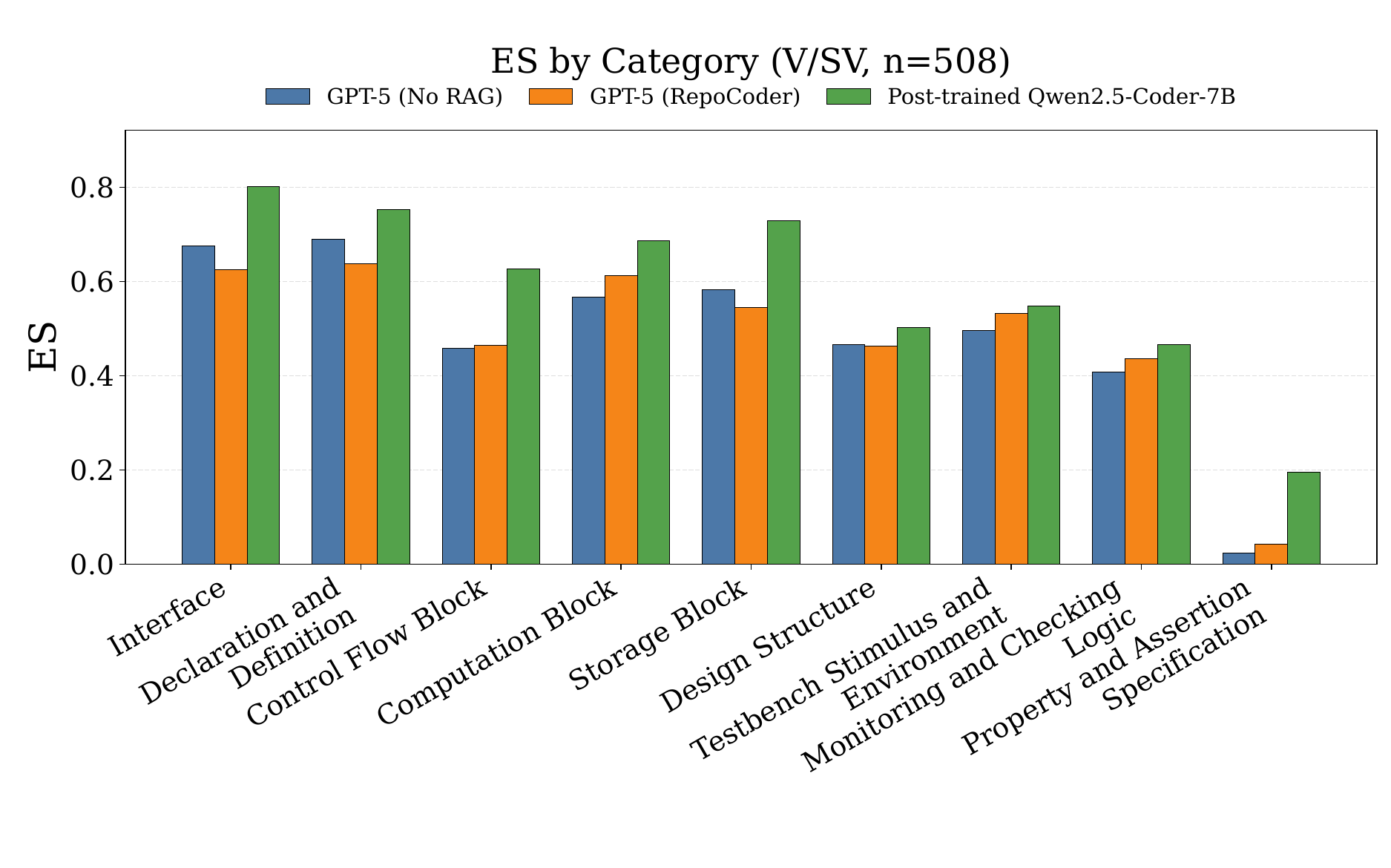}
    \caption{SystemVerilog: ES by Category}
    \label{fig:sv_cat_es}
  \end{subfigure}

  \begin{subfigure}[t]{\linewidth}
    \centering
    \includegraphics[width=0.6\linewidth]{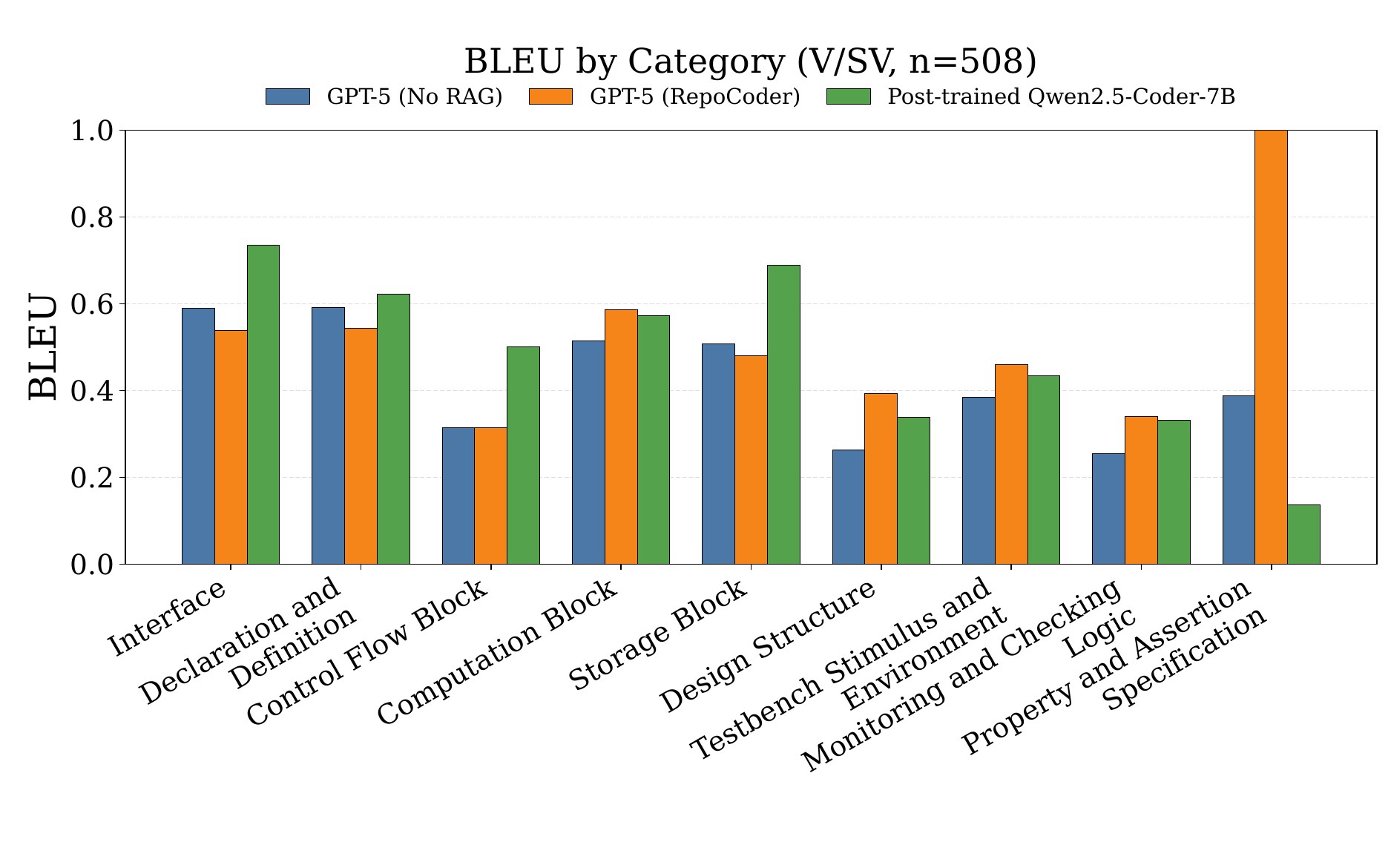}
    \caption{SystemVerilog: BLEU by Category}
    \label{fig:sv_cat_bleu}
  \end{subfigure}

  \caption{Per-category performance on SystemVerilog across EM, ES, and BLEU.}
  \label{fig:by_category_systemverilog}
\end{figure}

\begin{figure}[t]
  \centering

  \begin{subfigure}[t]{\linewidth}
    \centering
    \includegraphics[width=0.6\linewidth]{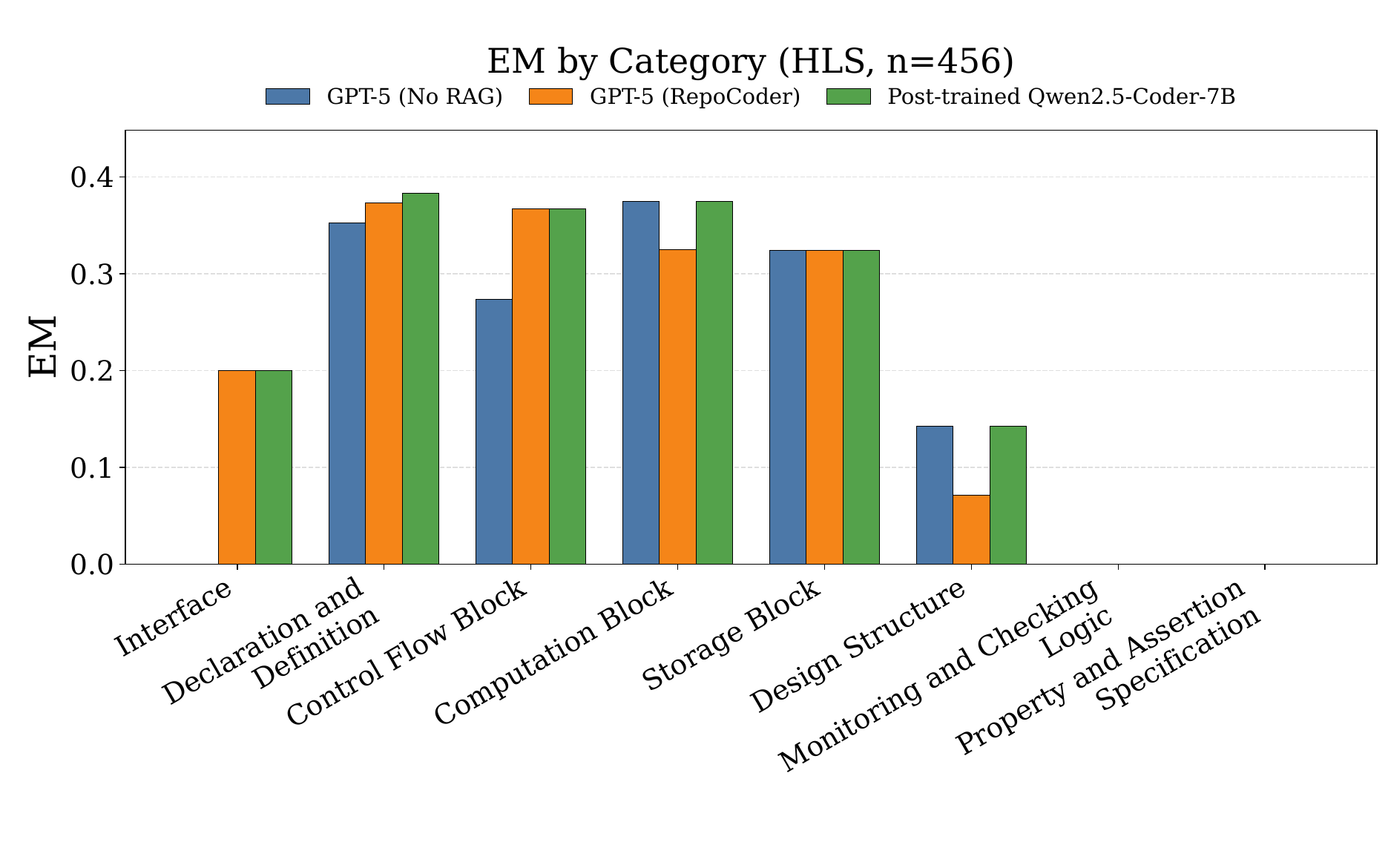}
    \caption{HLS: EM by Category}
    \label{fig:hls_cat_em}
  \end{subfigure}

  \begin{subfigure}[t]{\linewidth}
    \centering
    \includegraphics[width=0.6\linewidth]{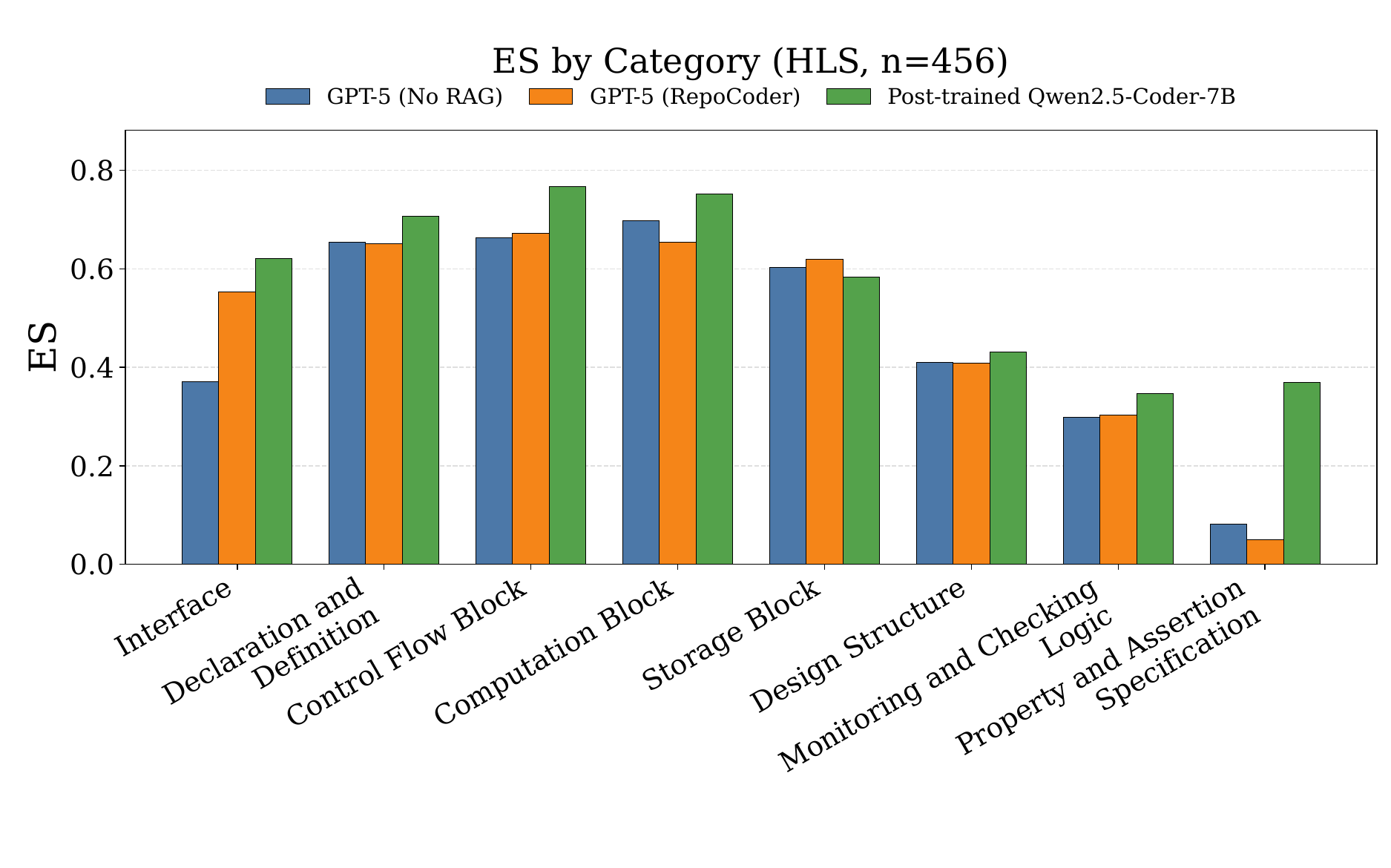}
    \caption{HLS: ES by Category}
    \label{fig:hls_cat_es}
  \end{subfigure}

  \begin{subfigure}[t]{\linewidth}
    \centering
    \includegraphics[width=0.6\linewidth]{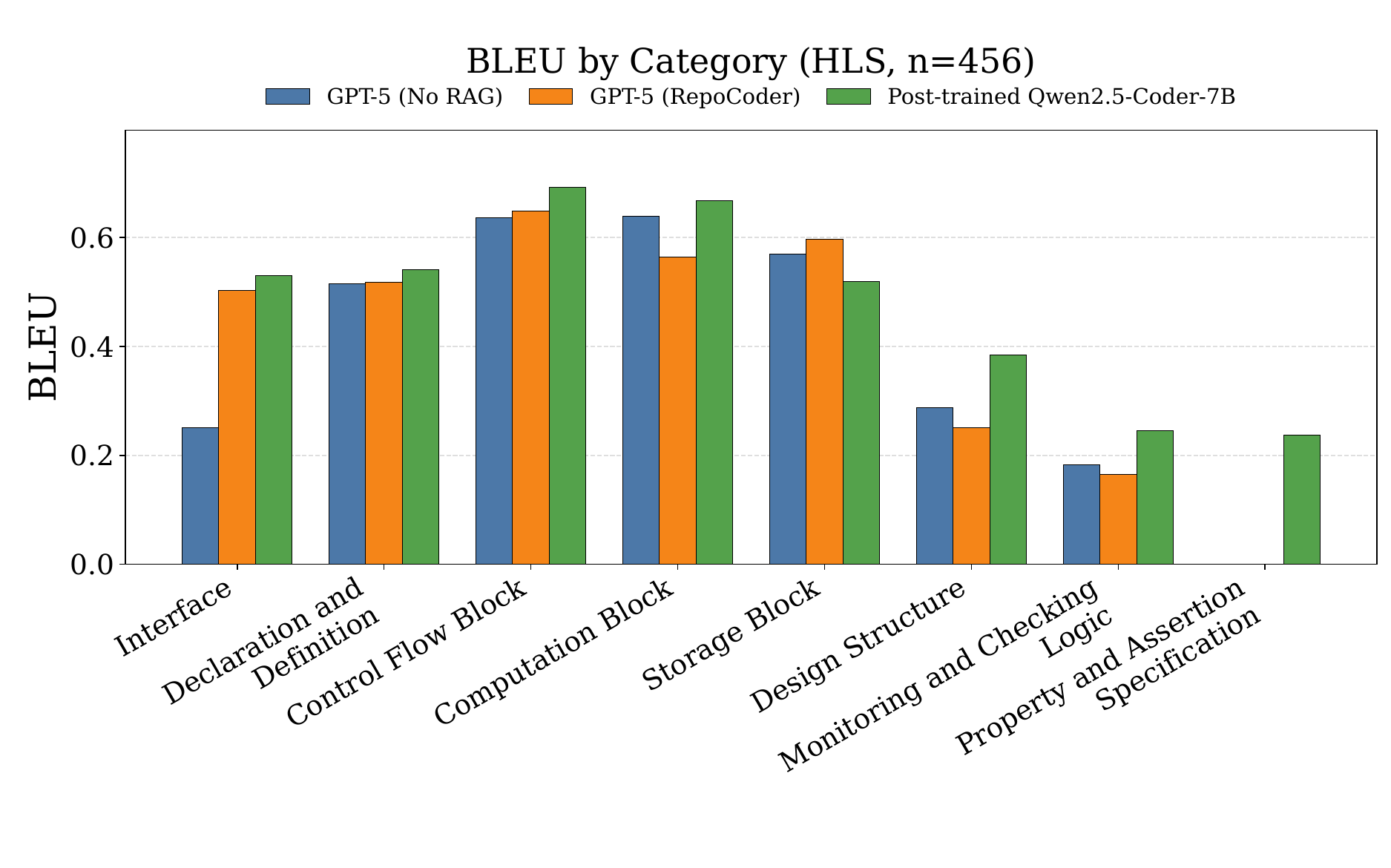}
    \caption{HLS: BLEU by Category}
    \label{fig:hls_cat_bleu}
  \end{subfigure}

  \caption{Per-category performance on HLS across EM, ES, and BLEU.}
  \label{fig:by_category_hls}
\end{figure}

\begin{figure}[t]
  \centering

  \begin{subfigure}[t]{\linewidth}
    \centering
    \includegraphics[width=0.6\linewidth]{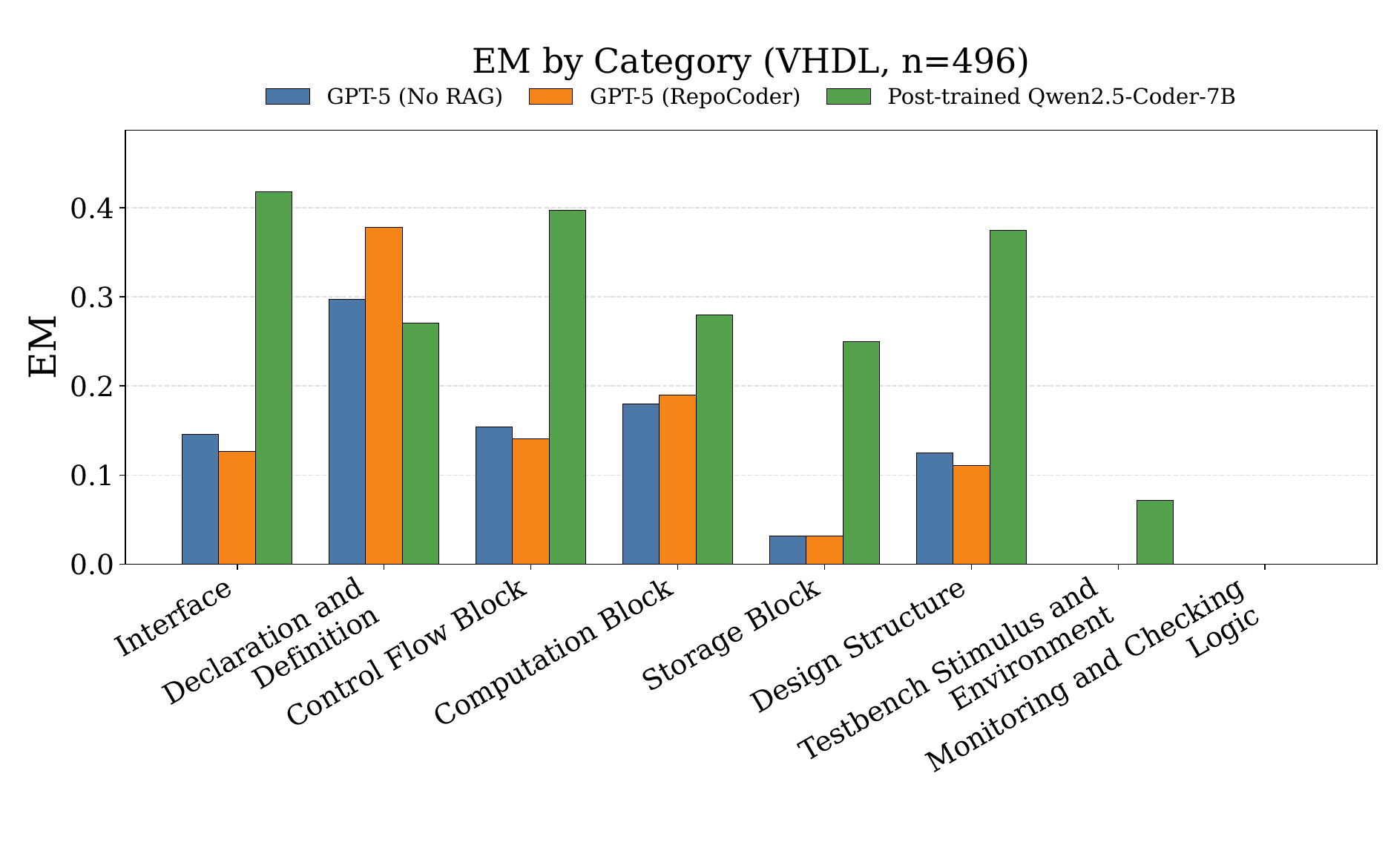}
    \caption{VHDL: EM by Category}
    \label{fig:vhdl_cat_em}
  \end{subfigure}

  \begin{subfigure}[t]{\linewidth}
    \centering
    \includegraphics[width=0.6\linewidth]{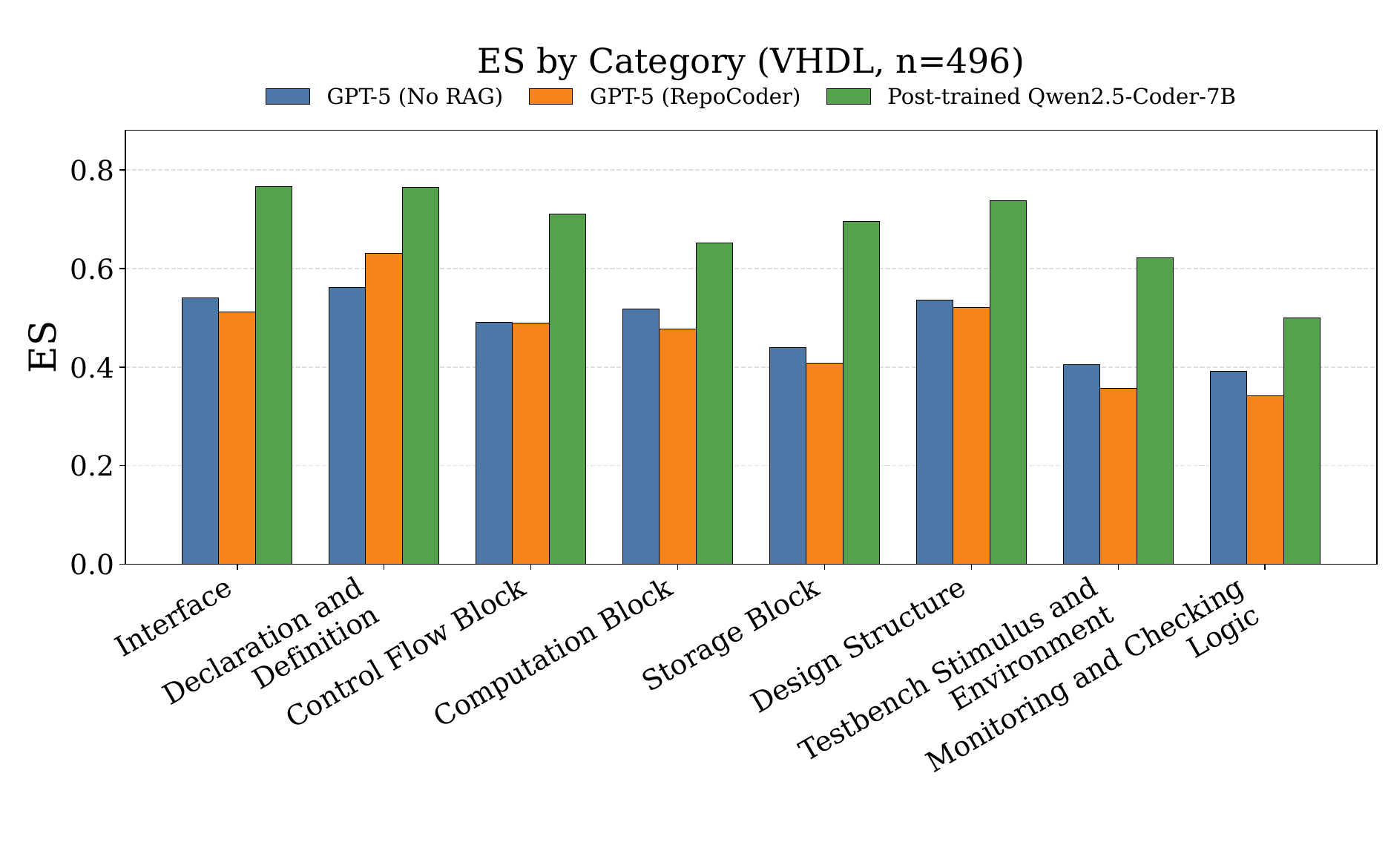}
    \caption{VHDL: ES by Category}
    \label{fig:vhdl_cat_es}
  \end{subfigure}

  \begin{subfigure}[t]{\linewidth}
    \centering
    \includegraphics[width=0.6\linewidth]{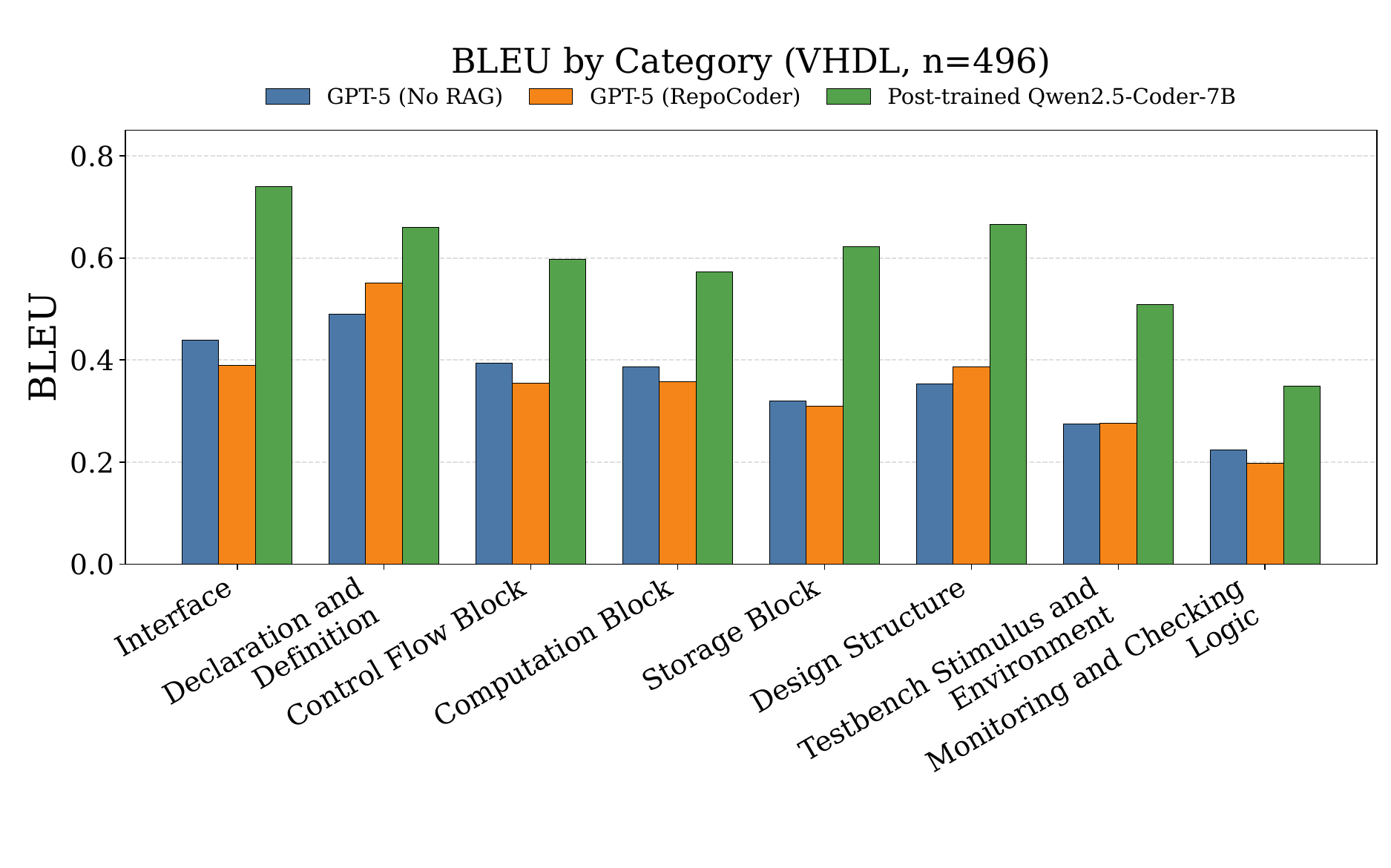}
    \caption{VHDL: BLEU by Category}
    \label{fig:vhdl_cat_bleu}
  \end{subfigure}

  \caption{Per-category performance on VHDL across EM, ES, and BLEU.}
  \label{fig:by_category_vhdl}
\end{figure}

\begin{figure*}[t]
  \centering
\includegraphics[width=\textwidth]{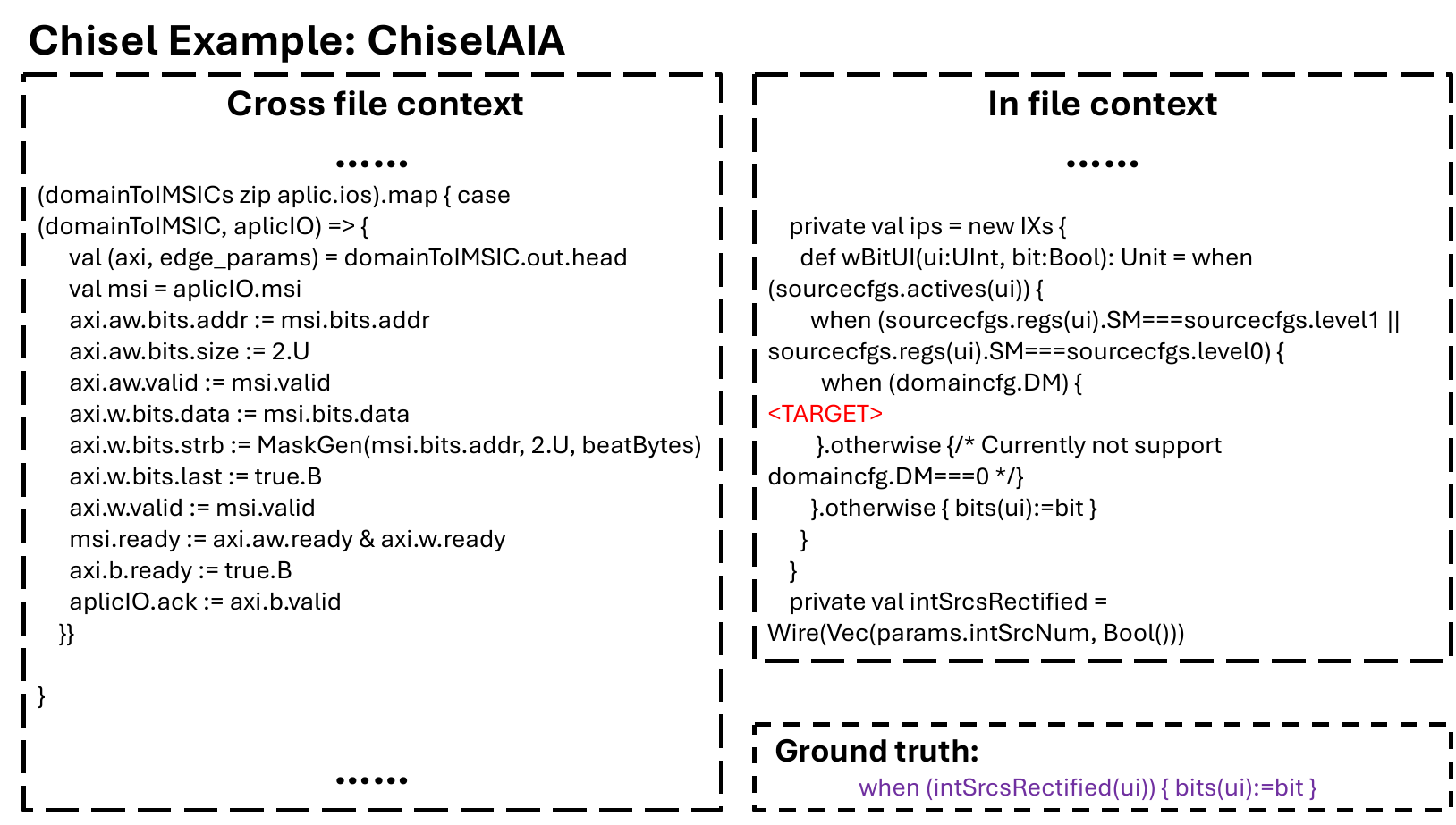}
  \caption{Example one of chisel code completion examples }
  \label{fig:chisel_example1}
\end{figure*}

\begin{figure*}[t]
  \centering
\includegraphics[width=\textwidth]{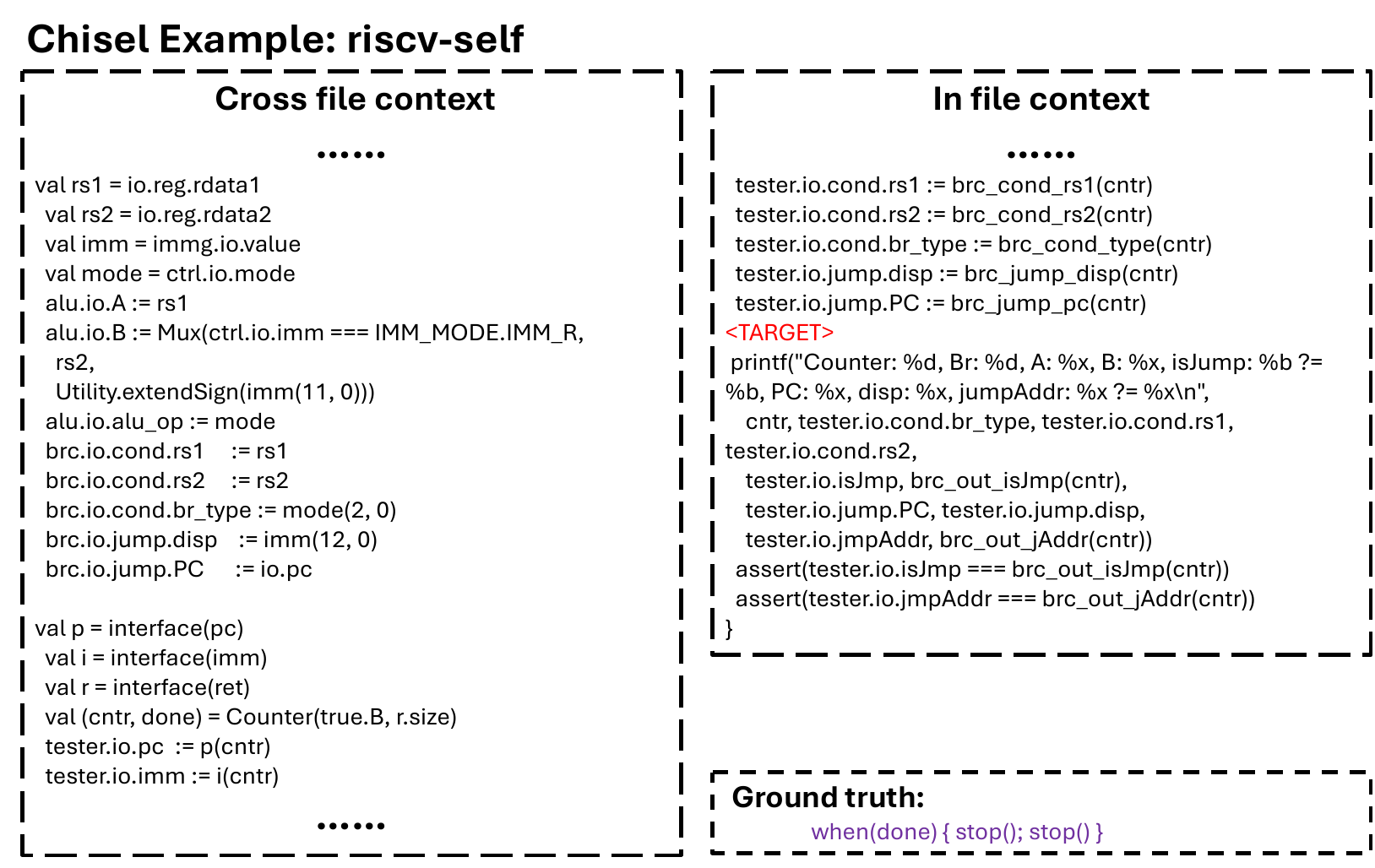}
  \caption{Example two of chisel code completion examples }
  \label{fig:chisel_example2}
\end{figure*}

\begin{figure*}[t]
  \centering
\includegraphics[width=\textwidth]{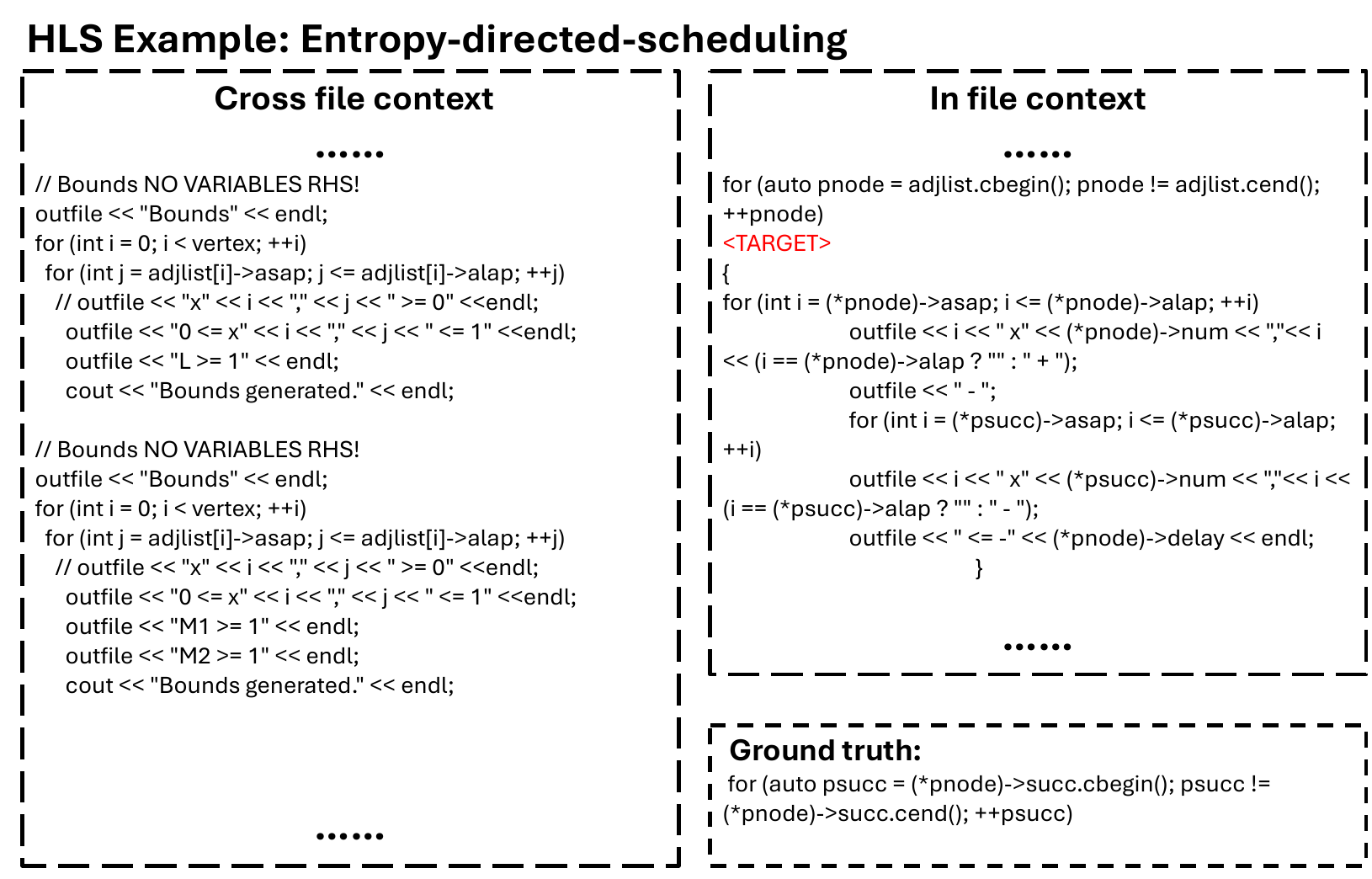}
  \caption{Example one of HLS code completion examples }
  \label{fig:HLS_example1}
\end{figure*}

\begin{figure*}[t]
  \centering
\includegraphics[width=\textwidth]{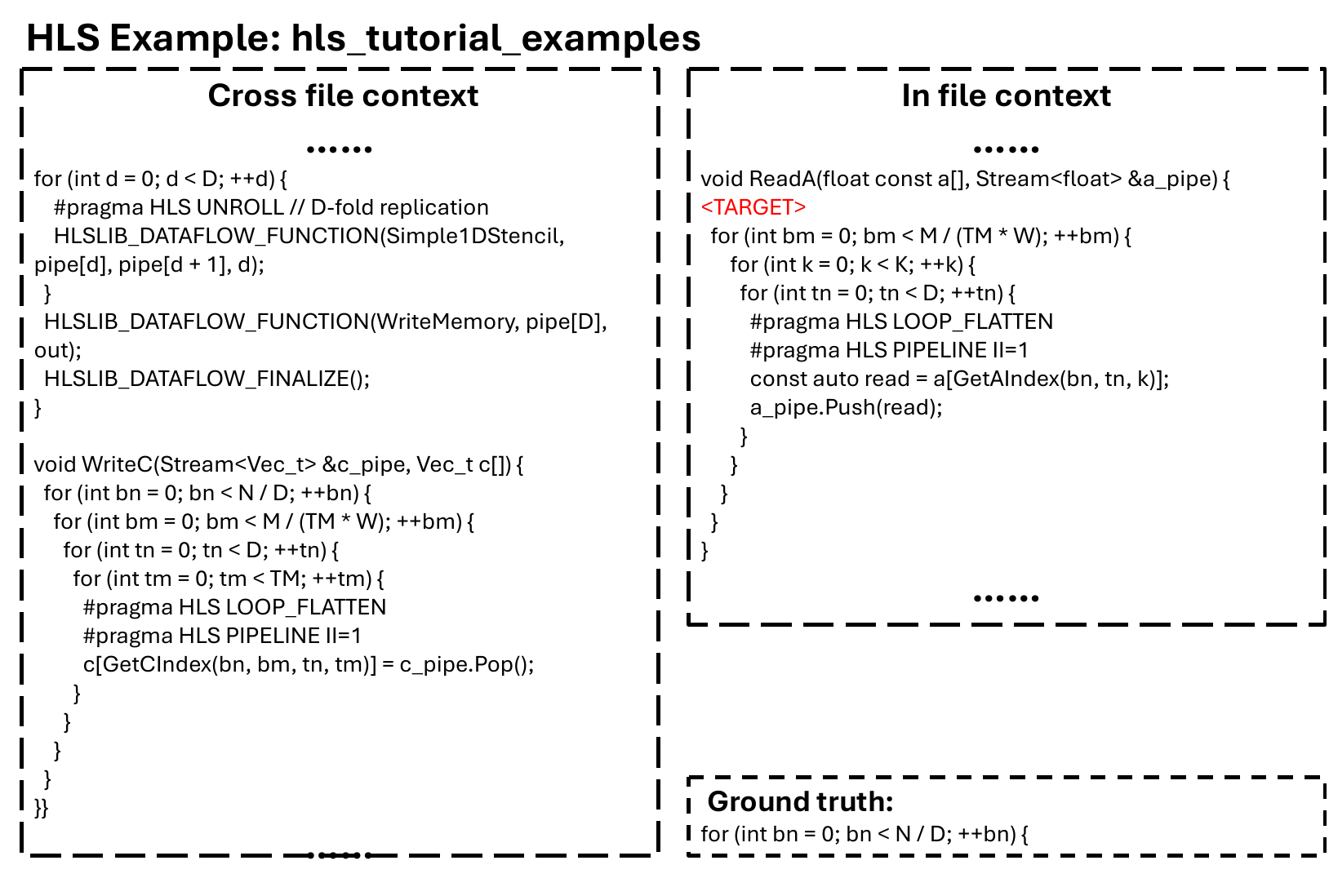}
  \caption{Example two of HLS code completion examples }
  \label{fig:HLS_example2}
\end{figure*}

\begin{figure*}[t]
  \centering
\includegraphics[width=\textwidth]{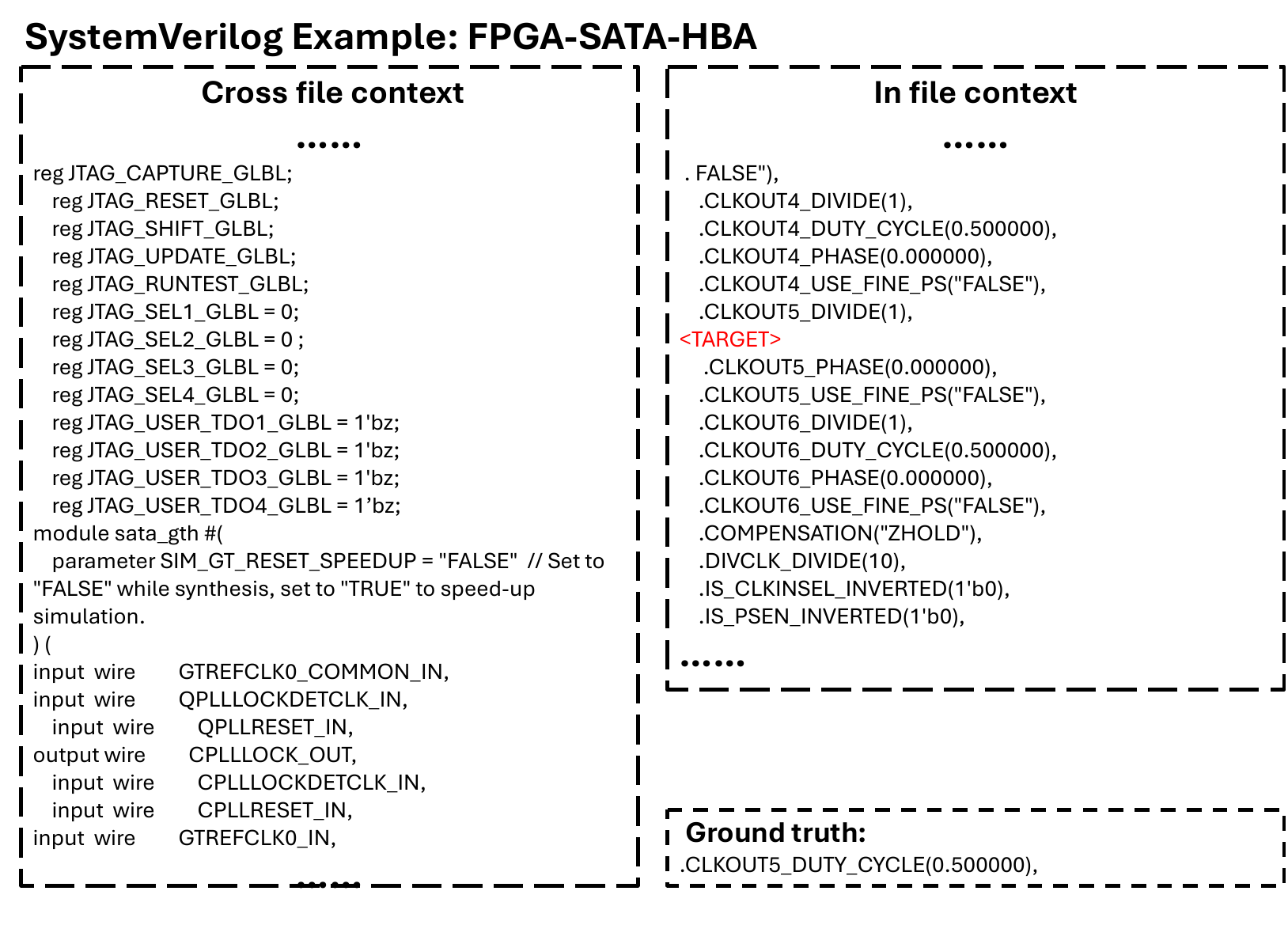}
  \caption{Example one of SystemVerilog code completion examples }
\label{fig:systemverilog_example1}
\end{figure*}

\begin{figure*}[t]
  \centering
\includegraphics[width=\textwidth]{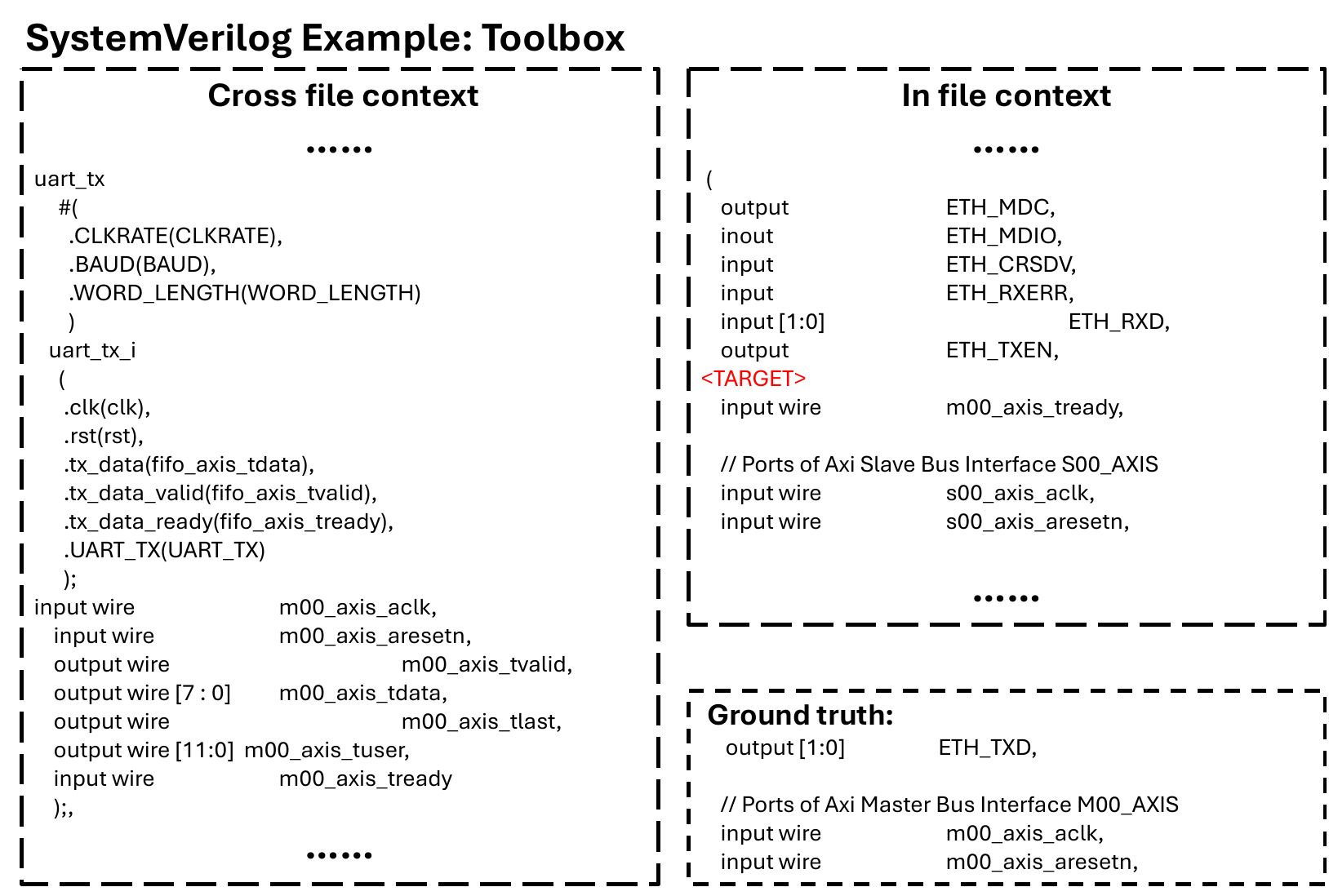}
  \caption{Example two of SystemVerilog code completion examples }
\label{fig:systemverilog_example2}
\end{figure*}

\begin{figure*}[t]
  \centering
\includegraphics[width=\textwidth]{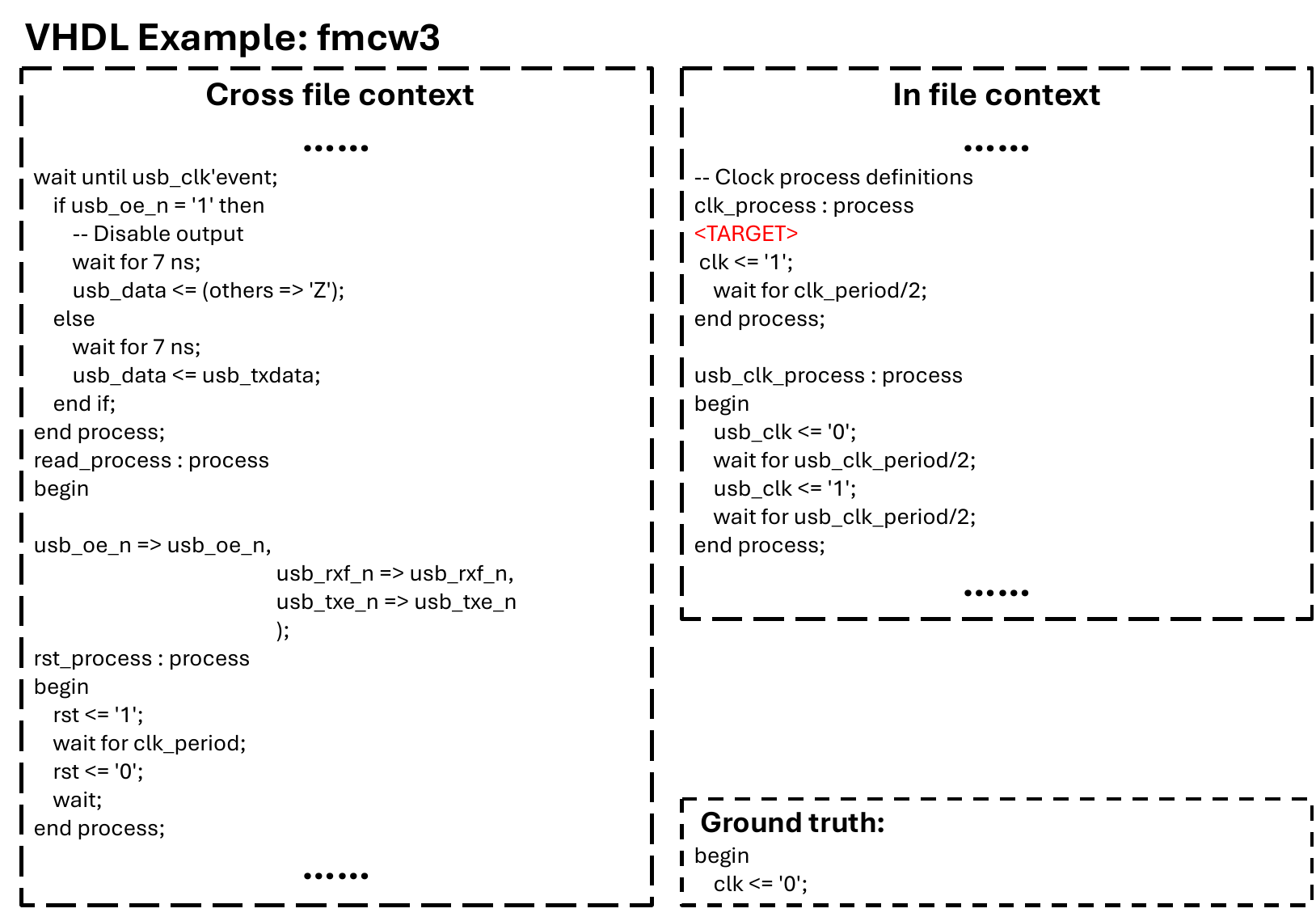}
  \caption{Example one of VHDL code completion examples }
\label{fig:vhdl_example1}
\end{figure*}

\begin{figure*}[t]
  \centering
\includegraphics[width=\textwidth]{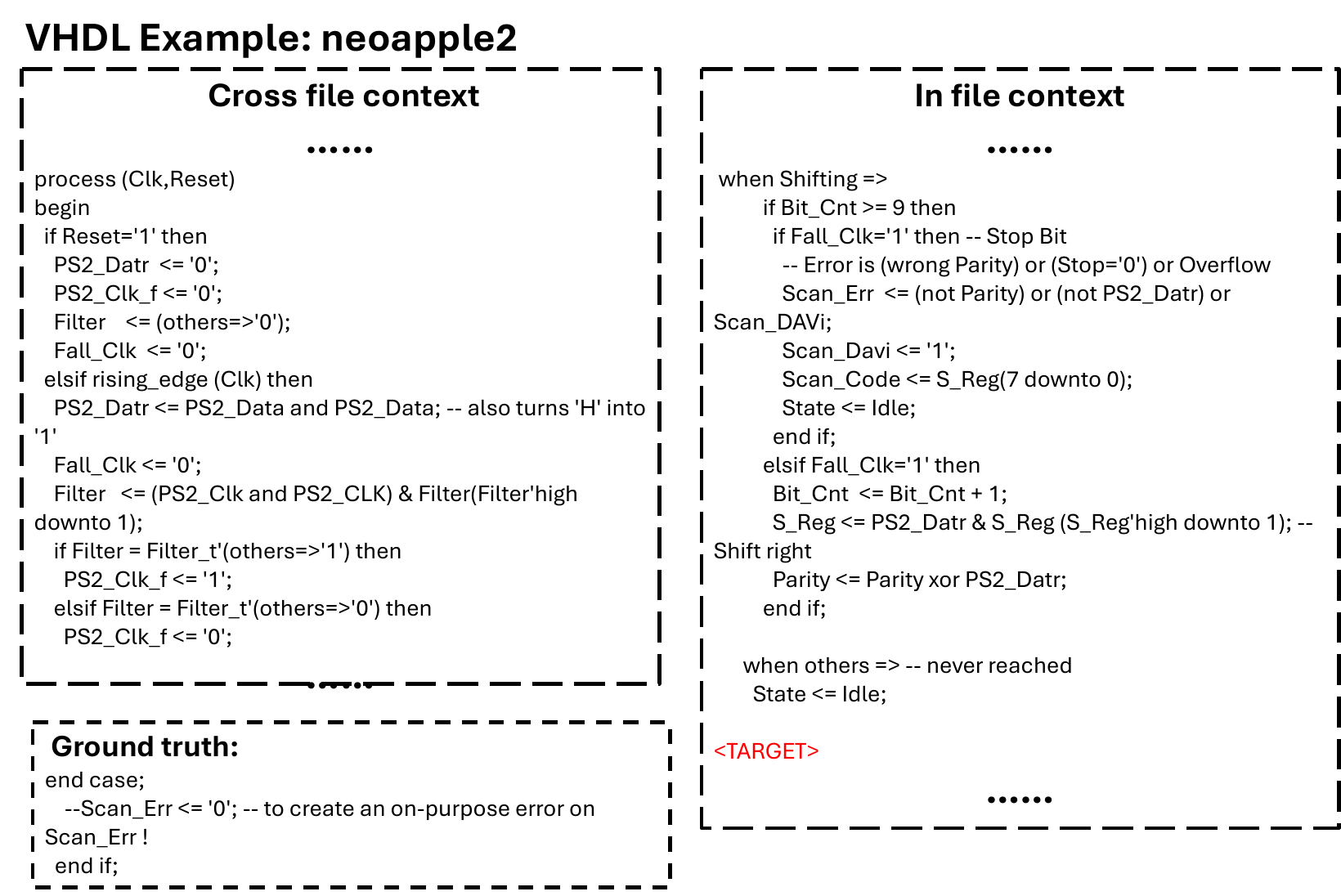}
  \caption{Example two of VHDL code completion examples }
\label{fig:vhdl_example2}
\end{figure*}

\subsection{Code Completion Examples}
\label{app:gemini_failure_example}
For each hardware description language, we select two representative repositories as examples to illustrate cross-file context, in-file context, the selected completion target, and the corresponding ground truth in Figures \ref{fig:chisel_example1} to \ref{fig:vhdl_example2}. 

\newcommand{\code}[1]{\texttt{\small #1}}

\begin{table*}[ht]
\centering
\small
\caption{Comparison of Ground Truth vs. Pre-Training Failures across Hardware Languages\\ (Examples extracted from pretrained Qwen2.5-Coder-7B output).}
\begin{tabularx}{\textwidth}{@{}l >{\raggedright\arraybackslash}p{0.24\textwidth} Y Y@{}}
\toprule
\textbf{Lang.} & \textbf{Repo Name} & \textbf{Ground Truth} & \textbf{Pre-Train Output (Failure)} \\
\midrule

Verilog & AXI-SDCard-High-Speed-Controller & 
\code{input logic[2:0] axilite\_\allowbreak arprot,} &
\code{.waddr(waddr), .wdata(wdata), .we(we)...} \textit{(Instantiation Hallucination)} \\
\addlinespace

Verilog & custom\_\allowbreak uvm\_\allowbreak report\_\allowbreak server &
\code{super.build\_\allowbreak phase(phase);} &
\code{// Copyright 2007-2010 Mentor Graphics...} \textit{(Header Regurgitation)} \\
\addlinespace

Verilog & SystemVerilog\allowbreak Assertions &
\code{logic gnt3;} & 
\code{// Write assertions for a simple Round Robin...} \textit{(Non-code Content Hallucination)} \\
\addlinespace

Verilog & virtio & 
\code{always\_\allowbreak ff @(posedge aclk...)} &
\code{\_\allowbreak EVENT\_\allowbreak IDX: begin tx.tvalid <= 1'b1; ...} \textit{(Logic Hallucination)} \\
\midrule

HLS & hls-spmv & 
\code{int col\_\allowbreak left = 0;} &
\code{\#include "spmv.h" \#include <hls\_\allowbreak stream.h>...} \textit{(Header Regurgitation)} \\
\addlinespace

HLS & hls-spmv & 
\code{L1: for (int i = 0; ...} & 
\code{void spmv(int rowPtr[NUM\_\allowbreak ROWS+1]...} \textit{(Function Definition Hallucination)} \\
\addlinespace

HLS & libllsm & 
\code{FP\_\allowbreak TYPE* x = calloc(ns...} &
\code{FP\_\allowbreak TYPE* llsm\_\allowbreak envelope(FP\_\allowbreak TYPE* spectrum...} \textit{(Function Definition Hallucination)} \\
\midrule

VHDL & FPGA-Speech-Recognition & 
\code{PARITY\_\allowbreak OUT <= parity\_\allowbreak temp;} &
\code{library IEEE; use IEEE.STD\_\allowbreak LOGIC\_\allowbreak 1164.ALL;...} \textit{(Header Regurgitation)} \\
\addlinespace

VHDL & cpu\_\allowbreak for\_\allowbreak nscscc2022\_\allowbreak single &
\code{C\_\allowbreak HAS\_\allowbreak WE => 1, C\_\allowbreak MEM\_\allowbreak ...} &
\code{--- FILE: dcache\_\allowbreak tag.vhd (END) ---} \textit{(Non-code Content Hallucination)} \\
\addlinespace

VHDL & FPGA-I2C-Minion & 
\code{if ENDSIM = false then...} & 
\code{i2c\_\allowbreak wait\_\allowbreak quarter\_\allowbreak clock; scl\_\allowbreak pre\_\allowbreak spike <= 'Z';} \textit{(Logic Hallucination)} \\
\midrule

Chisel & rjrouter & 
\code{c => new GCDUnitTester(c)} &
\code{<|fim\_\allowbreak middle|>} \textit{(Non-code Content Hallucination)} \\

\bottomrule
\end{tabularx}

\label{tab:pretrain_outputs}
\end{table*}

\clearpage
\section*{NeurIPS Paper Checklist}

\begin{enumerate}

\item {\bf Claims}
    \item[] Question: Do the main claims made in the abstract and introduction accurately reflect the paper's contributions and scope?
    \item[] Answer: \answerYes{}
    \item[] Justification: The abstract and the Introduction state our two contributions (MHRC-Bench-Eval and MHRC-Bench-Train) and four findings (language dependence, non-monotonic structural sensitivity, context-length plateau, software-tuned retrieval failure). Each is supported by the experimental results in Section~``Experiments'' and the analysis in Section~``Experimental Analysis''.
    \item[] Guidelines:
    \begin{itemize}
        \item The answer \answerNA{} means that the abstract and introduction do not include the claims made in the paper.
        \item The abstract and/or introduction should clearly state the claims made, including the contributions made in the paper and important assumptions and limitations. A \answerNo{} or \answerNA{} answer to this question will not be perceived well by the reviewers.
        \item The claims made should match theoretical and experimental results, and reflect how much the results can be expected to generalize to other settings.
        \item It is fine to include aspirational goals as motivation as long as it is clear that these goals are not attained by the paper.
    \end{itemize}

\item {\bf Limitations}
    \item[] Question: Does the paper discuss the limitations of the work performed by the authors?
    \item[] Answer: \answerYes{}
    \item[] Justification: The ``Conclusion and Limitations'' section notes that MHRC-Bench currently focuses on digital hardware description languages and does not cover analog/mixed-signal or physical-design languages. Appendix~\ref{subsec:graph_coder} additionally discusses limitations of software-tuned retrievers when applied to HDL.
    \item[] Guidelines:
    \begin{itemize}
        \item The answer \answerNA{} means that the paper has no limitation while the answer \answerNo{} means that the paper has limitations, but those are not discussed in the paper.
        \item The authors are encouraged to create a separate ``Limitations'' section in their paper.
        \item The paper should point out any strong assumptions and how robust the results are to violations of these assumptions (e.g., independence assumptions, noiseless settings, model well-specification, asymptotic approximations only holding locally). The authors should reflect on how these assumptions might be violated in practice and what the implications would be.
        \item The authors should reflect on the scope of the claims made, e.g., if the approach was only tested on a few datasets or with a few runs. In general, empirical results often depend on implicit assumptions, which should be articulated.
        \item The authors should reflect on the factors that influence the performance of the approach. For example, a facial recognition algorithm may perform poorly when image resolution is low or images are taken in low lighting. Or a speech-to-text system might not be used reliably to provide closed captions for online lectures because it fails to handle technical jargon.
        \item The authors should discuss the computational efficiency of the proposed algorithms and how they scale with dataset size.
        \item If applicable, the authors should discuss possible limitations of their approach to address problems of privacy and fairness.
        \item While the authors might fear that complete honesty about limitations might be used by reviewers as grounds for rejection, a worse outcome might be that reviewers discover limitations that aren't acknowledged in the paper. The authors should use their best judgment and recognize that individual actions in favor of transparency play an important role in developing norms that preserve the integrity of the community. Reviewers will be specifically instructed to not penalize honesty concerning limitations.
    \end{itemize}

\item {\bf Theory assumptions and proofs}
    \item[] Question: For each theoretical result, does the paper provide the full set of assumptions and a complete (and correct) proof?
    \item[] Answer: \answerNA{}
    \item[] Justification: The paper presents an empirical benchmark and empirical analysis; it does not include theoretical results requiring formal proofs.
    \item[] Guidelines:
    \begin{itemize}
        \item The answer \answerNA{} means that the paper does not include theoretical results.
        \item All the theorems, formulas, and proofs in the paper should be numbered and cross-referenced.
        \item All assumptions should be clearly stated or referenced in the statement of any theorems.
        \item The proofs can either appear in the main paper or the supplemental material, but if they appear in the supplemental material, the authors are encouraged to provide a short proof sketch to provide intuition.
        \item Inversely, any informal proof provided in the core of the paper should be complemented by formal proofs provided in appendix or supplemental material.
        \item Theorems and Lemmas that the proof relies upon should be properly referenced.
    \end{itemize}

    \item {\bf Experimental result reproducibility}
    \item[] Question: Does the paper fully disclose all the information needed to reproduce the main experimental results of the paper to the extent that it affects the main claims and/or conclusions of the paper (regardless of whether the code and data are provided or not)?
    \item[] Answer: \answerNo{}
    \item[] Justification: Training/inference settings and prompts are documented in Appendix~\ref{app:eval-config}, with statistical limitations in Appendix~\ref{app:seed_var}. The released functional testbench collection covers only a subset of the test cases (Appendix~\ref{app:functional_eval}), so complete functional-result reproduction is not established.
    \item[] Guidelines:
    \begin{itemize}
        \item The answer \answerNA{} means that the paper does not include experiments.
        \item If the paper includes experiments, a \answerNo{} answer to this question will not be perceived well by the reviewers: Making the paper reproducible is important, regardless of whether the code and data are provided or not.
        \item If the contribution is a dataset and\slash or model, the authors should describe the steps taken to make their results reproducible or verifiable.
        \item Depending on the contribution, reproducibility can be accomplished in various ways. For example, if the contribution is a novel architecture, describing the architecture fully might suffice, or if the contribution is a specific model and empirical evaluation, it may be necessary to either make it possible for others to replicate the model with the same dataset, or provide access to the model. In general. releasing code and data is often one good way to accomplish this, but reproducibility can also be provided via detailed instructions for how to replicate the results, access to a hosted model (e.g., in the case of a large language model), releasing of a model checkpoint, or other means that are appropriate to the research performed.
        \item While NeurIPS does not require releasing code, the conference does require all submissions to provide some reasonable avenue for reproducibility, which may depend on the nature of the contribution. For example
        \begin{enumerate}
            \item If the contribution is primarily a new algorithm, the paper should make it clear how to reproduce that algorithm.
            \item If the contribution is primarily a new model architecture, the paper should describe the architecture clearly and fully.
            \item If the contribution is a new model (e.g., a large language model), then there should either be a way to access this model for reproducing the results or a way to reproduce the model (e.g., with an open-source dataset or instructions for how to construct the dataset).
            \item We recognize that reproducibility may be tricky in some cases, in which case authors are welcome to describe the particular way they provide for reproducibility. In the case of closed-source models, it may be that access to the model is limited in some way (e.g., to registered users), but it should be possible for other researchers to have some path to reproducing or verifying the results.
        \end{enumerate}
    \end{itemize}

\item {\bf Open access to data and code}
    \item[] Question: Does the paper provide open access to the data and code, with sufficient instructions to faithfully reproduce the main experimental results, as described in supplemental material?
    \item[] Answer: \answerYes{}
    \item[] Justification: The benchmark is publicly available at \url{https://huggingface.co/datasets/Zanehpn/MHRC-Bench}, and training/evaluation code at \url{https://github.com/zanehpn/MHRC-Bench}. The data access guide documents the available functional testbenches and their partial coverage.
    \item[] Guidelines:
    \begin{itemize}
        \item The answer \answerNA{} means that paper does not include experiments requiring code.
        \item Please see the NeurIPS code and data submission guidelines (\url{https://neurips.cc/public/guides/CodeSubmissionPolicy}) for more details.
        \item While we encourage the release of code and data, we understand that this might not be possible, so \answerNo{} is an acceptable answer. Papers cannot be rejected simply for not including code, unless this is central to the contribution (e.g., for a new open-source benchmark).
        \item The instructions should contain the exact command and environment needed to run to reproduce the results. See the NeurIPS code and data submission guidelines (\url{https://neurips.cc/public/guides/CodeSubmissionPolicy}) for more details.
        \item The authors should provide instructions on data access and preparation, including how to access the raw data, preprocessed data, intermediate data, and generated data, etc.
        \item The authors should provide scripts to reproduce all experimental results for the new proposed method and baselines. If only a subset of experiments are reproducible, they should state which ones are omitted from the script and why.
        \item At submission time, to preserve anonymity, the authors should release anonymized versions (if applicable).
        \item Providing as much information as possible in supplemental material (appended to the paper) is recommended, but including URLs to data and code is permitted.
    \end{itemize}

\item {\bf Experimental setting/details}
    \item[] Question: Does the paper specify all the training and test details (e.g., data splits, hyperparameters, how they were chosen, type of optimizer) necessary to understand the results?
    \item[] Answer: \answerYes{}
    \item[] Justification: Data splits are described in the Dataset section and Appendix~``Dataset Splits''; LoRA fine-tuning hyperparameters (rank, scaling, learning rate, optimizer, schedule, epochs, context window) are listed in Appendix~\ref{app:eval-config}; prompt formats and FIM token usage are described both in the Experiments section and Appendix~\ref{app:eval-config}.
    \item[] Guidelines:
    \begin{itemize}
        \item The answer \answerNA{} means that the paper does not include experiments.
        \item The experimental setting should be presented in the core of the paper to a level of detail that is necessary to appreciate the results and make sense of them.
        \item The full details can be provided either with the code, in appendix, or as supplemental material.
    \end{itemize}

\item {\bf Experiment statistical significance}
    \item[] Question: Does the paper report error bars suitably and correctly defined or other appropriate information about the statistical significance of the experiments?
    \item[] Answer: \answerNo{}
    \item[] Justification: We report point estimates of EM, ES, and functional pass rates, without confidence intervals, significance tests, or statistics across independent fine-tuning seeds. Appendix~\ref{app:seed_var} describes these statistical limitations.
    \item[] Guidelines:
    \begin{itemize}
        \item The answer \answerNA{} means that the paper does not include experiments.
        \item The authors should answer \answerYes{} if the results are accompanied by error bars, confidence intervals, or statistical significance tests, at least for the experiments that support the main claims of the paper.
        \item The factors of variability that the error bars are capturing should be clearly stated (for example, train/test split, initialization, random drawing of some parameter, or overall run with given experimental conditions).
        \item The method for calculating the error bars should be explained (closed form formula, call to a library function, bootstrap, etc.)
        \item The assumptions made should be given (e.g., Normally distributed errors).
        \item It should be clear whether the error bar is the standard deviation or the standard error of the mean.
        \item It is OK to report 1-sigma error bars, but one should state it. The authors should preferably report a 2-sigma error bar than state that they have a 96\% CI, if the hypothesis of Normality of errors is not verified.
        \item For asymmetric distributions, the authors should be careful not to show in tables or figures symmetric error bars that would yield results that are out of range (e.g., negative error rates).
        \item If error bars are reported in tables or plots, the authors should explain in the text how they were calculated and reference the corresponding figures or tables in the text.
    \end{itemize}

\item {\bf Experiments compute resources}
    \item[] Question: For each experiment, does the paper provide sufficient information on the computer resources (type of compute workers, memory, time of execution) needed to reproduce the experiments?
    \item[] Answer: \answerNo{}
    \item[] Justification: The paper specifies all algorithmic and software-level training/inference hyperparameters in Appendix~\ref{app:eval-config}, but does not currently report the specific GPU type, number of workers, or wall-clock training/inference time.
    \item[] Guidelines:
    \begin{itemize}
        \item The answer \answerNA{} means that the paper does not include experiments.
        \item The paper should indicate the type of compute workers CPU or GPU, internal cluster, or cloud provider, including relevant memory and storage.
        \item The paper should provide the amount of compute required for each of the individual experimental runs as well as estimate the total compute.
        \item The paper should disclose whether the full research project required more compute than the experiments reported in the paper (e.g., preliminary or failed experiments that didn't make it into the paper).
    \end{itemize}

\item {\bf Code of ethics}
    \item[] Question: Does the research conducted in the paper conform, in every respect, with the NeurIPS Code of Ethics \url{https://neurips.cc/public/EthicsGuidelines}?
    \item[] Answer: \answerYes{}
    \item[] Justification: We have reviewed the NeurIPS Code of Ethics. The benchmark uses public GitHub source code and involves no human-subject study. Upstream attribution and license notices are retained; public names and email addresses may remain in source headers.
    \item[] Guidelines:
    \begin{itemize}
        \item The answer \answerNA{} means that the authors have not reviewed the NeurIPS Code of Ethics.
        \item If the authors answer \answerNo, they should explain the special circumstances that require a deviation from the Code of Ethics.
        \item The authors should make sure to preserve anonymity (e.g., if there is a special consideration due to laws or regulations in their jurisdiction).
    \end{itemize}

\item {\bf Broader impacts}
    \item[] Question: Does the paper discuss both potential positive societal impacts and negative societal impacts of the work performed?
    \item[] Answer: \answerNA{}
    \item[] Justification: The contribution is an evaluation benchmark and analysis for hardware code completion. We do not foresee a direct path to negative societal impacts beyond the general considerations applicable to all code-completion research; positive impacts include enabling reproducible measurement of LLM capability on hardware design tasks.
    \item[] Guidelines:
    \begin{itemize}
        \item The answer \answerNA{} means that there is no societal impact of the work performed.
        \item If the authors answer \answerNA{} or \answerNo, they should explain why their work has no societal impact or why the paper does not address societal impact.
        \item Examples of negative societal impacts include potential malicious or unintended uses (e.g., disinformation, generating fake profiles, surveillance), fairness considerations (e.g., deployment of technologies that could make decisions that unfairly impact specific groups), privacy considerations, and security considerations.
        \item The conference expects that many papers will be foundational research and not tied to particular applications, let alone deployments. However, if there is a direct path to any negative applications, the authors should point it out. For example, it is legitimate to point out that an improvement in the quality of generative models could be used to generate Deepfakes for disinformation. On the other hand, it is not needed to point out that a generic algorithm for optimizing neural networks could enable people to train models that generate Deepfakes faster.
        \item The authors should consider possible harms that could arise when the technology is being used as intended and functioning correctly, harms that could arise when the technology is being used as intended but gives incorrect results, and harms following from (intentional or unintentional) misuse of the technology.
        \item If there are negative societal impacts, the authors could also discuss possible mitigation strategies (e.g., gated release of models, providing defenses in addition to attacks, mechanisms for monitoring misuse, mechanisms to monitor how a system learns from feedback over time, improving the efficiency and accessibility of ML).
    \end{itemize}

\item {\bf Safeguards}
    \item[] Question: Does the paper describe safeguards that have been put in place for responsible release of data or models that have a high risk for misuse (e.g., pre-trained language models, image generators, or scraped datasets)?
    \item[] Answer: \answerNA{}
    \item[] Justification: The released artifacts comprise an HDL code-completion benchmark, training/evaluation code, and available testbench resources. They do not pose a high risk for misuse such as generating disinformation, malware, or sensitive imagery.
    \item[] Guidelines:
    \begin{itemize}
        \item The answer \answerNA{} means that the paper poses no such risks.
        \item Released models that have a high risk for misuse or dual-use should be released with necessary safeguards to allow for controlled use of the model, for example by requiring that users adhere to usage guidelines or restrictions to access the model or implementing safety filters.
        \item Datasets that have been scraped from the Internet could pose safety risks. The authors should describe how they avoided releasing unsafe images.
        \item We recognize that providing effective safeguards is challenging, and many papers do not require this, but we encourage authors to take this into account and make a best faith effort.
    \end{itemize}

\item {\bf Licenses for existing assets}
    \item[] Question: Are the creators or original owners of assets (e.g., code, data, models), used in the paper, properly credited and are the license and terms of use explicitly mentioned and properly respected?
    \item[] Answer: \answerNo{}
    \item[] Justification: Source code retains its upstream license terms, including permissive and copyleft licenses. The released repository inventory preserves located notices and records unresolved cases; complete license verification is not established. Author-created code, annotations, metadata, and documentation use MIT. Evaluated models and retrieval methods are cited where introduced.
    \item[] Guidelines:
    \begin{itemize}
        \item The answer \answerNA{} means that the paper does not use existing assets.
        \item The authors should cite the original paper that produced the code package or dataset.
        \item The authors should state which version of the asset is used and, if possible, include a URL.
        \item The name of the license (e.g., CC-BY 4.0) should be included for each asset.
        \item For scraped data from a particular source (e.g., website), the copyright and terms of service of that source should be provided.
        \item If assets are released, the license, copyright information, and terms of use in the package should be provided. For popular datasets, \url{paperswithcode.com/datasets} has curated licenses for some datasets. Their licensing guide can help determine the license of a dataset.
        \item For existing datasets that are re-packaged, both the original license and the license of the derived asset (if it has changed) should be provided.
        \item If this information is not available online, the authors are encouraged to reach out to the asset's creators.
    \end{itemize}

\item {\bf New assets}
    \item[] Question: Are new assets introduced in the paper well documented and is the documentation provided alongside the assets?
    \item[] Answer: \answerYes{}
    \item[] Justification: MHRC-Bench-Eval and MHRC-Bench-Train are documented in the Dataset section (collection, filtering, leakage check, and annotation procedure) and Appendix~\ref{app:eval-config} (evaluation protocol, hyperparameters, prompts). The construction rules and per-language semantic categories are fully specified in the appendix.
    \item[] Guidelines:
    \begin{itemize}
        \item The answer \answerNA{} means that the paper does not release new assets.
        \item Researchers should communicate the details of the dataset\slash code\slash model as part of their submissions via structured templates. This includes details about training, license, limitations, etc.
        \item The paper should discuss whether and how consent was obtained from people whose asset is used.
        \item At submission time, remember to anonymize your assets (if applicable). You can either create an anonymized URL or include an anonymized zip file.
    \end{itemize}

\item {\bf Crowdsourcing and research with human subjects}
    \item[] Question: For crowdsourcing experiments and research with human subjects, does the paper include the full text of instructions given to participants and screenshots, if applicable, as well as details about compensation (if any)?
    \item[] Answer: \answerNA{}
    \item[] Justification: The paper does not involve any crowdsourcing or research with human subjects. All data is sourced from public open-source repositories.
    \item[] Guidelines:
    \begin{itemize}
        \item The answer \answerNA{} means that the paper does not involve crowdsourcing nor research with human subjects.
        \item Including this information in the supplemental material is fine, but if the main contribution of the paper involves human subjects, then as much detail as possible should be included in the main paper.
        \item According to the NeurIPS Code of Ethics, workers involved in data collection, curation, or other labor should be paid at least the minimum wage in the country of the data collector.
    \end{itemize}

\item {\bf Institutional review board (IRB) approvals or equivalent for research with human subjects}
    \item[] Question: Does the paper describe potential risks incurred by study participants, whether such risks were disclosed to the subjects, and whether Institutional Review Board (IRB) approvals (or an equivalent approval/review based on the requirements of your country or institution) were obtained?
    \item[] Answer: \answerNA{}
    \item[] Justification: The research does not involve human subjects, so IRB approval is not required.
    \item[] Guidelines:
    \begin{itemize}
        \item The answer \answerNA{} means that the paper does not involve crowdsourcing nor research with human subjects.
        \item Depending on the country in which research is conducted, IRB approval (or equivalent) may be required for any human subjects research. If you obtained IRB approval, you should clearly state this in the paper.
        \item We recognize that the procedures for this may vary significantly between institutions and locations, and we expect authors to adhere to the NeurIPS Code of Ethics and the guidelines for their institution.
        \item For initial submissions, do not include any information that would break anonymity (if applicable), such as the institution conducting the review.
    \end{itemize}

\item {\bf Declaration of LLM usage}
    \item[] Question: Does the paper describe the usage of LLMs if it is an important, original, or non-standard component of the core methods in this research? Note that if the LLM is used only for writing, editing, or formatting purposes and does \emph{not} impact the core methodology, scientific rigor, or originality of the research, declaration is not required.
    \item[] Answer: \answerNA{}
    \item[] Justification: LLMs are the subject under evaluation in this paper, not a tool used in our core methodology. They are not used as a non-standard component of method development; the benchmark construction (data collection, CST parsing, expert-defined classification rules, leakage check) does not depend on LLM-generated content.
    \item[] Guidelines:
    \begin{itemize}
        \item The answer \answerNA{} means that the core method development in this research does not involve LLMs as any important, original, or non-standard components.
        \item Please refer to our LLM policy in the NeurIPS handbook for what should or should not be described.
    \end{itemize}

\end{enumerate}

\end{document}